\documentclass[hidelinks, print]{thesis}
\pdfoutput=1
\author{Rohit Kishan Ray}
\authorId{16PH91F02}
\supervisor{Prof. Sonjoy Majumder}
\department{Department of Physics}
\institute{Indian Institute of Technology Kharagpur}
\instituteShort{IIT Kharagpur}
\institutelogo{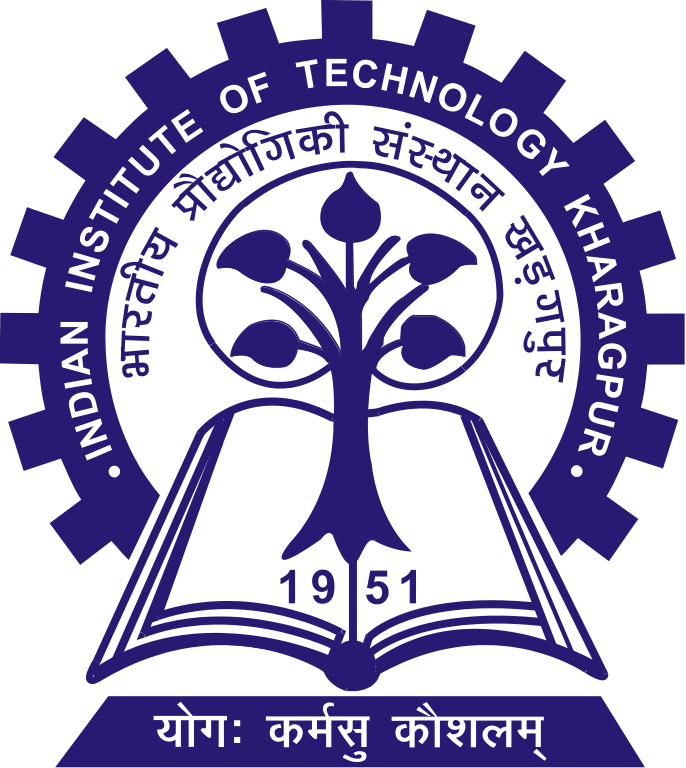}
\degree{Doctor of Philosophy}
\degreemonthyear{May}{2023}

\stackMath
\newcommand{\stretchedhat}[1]{%
    \savestack{\tmpbox}{\stretchto{%
            \scaleto{%
                \scalerel*[\widthof{\ensuremath{#1}}]{\kern.1pt\mathchar"0362\kern.1pt}%
                {\rule{0ex}{\textheight}}
            }{\textheight}%
        }{2.4ex}}%
    \stackon[-6.9pt]{#1}{\tmpbox}%
}
\newcommand{\stretchedtilde}[1]{%
    \savestack{\tmpbox}{\stretchto{%
            \scaleto{%
                \scalerel*[\widthof{\ensuremath{#1}}]{\kern.1pt\mathchar"307E\kern.1pt}%
                {\rule{0ex}{\textheight}}
            }{\textheight}%
        }{2.4ex}}%
    \stackon[-6.9pt]{#1}{\tmpbox}%
}
\newtheorem{theorem}{Theorem}[section]

\theoremstyle{definition}
\newtheorem{defRKR}{Definition}[section]
\newcommand{\tbar}[1]{\accentset{\rule{.5em}{.07em}}{#1}}
\newcommand{\itbar}[1]{\skew{2}{\tbar}{#1}}

\newcommand{\half}{\frac{1}{2}}

\newcommand{\kae}[1]{ \left|#1 \right)}
\newcommand{\bae}[1]{\left(#1\right| }
\newcommand{\inprc}[2]{\left(#1\mid#2\right)}

\newcommand{\seaexp}[3]{\left(#1\mid#2\mid#3\right)}

\newcommand{\imi}{\mathrm{i}}
\newcommand{\qexp}[3]{\vphantom{#2}\left\langle{#1}\middle\vert\smash{#2}\middle\vert{#3}\right\rangle}
\newcommand{\vstSpdv}[2]{\dfrac{\delta #1}{\delta #2}}
\newcommand{\vSpdv}[2]{\delta #1 /\delta #2}
\newcommand{\PTr}[2]{\text{tr}_{#1}\left(#2\right)}
\newcommand{\hamiltonian}{\mathit{H}}
\makeatletter
\let\origref\ref
\renewcommand{\eqref}[1]{\unskip\;\textup{\tagform@{\origref{#1}}}}
\makeatother
\newcommand{\Hil}{{\mathcal H}}
\newcommand{\J}{{J\vphantom{\overline{J}}}}
\newcommand{\Jbar}{{\overline{J}}}
\newcommand{\Boltz}{ k_{\rm\scriptscriptstyle B}}

\DeclareAcronym{SEA}{
    short = SEA ,
    long = steepest entropy ascent,
    tag = A
}
\DeclareAcronym{QM}{
    short = QM ,
    long = quantum mechanics,
    tag = A
}
\DeclareAcronym{EoM}{
    short = EoM ,
    long = equation of motion ,
    tag = A
}
\DeclareAcronym{BSEA}{
    short = BSEA ,
    long = Beretta SEA EoM,
    tag = A
}
\DeclareAcronym{BCSEA}{
    short = BCSEA ,
    long = Beretta composite SEA EoM,
    tag = A
}
\DeclareAcronym{bec}{
    short = BEC ,
    long = Bose-Einstein condensate,
    tag = A
}
\DeclareAcronym{QW}{
    short = QW ,
    long = quantum walk,
    tag = A
}
\DeclareAcronym{DTQW}{
    short = DTQW ,
    long = discrete-time quantum walk,
    tag = A
}
\DeclareAcronym{CTQW}{
    short = CTQW ,
    long = continuous-time quantum walk,
    tag = A
}
\DeclareAcronym{EPR}{
    short = EPR ,
    long = Einstein-Podolsky-Rosen,
    tag = A
}
\DeclareAcronym{GPB}{
    short = GPB ,
    long = Gian Paolo Beretta,
    tag = A
}
\DeclareAcronym{beta}{
    short = $\beta_i$,
    long = Lagrange's multiplier,
    tag = symb
}
\DeclareAcronym{tau}{
    short = $\tau$,
    long = system relaxation time,
    tag = symb
}
\DeclareAcronym{kB}{
    short = $k_\text{B}$,
    long = Boltzmann constant,
    tag = symb
}
\DeclareAcronym{hbar}{
    short = $\hbar$,
    long = Plank constant,
    tag = symb
}
\DeclareAcronym{rho}{
    short = $\rho$,
    long = Density matrix,
    tag = symb
}

\alltitlesformat
\captionsformat

\thesistitle{Finite-Dimensional Quantum Systems under the Fourth Law of Thermodynamics}

\begin{document}
\setcounter{page}{2}
\pagenumbering{roman}

\makethesistitlepage

\begin{dedication}
    \vspace*{2in}
    \begin{center}
        \centering
        \includegraphics[width=0.65\textwidth]{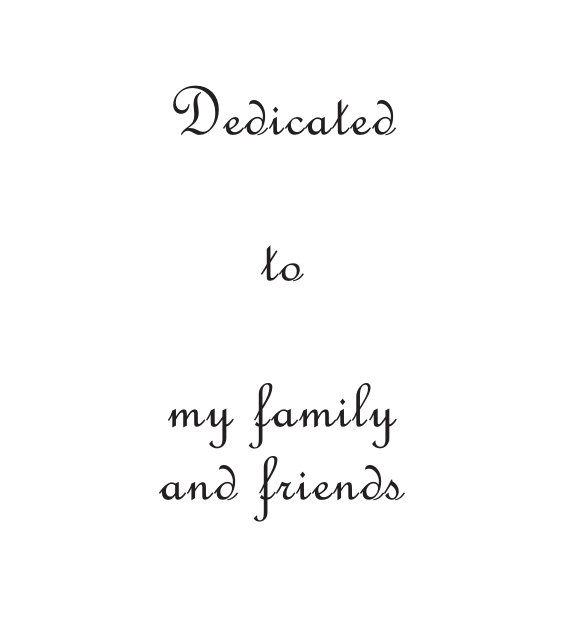}
    \end{center}

\end{dedication}

\begin{approval}
    Certified that the thesis entitled ``\textbf{\titleofthesis}", submitted by \textbf{\authorName} to the Indian Institute of Technology Kharagpur, for the award of the degree Doctor of Philosophy has been accepted by the external examiners and that the student has successfully defended the thesis in the viva-voce examination held today.
\end{approval}
\begin{certificate}
    This is to certify that the thesis entitled, ``\textbf{\titleofthesis}" submitted by \authorName~(\authorId) to the Indian Institute of Technology Kharagpur, is a record of bonafide research work carried under my supervision and is worthy of consideration for the award of the Doctor of Philosophy of the Institute.
\end{certificate}
\begin{declaration}
I certify that\\

\begin{tabular}{l p{13cm}}
a. & the work contained in the thesis is original and has been done by myself under the general supervision of my supervisor;\\
b & the work has not been submitted to any other Institute for any degree or diploma;\\
c. & I have followed the guidelines provided by the Institute in preparing the thesis;\\
d. & I have conformed to ethical norms and guidelines while writing the thesis and;\\
e. & whenever I have used materials (data, models, figures, and text) from other sources, I have given due credit to them by citing them in the text of the thesis, giving their details in the references, and taking permission from the copyright owners of the sources, whenever necessary.\\
\end{tabular}
\end{declaration}
\begin{acknowledgment}
    The journey to Ph.D. is hardly simple. Although the degree is awarded to one person, it is really not the case. The road to Ph.D. is befitted with numerous cameos and unsolicited help. A hand of support extended from an unbeknownst person, and sometimes the cozy comforts from the ones we have known through heart. This is where I would like to acknowledge all those entities, without whom this journey would have been different, and in some cases not even possible.\par

    First and foremost, I express my gratitude to my supervisor, Prof. Sonjoy Majumder. Not because it is customary, but because he has a unique relationship with me, unlike the usual supervisor-student dynamics. He is Sonjoyda (da for elder brother) to me. He took me in as a doctoral student when others were rejecting me on various grounds. He showed exceptional trust in me by allowing me to have free reign over the choice of problem, and selection of topic for my Ph.D.. Freedom is a different type of responsibility, and allowing one’s subjects freedom requires shedding of one’s ego. Sonjoyda provided me with the necessary emotional and time-upon-time financial support so that I can seek my own path and walk in it. In this context, I must thank Dr. Aritra Banerjee who introduced me to Sonjoyda.\par

    While Sonjoyda bestowed my freedom, it was Bibhasda (Prof. Bibhas Adhikary) who inspired me to study quantum information and related topics which eventually became subjects of my research. Bibhasda has always been there to answer my questions and address my doubts. In the same breath, I acknowledge Prof. Alok Pan of IIT Hyderabad. Alokda has taught me the fundamental aspects of the quantum foundation. He has also been, more or less like Bibhasda, my go-to person for any doubts. We have had discussions over many topics, and he has always inspired me with his actions and posed intriguing questions to keep on challenging me. \par

    Finally, if I do not acknowledge Prof. Gian Paolo Beretta of the University of Brescia, Italy, it would be a sin by omission. Prof. Beretta has taught me the nuances of SEA. He has mentored me. As a collaborator, he has worked with me (sometimes for long hours, and through even longer emails), in tune with the rush of deadlines despite having health and other obligations. He has reached out to others for me, and he has been an inspirational figure who I have looked up to. Without his active guidance, this thesis may not have been a reality. \par

    I thank my Doctoral Scrutiny Committee (DSC) members Prof. Shivakiran B N Bhaktha, Prof. Partha Roy Chaudhuri, Prof. Sabyashachi Mishra, and chairman Prof. Sudhansu S. Mandal for their help and inputs whenever required.\par

    As guidance is essential to Ph.D., so is proper funding. I acknowledge the Department of Science and Technology (DST), Govt. of India for awarding me the INSPIRE-Fellowship, because of which I got the privilege of doing a Ph.D. from IIT Kharagpur. This fellowship has been my lifeline throughout this program. Although, just like the monsoon in India, the flow of stipends was not a timely affair. It was my friends who kept me afloat during those dry patches. I would like to thank Arnab Priya, Bankim, Sujoyda, Meghadeepa, Supriyo, Nilotpal aka Nilu, Analda, Saptarshi, Aman, Singla, Aritra aka Gol, Sonjoyda, Soumyadeep, Arpana, Hari, Himanshi for being my go-to financiers whenever I felt cash-crunch.\par

    Once finances and academic needs are satisfied, a person realizes Ph.D. becomes intolerable and insufferable imprisonment without the constant support, encouragement, and empathy of friends. My school friends who I am friends with for more than two decades have been my pillars of support on the home front. Arani (Rubai), Anirban, Dhritiman, Saptarshi, Titas, Prosenjit, Sayandip (Bapin), Sangita, Indranil, Sumit have been my support through thick and thin. In the same spirit Manprit (Mini), Trina, and Priyanka have also been there. \par

    From my college days, I have Arnab Priya, Anirban (Mampu), Supriyo, Gol, Nilu, Koushik and Kingshuk who has always been part of my advisory committee outside of Kharagpur. Augniva and Pallabi from my Masters days have formed the \emph{Three Idiots} group (including me, of course), which has been my safe place to rant and talk and hang out. I remember Sujoyda, Rashmi, Shatavisha, and Silpa in the same spirit as they have been blessings in my life.\par

    These friends outside campus have been my support away from home, it was the buggers from campus who made me call IIT Kharagpur my home for more than six years. My lab seniors who befriended me: Analda, Arghyada, Pradeepda, Narenda, Koushik, Dilip; and my lab juniors who I befriended, Anirban, Soumyadeep, Shainee, Tanima, Arpana, Hari, Sabyasachi, Harshdeep made my lab a comfortable space to work year after years. They empathized with my struggle, and they helped me whenever they could. We had fun together, we celebrated together, and we shared more than just office space and computational resources. I have learned, through them, if one does not have a healthy relationship with one’s lab-mates, one may find oneself in a lot more difficulty in finishing the arduous journey of Ph.D. I must specially thank Soumyadeep, Anirban, and Arpana for they have played a very crucial supportive role in my journey. Hari for being the perfect little kid who is always enthusiastic. Sincere thanks to Arpana for being my go-to Grammarly. \par
    I am a very fortunate person that I have found good friends in my hostel also. Friends like Aritra, Jagan, Nilesh, Hari Priyo, Mahesh, Eeshan have been my buddies who I have hung around with, got intoxicated with, partied with and shared bonds of brotherhood. I miss my fun days with Sayan, Roop, Jit, Raj, Nilesh. My buds from the department, Anuj, Debdutta, Abhijit, Debjyoti, Anang, Anurag, Abinash, Tara, Sayan, Sudipto, Snehashish, Dibyenduda, Arnabda, Ambaresh, Sambo, Abhirup made me feel wanted through our nods of acknowledgment, random discussions on the corridors, organizing different departmental activities. A list like this is incomplete if the wonderful women of physics are not mentioned. My heart goes to Urmi, my best friend and my favorite TA. Sahana, Abyaya, and Anasuya of the \emph{Bawal} group have been with me from day one. I can’t imagine navigating the first couple of years without them. Manobina has been a colleague during my research scholar representative days, and she has been amazing.\par

    I would like to devote at least a page to some of the following individuals and express my deepest gratitude to them, but they must forgive me as I run out of space. My first and foremost declaration of gratitude goes to Subrata. My lab-mate since the very first day, my brother in Ph.D.. He and I began this journey together, shared our darkness with each other over pints of beer. We have pushed each other through tough times, and I must express what I have always felt, `had it not been for Subrata, I would not have finished my Ph.D.'. \par

    Now cometh the turn of the \emph{Bund} group. Our gang comprising Aman, Guru, Gaurav, Himanshi, Singla, Zenia and I was our little family. We had a great time together. They helped me become who I am in more than one dimension. While writing this piece I am feeling overwhelmed by the sheer volume of memories generated over numerous trips, eating outs, partying, arguments and other interactions of a similar kind. My calls for help to Himanshi, Aman, or Sinlga has been answered with love and care that a family provides, and I am immensely grateful for that. \par

    Upon returning to the campus post-COVID-lockdown, and finding most of the known faces having departed, I felt an indescribable sense of emptiness. I am grateful beyond mere words to the friends of the Biryani lovers group. Dwiti, Leena, Uddhav, Jun, Supriya, and then Loknath, Plawana, Nilotpal provided me the necessary support to glide through the final stages of my Ph.D.. A humongous thanks to Kalpana aka Kimi for being the elder sister I never knew I needed. She kept encouraging me through my difficult times, albeit from afar, but a phone call to her never went unanswered. In the same spirit, Ayan Bhinedeu has also become my elder brother, a relationship I will cherish beyond the years of my Ph.D.\par

    Partha Pratim Saikia, the man I did not think will be friends with me, has become now a part of me. It is because of him, I got a breather during my moments of stress. He and I along with Loknath cooked delicacies on Sundays to get the necessary break from ever-difficult hostel food. It is just not possible to think of the last year and not see him, he covers the entire vision. \par

    Thinking of the last one and a half years, I finally convey my heartfelt gratitude to Jigyasa. This woman has been the key aspect of my life in the most stressful time of my career so far. Her therapy, her company, and her affection have steered me through the valley of negative reminiscence to the positive peaks of a hopeful tomorrow. I can never thank her enough.\par

    Last, but not the least, I thank the Department of Physics, and Sureshji. I am thankful for the continuous tea supply at the ECE canteen, due to Pintuda, Anirban, and Bijoyda. And for the morning teas at Avijitda I am grateful. The delicious homely food of Chainadi kept me nourished. I thank the folks at the accounts section students unit, Ullas Sir, Dhruboda, people who constituted my day-to-day life here at IIT Kharagpur. I must thank the Paramshakti Supercomputing facility, StackExchange, OpenAI, SciHub, Libgen and other such resources.\par

    Finally, a note of gratitude to my mother, who fought against ignorant relatives, and struggled through immeasurable hardship, to see me be anointed as a Dr. My father deserves my thanks, for he kept working hard, toil constantly in sickness and in health so that the family is supported while I pursued my dreams. And my lovely sister, who has shown compassion and support beyond rationality. She is my strongest advocate, she has cheered for me the most and campaigned for me the hardest. I have had the fortune of having her as my sister, and this family, and this life full of such beautiful souls, and for that, I am grateful.\par

\end{acknowledgment}
\begin{publications}
    \begin{enumerate}
        \item \textbf{R. K. Ray}, \textit{Steepest entropy ascent solution for a continuous-time quantum walker}, \href{https://link.aps.org/doi/10.1103/PhysRevE.106.024115}{\color{blue} Phys. Rev. E 106, 024115 (2022)}.
              \vspace{0.3em}

        \item \textbf{R. K. Ray}, G. P. Beretta, \textit{No-Signaling in Steepest Entropy Ascent: A Nonlinear Non-local Non-equilibrium Quantum Dynamics of Composite Systems}, \href{https://arxiv.org/abs/2301.11548}{\color{blue}{arXiv:2301.11548v4}}.
    \end{enumerate}
\end{publications}

\begin{abstract}
    The ansatz of steepest entropy ascent (SEA) has been recently identified as the fourth law of thermodynamics. The law describes a system's evolution from an out-of-equilibrium state toward the globally unique stable equilibrium state of maximum entropy. The SEA ansatz sets the second law of thermodynamics as a foundation to merge mechanics and thermodynamics. We present a brief introduction to the fundamental tenets of the theory and provide the underlying principles contributing to formalism. SEA equation of motion is highly nonlinear; its exact analytical solutions are limited and available only for some very special cases. We have successfully developed an approximate analytical tool called the fixed Lagrange's multiplier (FLM) method to help us analytically solve the two-level and higher dimensional systems. \\
    Quantum walks are used as a universal model of computation. Using this model, we analyze a single component $N-$level system and apply our FLM scheme to solve the SEA equation of motion analytically. A comparison of the solution obtained using FLM, and the complete numerical solution is presented, and we notice strong agreement. Regions of maximum entropy production rate in agreement with the SEA have been identified.\\
    To extend the SEA analysis to simple composites involving two qubits, we need analytical roots and relevant results for the case of four-level Bloch vector formalism. We present a general framework for the characterization of $N-$level Bloch parametrization. We provide analytical roots for the $N=3$ level and completely parametrized roots for the $N=4$ level. We also provide a framework for finding an analytical trace of general operators in this representation.\\
    Lastly, we address the problem of no-signaling in a nonlinear quantum theory. It has been well established in the literature that a nonlinear theory of quantum mechanics allows for faster-than-light communication (signaling) between two noninteracting parts of a composite system. However, we show that SEA is built to respect no-signaling. We present the equation of motion for composite systems. We consider the cases of separable composites and nonseparable entangled/mixed composites in the form of Bell diagonal states. Our results confirm that the SEA is a valid theory involving nonlinear dynamics that respects no-signaling criteria and presents a fundamental approach to the problem of decoherence modeling for open and closed quantum systems.
    \par
    \vspace{2cm}
\end{abstract}

\generateAbbreviations
\generateContents
\generateListOfFigures

\doublespacing
\setcounter{page}{1}
\pagenumbering{arabic}


\chapter{Introduction}\label{ch:intro}
\blankpage
\newpage
\begin{figure}[H]
    \centering
    \captionsetup{justification=centering}
    \includegraphics[height=0.75\textheight, width=0.75\textwidth]{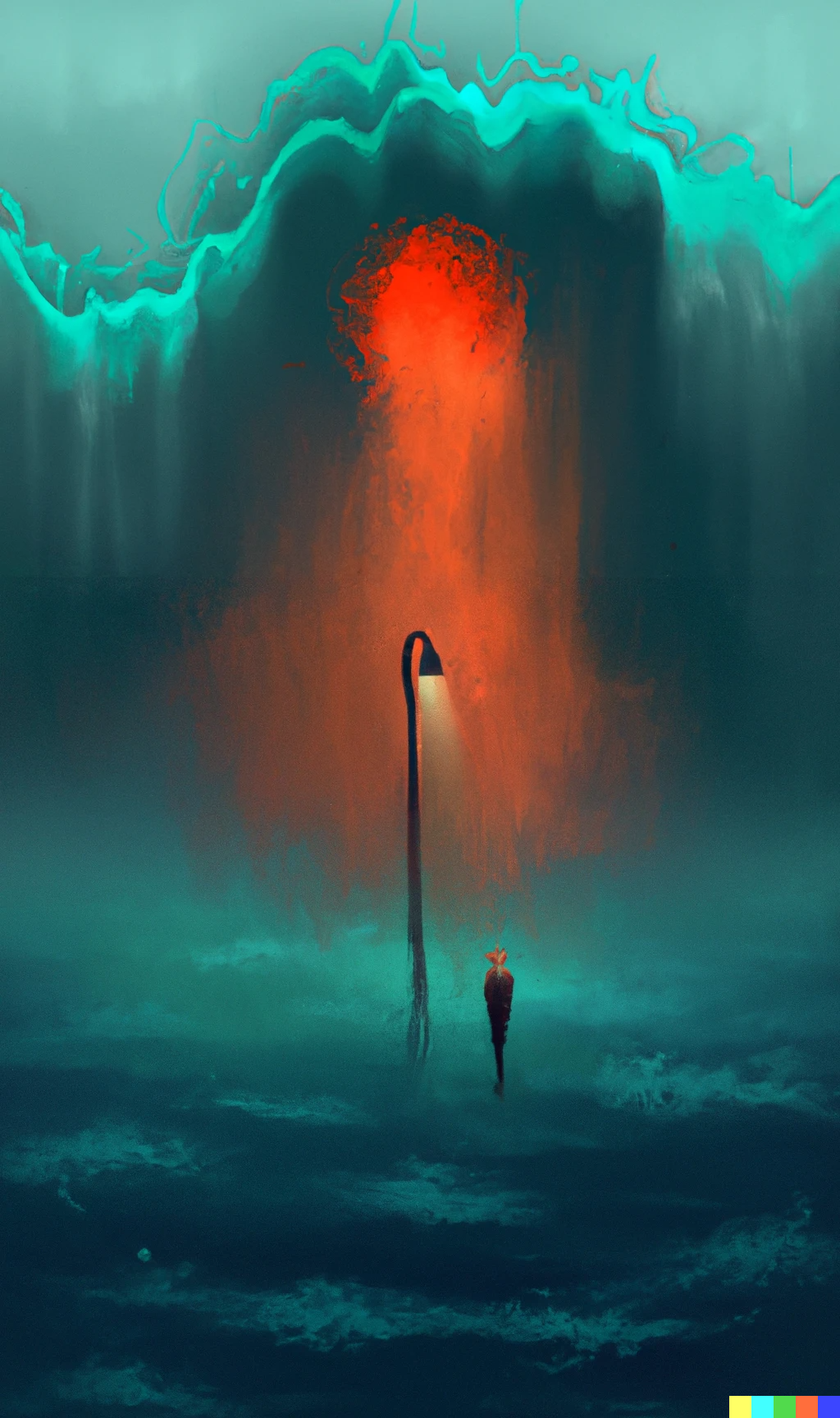}
    \caption*{\emph{I stood by the edge,}\\
        \emph{of the sea of knowledge}\\
        \emph{the dark of the unknown abyss,}\\
        \emph{was terrifying.}
    }
\end{figure}

\thispagestyle{empty}
\newpage
\blankpage
\lettrine[lines=3]{S}{ince} the introduction of \ac{QM} into the arena of physics, every new revelation of the theory has also produced new challenges to our classical intuition of the physical world. The pile of research works invested in understanding each of these complex ideas- be it wave-particle duality, entanglement, the meaning of state, the measurement problem, the interpretation of QM, or the philosophy of QM, has only grown higher and reached deeper. Yet, the subjects still elude a clear understanding (see Ref. \cite{turchi_2001_decoherence}). All of these are in the context of linear quantum mechanics. When we consider the case of nonlinear extensions of QM, we find ourselves in a neverending, ever-expanding canyon of articles, reports, journals, books, and blogs, which can easily overwhelm the noob of this field. This vast, and expansive knowledgebase can also fog one's field of vision to the pre-existing literature so much so that some theories become obscured by time despite being interesting and intriguing, while others become re-discovered and so on. One such theory of nonlinear extension of quantum mechanics involved seeking an attempt to merge mechanics and thermodynamics through setting entropy as a fundamental theory of nature via introducing a stability postulate in the framework of QM (\cite{hatsopoulos_1976_unified,hatsopoulos_1976_unifieda,hatsopoulos_1976_unifiedb,hatsopoulos_1976_unifiedc}). Consequently the stage was set for merging mechanics and thermodynamics which gave birth to the quantum thermodynamics formalism \cite{beretta_1984_quantum,beretta_1985_entropy,beretta_1985_quantum}. This theory eventually was called the \ac{SEA} ansatz \cite{beretta_1987_steepest}. Albeit radical in nature, some of the principle ideas of this formalism have been overlooked by the larger physics community and the theory remained dormant for quite some time. Eventually, a researcher from MIT re-discovered the results using some similar fundamental tenets of the SEA \cite{gheorghiu-svirschevski_2001_nonlinear,gheorghiu-svirschevski_2001_addendum}. This publication became the key moment that initiated a flurry of publication by the original author and began a new generation of research in SEA \cite{beretta_2005_general,beretta_2005_nonlinear,beretta_2006_nonlinear,beretta_2009_nonlinear,beretta_2010_maximum,beretta_2014_steepest,li_2014_atomisticlevel,vonspakovsky_2014_trends,cano-andrade_2015_steepestentropyascent,li_2016_generalized,li_2018_steepest,beretta_2019_time}. All of these works culminated into the confirmation that SEA can be called as the fourth law of thermodynamics \cite{beretta_2020_fourth}. \par

The goal of this thesis is to study the effect of the fourth law of thermodynamics on finite-dimensional single and composite quantum systems. Finite-dimensional quantum systems lie at the heart of modern quantum applications, and the fourth law of thermodynamics provides a fundamental basis for understanding decoherence in those systems. Before we proceed with our agenda, it is necessary that we retrace our path through the historical development of SEA. We can then discuss the current state of the literature regarding the application of decoherence modeling in the context of single and composite quantum systems. As SEA theory produces a strongly nonlinear equation of motion for the system under observation, certain physical and philosophical implications become difficult to look away from. We provide a backdrop of such development. Finally, we discuss how this thesis is structured.

\section{A brief history of SEA}\label{sec:intro_history}

The history of SEA begins with the debate regarding the status of the second law of thermodynamics as a fundamental law of nature. As opposed to the prominent notion of the time, Margenau, Hatsopulous, Keenan, Park, Gyftoplulous and like-minded physicists desired the second law to be as fundamental as the conservation of energy, and not merely of statistical nature \cite{hatsopoulos_1965_principles}. This claim was justified from a historical perspective, and also from the perspective of various stability criteria, and Gibbs's principle of general inertia \cite{hatsopoulos_1965_principles}. Later on, Park questioned the very foundation of the von Neumann formalism in Ref. \cite{park_1968_nature}. Here the character of the mixed state is pondered upon and claimed that the concept of a mixed state does not arise from a mixture of pure states, but rather has an interpretation of its own. In follow-up work, Park went on to prove an earlier version of the no-go theorem (arising from the linear structure of QM) \cite{park_1970_concept}. Using these results, Hatsopulous and Gyftopulous, in their series of publications, introduced the stability postulate in the realm of quantum mechanics, gave a new interpretation to the entropy functional, and introduced the concept of preparation contextuality through the terms `unambiguous preparation' \cite{hatsopoulos_1976_unified,hatsopoulos_1976_unifieda,hatsopoulos_1976_unifiedb,hatsopoulos_1976_unifiedc}. Park also discussed how a quantal theory compatible with thermodynamics will essentially be non-linear \cite{park_1983_knots}. Based on these works, in 1984, Beretta introduced a thermodynamically compatible nonlinear equation of motion for a single constituent of matter \cite{beretta_1984_quantuma}. In the subsequent year, he provided exact analytical results for a qubit under certain constraints \cite{beretta_1985_entropy}, followed by providing a general equation of motion under similar dynamics for a composite \cite{beretta_1985_quantuma}. Following this, he showed that a nonlinear evolution governed by entropy maximization is physically acceptable if we let go of the notion of equilibrium based on the Lyapunov concept only, as it is not sufficient for the existence of a globally stable equilibrium \cite{beretta_1986_theorem}. Beretta finally named this theory as the steepest entropy ascent (SEA) formalism \cite{beretta_1987_steepest}.

\section{The second age of SEA}\label{sec:intro_second_age}
The initial phase of SEA development was suppressed by a sheer lack of enthusiasm and discussion. In 2001, Gheorghiu-Svirchevski independently got the idea of maximization of entropy production subject to certain constraints \cite{gheorghiu-svirschevski_2001_nonlinear}. Soon after he realized that his theory is a variation of the SEA ansatz \cite{gheorghiu-svirschevski_2001_addendum}. However, this result created a new splash in the otherwise calm pool of SEA research. Beretta reinitiated the discourse with renewed vigor. He first showed how this formalism is compatible with thermodynamics \cite{beretta_2005_general}. Then it was shown that the non-linear equation of motion thus derived follows a set of necessary and sufficient conditions \cite{beretta_2005_nonlinear}. In Ref. \cite{beretta_2009_nonlinear} Beretta further formalized SEA, polished some arguments, introduced the ontological hypothesis, and explained the underlying geometric construction with vivid details \cite{beretta_2009_nonlinear}. In the work published in Ref. \cite{beretta_2010_maximum}, he further worked out the detailed structure of SEA for composite systems and in essence completed the formalism, suitable for application to modern problems of contemporary physics. This flurry of publications did not go unnoticed this time, and von Spakovsky picked up the mantle and started applying SEA to various scenarios. First, it was shown how SEA leads to typicality and the agreement is not merely pedantic \cite{vonspakovsky_2014_trends}. In the same year, Beretta showed that SEA has a structure similar to other non-linear models of studying nonequilibrium thermodynamics and presented a uniform notation to the problem \cite{beretta_2014_steepest}. This thesis will heavily rely on those standardized notations. G. Li and Spakovsky studied SEA for chemically reactive systems and performed atomistic modeling  \cite{li_2014_atomisticlevel}. Then they studied heat and mass diffusion in the far-from-equilibrium region using SEA \cite{li_2016_generalized}. In Ref. \cite{li_2016_steepestentropyascent} they introduced the concept of hypoequilibrium to define temperature for out-of-equilibrium systems, and in this regard, studied the effect of SEA on the relaxation process of isolated chemically reactive systems.\\
Meanwhile, Beretta has established the equivalence of the SEA formalism to the GENERIC formalism of intrinsic quantum thermodynamics \cite{montefusco_2015_essential}. A note on GENERIC. It is an abbreviation for the general equation for the non-equilibrium reversible-irreversible coupling. In this scheme, the thermodynamic evolution of quantum states is modeled through \emph{different levels of description} \cite{grmela_1997_dynamics,ottinger_1997_dynamics}. In this formalism, macroscopic dynamics is favored over the microscopic structure of the primitive level, thus setting the constraints as global invariants of motion. By showing equivalence with this formalism, Beretta showed the diversity of SEA and settled many doubts regarding its applicability. Cano-Andrade \emph{et. al.,} showed how SEA can be used to study two-electron composites, alongside showing better agreement with the experimental results regarding cavity-QED correlation studies compared to existing phenomenological modeling \cite{cano-andrade_2015_steepestentropyascent}. Beretta continued pushing the scope of SEA, and very recently showed that this theory can be called the fourth law of thermodynamics \cite{beretta_2020_fourth}.

\section{Decoherence study in quantum walks}\label{sec:intro_decoherence}
The SEA approach is the first principled take on decoherence. However, the paradigm of decoherence studies has been dominantly enriched by involving master equations and phenomenologically modeling decoherence. In this approach, Lindblad-type master equations are employed to study the relaxation of an out-of-equilibrium system \cite{beretta_2009_nonlinear,gisin_1995_relevant}. As a modeling tool for studying such phenomena, \ac{QW}, which is the quantum analog of the classical random walk, are used \cite{aharonov_1993_quantum,aharonov_1998_quantum,aharonov_2001_quantum}. They can be mainly categorized into \ac{DTQW}, and \ac{CTQW} (see Refs. \cite{kendon_2007_decoherence,manouchehri_2014_physical,fedichkin_2021_analysis} for review). The reasons for the use of QWs are versatile and widespread. They present a universal model for computation \cite{childs_2009_universal}, and are used for spatial search algorithms \cite{childs_2004_spatial}. QWs have also been applied to study and understand the nature of entanglement in many-body systems \cite{nayak_2000_quantum,omar_2006_quantum,pathak_2007_quantum,rohde_2011_multiwalker,rohde_2012_entanglement}. In this thesis, we are interested in modeling the dynamics of the single finite-level system.

Decoherence, being a fundamental aspect of reality, is studied in various contexts. QWs being universal in nature provide some elementary yet insightful testbeds for such studies. The first studies were done by Viv Kendon and group in Ref. \cite{kendon_2003_decoherence} where they showed that introducing little decoherence in QWs produces faster spread and rapid mixing. Their results were for DTQWs. In Ref. \cite{brun_2003_quantum} decoherence was observed for QWs driven by many coins. For the CTQW mixing and decoherence studies were performed by Fedichkin \emph{et. al.,} in Ref. \cite{fedichkin_2005_mixing}. Even now there is active research in understanding decoherence and mixing in the case of QWs via various phenomenological modeling \cite{chakraborty_2016_spatial,chakraborty_2020_analog,chakraborty_2020_finding,chakraborty_2020_how}.

Besides studying decoherence, thermodynamic studies on QWs are also active areas of research. Consider the work of Romanelli and colleagues in Refs. \cite{romanelli_2012_thermodynamic,romanelli_2014_thermodynamics}, where they have studied thermodynamic properties of QWs by varying the contribution due to interference factor. This interference factor was introduced by Romanelli \emph{et. al.,} in Ref. \cite{romanelli_2010_distribution} where they split the QW evolution into two parts- one consisting of a Markovian process, and the other being the above-mentioned interference term which induces unitary evolution. Recently CTQW in the presence of quadratic Hamiltonians has been studied \cite{candeloro_2020_continuoustime}, and the results somewhat agree with SEA results. However, it appears that this formalism may attract some unwanted non-physical effects should it be applied to composite walks.

\section{The Bloch representation}\label{sec:intro_bloch}
To study finite-dimensional quantum systems, a proper representation seems to be essential. The representation should be scalable, should help us exploit symmetries in the system, and may not hinder access to the underlying structures embedded in the description. The Bloch representation seems to be the just candidate in this case. Most of the success of the representation lies in the two-level case, where the available state space is a Reimann sphere in three dimensions \cite{nielsen_2000_quantum,jakobczyk_2001_geometrya}. The case of the general representation is not so straightforward either geometrically or algebraically. In 1971, Park and Band introduced the multipole expansion concept to the scheme of Bloch representation and showed a general formalism exists \cite{park_1971_general,band_1971_general}. Their formalism equated the Bloch representation to the angular momentum representation and used properties of SU($N$) algebra, although this result was way ahead of its time. After a long gap, as the demand of the times increased, a set of works in this arena began to surface. Byrd and Khaneja used the concept of coherence vector instead of Bloch vector to parametrize the density matrix, and consequently described the characterization of positivity of such matrices \cite{byrd_2003_characterization}. In the same year, Kimura described the structure of $N$-level systems using Bloch vectors and introduced some nice trace invariants in the foray \cite{kimura_2003_blocha}. A few years down the line, Boya and Dixit discussed the geometry of the $N-$level Bloch representation using Casimir invariants \cite{boya_2008_geometry}. In the same year, Bertlmann \emph{et. al.,} re-introduced Bloch representation for qudits using generalized Gell-Mann matrices \cite{bertlmann_2008_bloch}. A good review of the geometry and structure of general Bloch representation was given by Br\"{u}ning \cite{bruning_2012_parametrizations}.

\section{A problem of signaling}\label{sec:intro_signaling}
In the year 1989, the late Prof. Weinberg enquired whether QM is truly linear, and suggested some ideas for testing that non-linearity, if present \cite{weinberg_1989_precision,weinberg_1989_testing}. This was a very radical proposition at the time and immediately garnered attention. In an attempt to meet Weinberg's criteria, Gisin \cite{gisin_1990_weinberg} and Polchinski \cite{polchinski_1991_weinberg} showed that a non-linear Hamiltonian formalism leads to supraluminal communication (signaling) between two non-interacting subsystems of a composite, thus establishing something known as an \ac{EPR} telephone \cite{gisin_1990_weinberg}. This immediately violates causality and allows the EPR channel to be used as a resource for faster-than-light communication. Despite the negative result, many researchers devoted a significant amount of resources to finding a non-linear formalism of QM that will respect no-signaling. Gisin continued the search and established that non-linearity introduced via the stochastic formalism involving Lindblad-type master equations does respect no-signaling \cite{gisin_1995_relevant}. However, in this formalism, the mixed states can go to pure states and vice versa, implying there is a non-unital underlying process involved. Later, Ferrero \emph{et. al.} showed, that the only type of non-linearity that can be accommodated in the framework of no-signaling respecting irreversible quantum theories is the one involving non-linear temporal evolution \cite{ferrero_2004_nonlinear}. They also raised some pertinent philosophical questions by asking the validity of convex linear maps in this regard, and how the processes that involve mixed state to pure state evolution can coexist in a thermodynamically compatible framework. Citing this, Spakovsky claimed that SEA is also no-signaling but did not provide definitive proof \cite{vonspakovsky_2014_trends}. Beretta had also considered the no-signaling problem but did not conclusively give proof of it either \cite{beretta_2010_maximum,cano-andrade_2015_steepestentropyascent}. Very recently, Rembieli\'{n}ski and Caban showed that the minimal non-linearity that can be no-signaling must be convex quasilinear maps \cite{rembielinski_2020_nonlinear,rembielinski_2021_nonlinear}.

\section{Motivation for the thesis}\label{sec:intro_motiv}
The previous sections provided a gist of the backdrop upon which this thesis is being to be presented. As we saw, since the development of SEA, despite having a rich structure, disruptive physical intuition in its conception, and myriad applications, SEA has not gained the kind of attention it is due. In our opinion, one principal reason could be the apparent non-trivial and difficult-to-solve equation of motion that SEA produces. Alongside this, the limited cases (one in this case) of exact analytical results make this theory difficult to grasp. Numerical results exist for different scenarios, but not having an analytical structure to it produces a sense of ambiguity regarding what physical principles are at play. Motivated by this, we present this thesis with the following objectives:
\begin{enumerate}
    \item one must try to solve more cases that produce analytical or at least approximately analytical results.
    \item That the solution thus presented should be scalable, so that overall dynamical features can be easily identified, thus making the theory easier.
    \item To present definitive proof that SEA is a truly non-linear extension of QM that does not signal.
\end{enumerate}
As we can see, objectives 1 and 2 can be used to simplify some aspects of the SEA without losing its essential non-linearity, thus opening up the scope for analytical study and semi-analytical approximation in various cases involving finite-level quantum systems. We aim to introduce an approximate method to solve various finite-level systems. In the discussion of sec \ref{sec:intro_decoherence} we have seen how the QW case can be used as a good model to study decoherence for $N-$level systems. We also noticed a research gap that exists as there has not been any study of SEA on QWs. Motivated by this, we desire to apply SEA on QW using our approximation tool to fulfill objectives 1 and 2. Objective 3 aims to grant SEA the status of a valid non-linear theory of QM that precludes signaling. This result will open philosophical implications hitherto undiscussed. Furthermore, the apparent skepticism regarding the validity of SEA in the physics community at large can also be addressed and probably erased.

\section{The outline of the thesis}\label{sec:intro_outro}
This thesis is outlined as follows. In chapter \ref{ch:litsurv}, we explain the SEA formalism as it appears in the literature. We elaborate on the underlying principle of stability, rephrase the second law accordingly, and show how the entropy emerges as a functinoal that can be used to characterize stability. Thereafter, we present the variational approach via which the SEA equation of motion (EoM) is proposed by Beretta. Hereafter will be denoted as Beretta SEA (BSEA) EoM. We discuss the geometric structure upon which SEA motion is embedded and conclude with a discussion on the system relaxation time.

In chapter \ref{ch:qubitSEA}, we present the Beretta-given exact solution for a two-level system. Thence we introduce our approximate analytical formalism. We show how the intrinsic non-linearity of BSEA is reduced while retaining the basic features of the evolution. We present the solution of qubit evolution using our approximate scheme and comment on the nature of the same. Thus, we partially address objectives 1. and 2.

The chapter \ref{ch:qwalker}, focuses on solving the CTQW problem using SEA formalism using our approximation tool developed in the previous chapter. We discuss the quality of the approximation benchmarked against the full numerical solution for various analyses. By presenting this work, we complete objectives 1. and 2.

Objective 3. requires the development of some mathematical formalism in the form of analytical roots of Bloch vector representation for $N>2$ level systems. As we saw in section \ref{sec:intro_bloch}, there is a need for such analytical results. This research gap has been addressed in chapter \ref{ch:bloch_SEA}. We present the general Bloch parametrization and then show some analytical roots of density matrices under the same (for the cases $N=3$, and $N=4$).

In chapter \ref{ch:comSEA}, we use the results developed in chapter \ref{ch:bloch_SEA} to serve the following purposes. We first present the Beretta composite SEA (BCSEA) EoM. As seen in the historical developments of SEA in sections \ref{sec:intro_history} and \ref{sec:intro_second_age}, there is a need for analytically studying the SEA in the case of simple composites also. To serve this purpose, we present our solution for separable and mixed composites. We present our criteria for no-signaling and provide definitive proof of how SEA respects no-signaling in those cases, thereby completing objective 3.

In the end, we summarize our findings, discuss the possible limitations, and present possible future directions for research on SEA in the final concluding chapter \ref{ch:conclusion}.\conditionalBlankPage

\chapter{Steepest Entropy Ascent}\label{ch:litsurv}
\blankpage
\newpage
\thispagestyle{empty}
\begin{figure}[H]
    \centering
    \captionsetup{justification=centering}
    \includegraphics[height=0.75\textheight, width=0.75\textwidth]{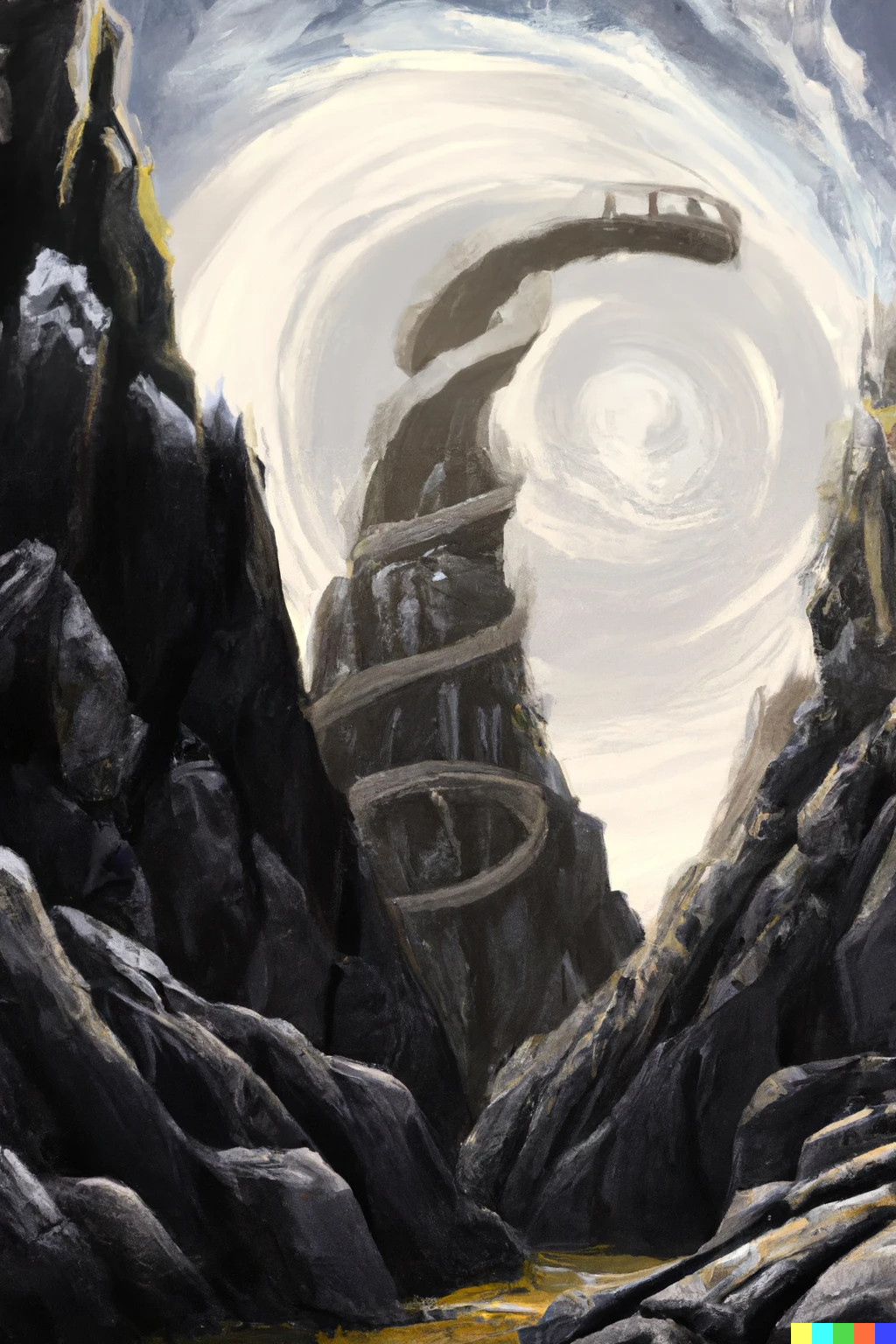}
    \caption*{\emph{I looked up towards the ledge,}\\
        \emph{beyond a very steep ascent}\\
        \emph{that lay yonder,}\\
        \emph{impossibiliy personifying.}}
\end{figure}
\newpage
\blankpage
\lettrine[lines=3]{Q}{uantum} mechanics (QM) is a linear theory. By linear we mean, that linear operators act on a linear state-space (Hilbert space) and evolve linearly in time. In such a theory, a closed system evolves via unitary transformation, following the Schr\"{o}dinger-von Neumann formalism. As a consequence, incorporating non-reversible processes in the scheme of QM becomes a non-trivial task. One can consider a system $\mathbf{S}$ as a part of some larger system, $\mathbf{M}$ (see Fig. \ref{fig:litsurv_decoherenceM}). Then the interaction with the environment ($\mathbf{E}$) is considered, which is $\mathbf{M}-\mathbf{S}$. In this formalism, although $\mathbf{M}$ evolves unitarily, the evolution of the reduced subsystem $\mathbf{S}$ can become non-unitary. The correlations that build up between $\mathbf{S}$ and $\mathbf{E}$ are destroyed during the coarse-graining process (while partially tracing out, \textit{i.e.,} averaging over one of the subsystems). At the heart of this formalism lies the phenomenological reasoning of decoherence: that the destruction of entanglement happens as soon as correlations are created, thus averaging over a long time does not contain any information about the initial states. However, another approach would be to consider decoherence to be a fundamental process in competition with the building up of correlation due to the Hamiltonian evolution. Such a mechanism would consider the second law of thermodynamics to be of a fundamental status, not just a statistical theory. The steepest entropy ascent (SEA) formalism finds its roots in this fundamental approach to decoherence. Here in this chapter, we first discuss the phenomenological modeling of decoherence to highlight the basic structure of the theory and look at its drawbacks. Then we take a look at the review of the literature on SEA so that we have a grasp of what the theory entails as we move on to the rest of the thesis.
\section{Decoherence in phenomenological models}\label{sec:litsurv_decoherence}
\begin{figure}
    \centering
    \includegraphics[]{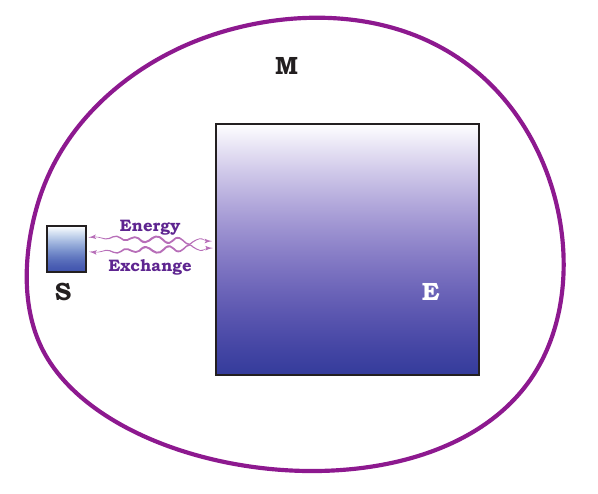}
    \caption[Schematic of phenomenological decoherence model]{\label{fig:litsurv_decoherenceM}A subsystem $\mathbf{S}$ interacting with its environment $\mathbf{E}$. $\mathbf{M} = \mathbf{E}+\mathbf{S}$ evolves unitarily. The weak coupling between $\mathbf{S}$ and $\mathbf{E}$ builds up entanglement between the subsystems.}

\end{figure}
We begin with the idea that a system is never truly isolated, but is in interaction with some environment, which leads to the evolution of pure states to mixed states. \ac{rho} is used in this regard. We identify pure states via the condition $\rho^2 = \rho$, and mixed states satisfy $\rho^2\ne\rho$. The system $\mathbf{S}$, and the environment $\mathbf{E}$ undergo a joint unitary evolution (see the schematic diagram Fig. \ref{fig:litsurv_decoherenceM}). During this evolution, entanglement builds up between $\mathbf{S}$ and $\mathbf{E}$. So the necessary question arises, how do we not observe so? How is it that the reduced density matrices are effectively decorrelated? The answer lies in the key assumption made in this modeling. The so-called \textit{Markovian} assumption \cite{beretta_2009_nonlinear}, ensures that the correlations smear out as fast as they are built. The whole process can be summed up as given -
\begin{itemize}
    \item $\mathbf{S}$ and $\mathbf{E}$ undergo joint unitary evolution.
    \item We can assume a weak coupling between $\mathbf{S}$ and $\mathbf{E}$ resulting in the building of entanglement between the two.
    \item \textit{Markovian condition-} the correlations die out as soon as they are formed.
    \item The resulting time-averaged dynamics remain uncorrelated.
\end{itemize}
One standard justification for using the Markovian assumption lies in the fact that, if we consider a coarse-graining (time averaging) process over a sufficiently large time interval, yet small enough compared to that of the system as a whole, the average correlation is usually negligible. However, this approximation is at odds with the underlying unitary dynamics of quantum mechanics, and thus presents itself as an \textit{ad hoc} addition to the theory of open quantum systems. The reduced probability distribution follows the Kossakowski-Sudarshan-Gorini-Lindblad (KSGL) form of master equation used to understand stochastic as well as open quantum systems \cite{gisin_1995_relevant}, and is given as follows-
\begin{equation}\label{eq:litsurv_KSGL1}
    \dv{\rho}{t} = -\dfrac{\imi}{\hbar}\comm{\mathit{H}}{\rho} + \half\sum_j\left(2V^\dagger_j\rho V_j-\acomm{V_j^\dagger V_j}{\rho}\right).
\end{equation}
The summands in the r.h.s are traceless. An alternative way of writing the Eq. \eqref{eq:litsurv_KSGL1} is as follows
\begin{equation}\label{eq:litsurv_KSGL2}
    \dv{\rho}{t} = -\dfrac{\imi}{\hbar}\comm{\mathit{H}}{\rho} + \half\sum_j\left(\comm{V_j}{\rho V_j^\dagger}+\comm{V_j \rho}{V_j^\dagger}\right).
\end{equation}
These operators $V_j$'s are either creation-annihilation operators or transition operators, effectively allowing mixing from pure states, and \ac{hbar}. It is to be noted, KSGL provides a nonunitary, linear, completely positive map. Despite being applied in many studies involving decoherence, the KSGL formalism has some of the following drawbacks \cite{beretta_2009_nonlinear,beretta_2010_maximum,beretta_2020_fourth,gheorghiu-svirschevski_2001_nonlinear,gheorghiu-svirschevski_2001_addendum,vonspakovsky_2014_trends,li_2018_steepest,ray_2023_nosignalinga}.
\begin{enumerate}
    \item The Markovian assumption, also known as the `erasure of correlations' lies at the heart of KSGL formalism. This assumption finds itself in contradiction with the underlying unitary quantum dynamics given that no modification of the existing QM formalism is being considered.
    \item The Markovian assumption is also responsible for the generation of entropy in this formalism, which makes sense if we assume the second law of thermodynamics to be of statistical nature. The debate on the nature of the second law not yet being settled \cite{hatsopoulos_1976_unified,hatsopoulos_1976_unifieda,hatsopoulos_1976_unifiedb,hatsopoulos_1976_unifiedc}, it does not seem prudent to rely on such \textit{ad hoc} origin of entropy production only.
    \item As quoted in \cite{beretta_2009_nonlinear}, that a `loss of \textit{information} on the time scale of the observer leading to a rapid decoherence from the entanglement which continuously builds up by (at least) weak coupling with environmental degrees of freedom,' cannot properly explain diffusion through the transport of various physical properties such as mass, momentum, energy to name a few.
    \item This formalism being phenomenological in origin is mainly supported by empirical results. However, if a fundamental theory can explain the experimental results with better precision, the new theory should be preferable to KSGL.
    \item Due to interaction with the environment, zero eigenvalued states of $\rho$ evolve to states with non-zero eigenvalues, which leads to mixing. This presents itself in a position of problem, as we know due to unitary evolution in QM pure states evolve to pure states. Also, in many situations, physicists restrain themselves to the \textit{effective} subspace of the density matrix in the context of large systems to study some evolution, and KSGL formalism presents itself as a challenge in those cases as well.
    \item The positivity of the $\rho$ is ensured in forward time but is not guaranteed in the time-reversed scenario. Consequently, the dynamical semigroup nature of evolution can be a matter of concern to physicists who expect a theory consistent with thermodynamics and mechanics to display strong causal behavior. By this, we mean that if there exists a bijective map between the initial and final states, we have established strong causality.
\end{enumerate}

Considering these, we are motivated to consult a theory that attempts to address some of the drawbacks of the KSGL formalism while involving a more fundamental approach to decoherence.
\section{The entropy functional}\label{sec:litsurv_entropy}
The stability principle \cite{hatsopoulos_1965_principles}, the fourth postulate of QM as introduced in \cite{hatsopoulos_1976_unified}, and many other postulates regarding stability and equilibrium are derived from the concept of entropy. But the direct use of von Neumann form or similar forms of the definition of entropy may lead to foundational errors among other things. The definition of state and its philosophical interpretation not being on the same footing creates most of the issues \cite{park_1968_nature}. Hence, it is prudent to arrive at a definition of entropy that encapsulates the concept of stability as well as satisfies the desired properties of a state functional to be used in the SEA formalism. Consider the following statement \cite{hatsopoulos_1976_unified},
\begin{theorem}\label{th:litsurv_entropy1}
    Consider a separable system in a state $\rho$. There exists a property $\mathit{I}$, such that it is invariant during \textit{all} unitary processes, additive for independent systems, given by the expectation value of the measurement results, and does not explicitly depend on the number of particles of the constituent species. For a fixed constant $c$, this property is given by
    \begin{equation*}
        \mathit{I} = c\tr(\rho\ln(\rho)).
    \end{equation*}
\end{theorem}
\begin{proof}
    This proof is taken from Ref. \cite{hatsopoulos_1976_unified}, here it is reproduced for the sake of continuity. \\
    Consider a finite-dimensional ($N$) separable system $\mathbf{A}$ with density matrix $\rho$ having a complete set of eigenvalues and eigenvectors $\{\lambda_k\}$, and $\{v_k\}$, respectively. For unitary processes, employing the Schr\"{o}dinger-von Neumann evolution, besides the number of particles, eigenvalues remain invariant during unitary time evolution. Consequently, any property that is invariant during unitary evolution should be dependent on the $\{\lambda_k\}$ only. Since by theorem, $\mathit{I}$ must also be an expectation value, it is given by
    \begin{equation}\label{eq:litsurv_ent1_the1}
        \mathit{I} = \sum_k \lambda_k \mathit{p}_k\left(\{\lambda_i\}\right).
    \end{equation}
    Where, $\mathit{p}_k\left(\{\lambda_i\}\right) = \qexp{v_k}{\mathit{P}}{v_k}$, for some property $\mathit{P}(\mathbf{A})$. A unitary process $\mathcal{U}_{12}$ between two states $\mathbf{A}_1$, and $\mathbf{A}_2$ of the separable system $\mathbf{A}$ involves a cyclic change in parameters. Consequently, the eigenvalues $\{\lambda_k\}$ remain unchanged but the associated eigenvectors $\{v_k\}$ undergo a permutation with respect to the fixed ordering of the eigenvalues or vice versa. Considering this invariance property of the eigenvalue-eigenvector combination, the quantity $\mathit{I}$ becomes a symmetric function of $\{\lambda_k\}$ of the form,
    \begin{equation}\label{eq:litsurv_ent1_the2}
        \mathit{I} = \sum_k \lambda_k \mathit{p}_k\left(\{\lambda_i\}\right) = \sum_k f\left(\lambda_k\right).
    \end{equation}\\
    Where, $f(\lambda_k)$ is a function of $\lambda$ only.
    Now, let us consider two independent systems $\mathbf{A}$ and $\mathbf{B}$ with associated states $\mathbf{A}_1$ and $\mathbf{B}_1$ respectively. The corresponding eigenvalues of the density operators of $\mathbf{A}$ and $\mathbf{B}$ are $\{\lambda_k\}$ ($k=1\cdots M$) and $\{\mu_\ell\}$ ($\ell=1\cdots N$) respectively. To satisfy the condition of additivity set in the theorem for $\mathit{I}$, we can write
    \begin{equation}\label{eq:litsurv_ent1_the3}
        \mathit{I}_{\mathbf{AB}} = \sum_{k,\ell}f\left(\lambda_k\mu_\ell\right) = \sum_k f(\lambda_k)+\sum_\ell f(\mu_\ell).
    \end{equation}
    Consider for some fixed $m$ and $n$ we have, $\lambda_m,\mu_n = 1$, and $\lambda_k,\mu_\ell = 0$ for $k\ne m,~ \ell\ne n $. Then Eq. \eqref{eq:litsurv_ent1_the3} implies the following,
    \begin{equation}\label{eq:litsurv_ent1_the4}
        -f(1)+\left(M\times N-M -N +1\right)f(0) = 0.
    \end{equation}
    Since system sizes are independent of each other, we have $f(1)=0$, and $f(0) = 0$. \\

    Let us now focus on the variation of $f$ with respect to variations in eigenvalues. Keeping all $\lambda, ~\mu$ fixed except for $\lambda_m$ and $\lambda_n$ so that the following is satisfied,
    \begin{equation}\label{eq:litsurv_ent1_the5}
        \dd{\lambda_m}+\dd{\lambda_n} = 0.
    \end{equation}
    Using this we get from Eq. \eqref{eq:litsurv_ent1_the3}
    \begin{equation}\label{eq:litsurv_ent1_the6}
        \sum_\ell \mu_\ell\dv{f(\lambda_m\mu_\ell)}{(\lambda_m\mu_\ell)} - \dv{f(\lambda_m)}{\lambda_m} = \sum_\ell \mu_\ell\dv{f(\lambda_n\mu_\ell)}{(\lambda_n\mu_\ell)} - \dv{f(\lambda_n)}{\lambda_n}.
    \end{equation}
    Suggesting, given the choice of $m,n$ is arbitrary, each side of Eq. \eqref{eq:litsurv_ent1_the6} must be dependent on $\mu_\ell$ only. Hence, differentiation w.r.t. $\lambda_m$ gives,
    \begin{equation}\label{eq:litsurv_ent1_the7}
        \sum_\ell\mu_\ell^2\dv[2]{f(\lambda_m\mu_\ell)}{(\lambda_m\mu_\ell)}-\dv[2]{f(\lambda_m)}{\lambda_m} = 0.
    \end{equation}
    Now keeping all $\mu$ fixed except for $\mu_r$ and $\mu_s$ (for arbitrary $r$ and $s$) so that,
    \begin{equation}\label{eq:litsurv_ent1_the8}
        \dd{\mu_r}+\dd{\mu_s} = 0,
    \end{equation}
    we arrive from Eq. \eqref{eq:litsurv_ent1_the7},
    \begin{equation}\label{eq:litsurv_ent1_the9}
        2\mu_r\dv[2]{f(\lambda_m\mu_r)}{(\lambda_m\mu_r)}+\mu_r^2\lambda_m\dv[3]{f(\lambda_m\mu_r)}{(\lambda_m\mu_r)} = 2\mu_s\dv[2]{f(\lambda_m\mu_s)}{(\lambda_m\mu_s)}+\mu_s^2\lambda_m\dv[3]{f(\lambda_m\mu_s)}{(\lambda_m\mu_s)}.
    \end{equation}
    This can be re-written as,
    \begin{equation}\label{eq:litsurv_ent1_the10}
        2x\dv[2]{f(x)}{x}+x^2\dv[3]{f(x)}{x} = c
    \end{equation}
    for some fixed constant $c$. The solution to Eq. \eqref{eq:litsurv_ent1_the10} along with the conditions of $f(0),~f(1)=0$ gives,
    \begin{equation}\label{eq:litsurv_ent1_the11}
        f(x) = cx\ln(x).
    \end{equation}
    Thus, we have,
    \begin{equation}\label{eq:litsurv_ent1_the_pr}
        \mathit{I} = \sum_k f(\lambda_k) = c\sum_k \lambda_k\ln(\lambda_k) = c\tr(\rho\ln\rho).
    \end{equation}
\end{proof}

Through Th. (\ref{th:litsurv_entropy1}), we found $\mathit{I}$ to be a functional which remains invariant during a unitary process and holds additive property when it comes to a separable composition of systems. However, we have not yet discussed the case of nonunital irreversible evolution. To do this, consider the following theorem \cite{hatsopoulos_1976_unifieda},
\begin{theorem}\label{th:litsurv_entropy2}
    A property $\mathit{S}$ exists for every system in any state that is invariant in any reversible adiabatic process, increases in any irreversible adiabatic process, and is additive for independent separable systems.
\end{theorem}
\begin{proof}
    Following Ref. \cite{hatsopoulos_1976_unifieda}, we present the proof here. \\
    Before we proceed, a definition is in order
    \begin{defRKR}
        The work done in any reversible adiabatic process for a system $\mathbf{A}$ in contact with a reservoir $\mathbf{R}$, where $\mathbf{A}$ starts from some state $\mathbf{A}_1$ and ends in the mutual equilibrium state $\mathbf{A}_0$ with $\mathbf{R}$, is the maximum amount of work that can be extracted through such interaction, and is called \textit{available energy} of the system $\mathbf{AR}$.
    \end{defRKR}
    Hamiltonian operators ($\mathit{H}$) for a separable system are expressed as a direct sum implying energy ($\mathtt{E} = \tr(\rho\mathit{H})$) is additive, so is true for the available energy ($\mathtt{E}^a$) just defined above. Moreover, the density matrix for such a combination can be expressed in terms of products. Let's denote finite changes in any property of a system by $\Delta$. Given all these, it follows that to define an additive property such as $\mathit{S}$, we can use the following relation,
    \begin{equation}\label{eq:litsurv_ent2_the1}
        \Delta\mathit{S} = c_s\Delta(\mathtt{E}-\mathtt{E}^a).
    \end{equation}
    Where $c_s$ is a constant to be fixed later. As we can see, because $\mathtt{E}$ and $\mathtt{E}^a$ are expectation values, following Eq. \eqref{eq:litsurv_ent2_the1}, $\mathit{S}$ is also an expectation value. It can be shown for an adiabatic reversible process ($\mathcal{U}_{12}$) between two states of the system $\mathbf{A}_1$ and $\mathbf{A}_2$, $\Delta_{12}\mathtt{E} = \Delta_{12}\mathtt{E}^a$ \cite{hatsopoulos_1976_unifieda}. Hence, for such $\mathcal{U}_{12}$ we have,
    \begin{equation}\label{eq:litsurv_ent2_the2}
        \left(\Delta_{12}\mathit{S}\right)_\text{rev} = 0.
    \end{equation}
    Whereas for nonreversible adiabatic changes, the work done is equal to the change in energy, \textit{i.e.,} $-\Delta_{12}\mathtt{E}$, but is less than $-\Delta_{12}\mathtt{E}^a$, so for such a system,
    \begin{equation}\label{eq:litsurv_ent2_the3}
        \left(\Delta_{12}\mathit{S}\right)_\text{irr} > 0.
    \end{equation}
\end{proof}
As we can see, although Th. (\ref{th:litsurv_entropy2}) establishes $\mathit{S}$ as a property that is additive for a separable system, and is increasing for an irreversible sense, it does not guarantee that $\mathit{S}$ has any relation to $\mathit{I}$, but for the fact that the former seems to be a generalization of the latter. To establish the form of $\mathit{S}$, let us consider the contradictory proposition, that
\begin{equation}\label{eq:litsurv_ent_Cform}
    \Delta\mathit{S} \ne c_s\Delta\tr(\rho\ln\rho).
\end{equation}
This assumption does not hold by Th. (\ref{th:litsurv_entropy1}) for unital evolutions, as either $\Delta\mathit{S}$ is no more invariant, or it loses the additive property for independent systems or both. However, as seen in Eq. \eqref{eq:litsurv_ent2_the2}, $\Delta\mathit{S}$ for such processes is zero, thus contradicts our assumption in Eq. \eqref{eq:litsurv_ent_Cform}. To establish reason further, let's note that the processes used in the definition of $\Delta\mathit{S}$ do not change the degrees of freedom of the system, and $\mathit{S}$ is an additive property. Therefore, it can be at most a sum of $c_s\tr(\rho\ln\rho)$ and some linear form containing the particle number of the constituent species. Let us venture into finding the coefficients for such terms. Consider the case where there exists no reversible process which connects any two states of the system with a different number of degrees of freedom. This allows us to choose for ourselves whatever value of the coefficient we like, including zero. On the other hand, if such a connection does exist, we would have to determine each such coefficient experimentally \cite{hatsopoulos_1976_unifieda}, and this is where the third law of thermodynamics plays a key role. By this law, each of these coefficients becomes zero. Hence, we are compelled to conclude that $\mathit{S}$ must be of the form of $\mathit{I}$, \textit{i.e.,}
\begin{equation}\label{eq:litsurv_ent_form}
    \mathit{S} = -k_\text{B}\tr(\rho\ln\rho),
\end{equation}
where $c_s $ is \ac{kB} and is determined from experiments on simple gaseous systems passing through stable equilibrium states \cite{hatsopoulos_1976_unifiedb}.

So far, we have established that $\mathit{S}$ is a property of the system that behaves like entropy, and looks like entropy, hence we can call it the entropy functional. However, this description is more general and considers reversible and irreversible processes as originally thought upon by Clausius. This representation also allows us to formulate the following statement on stable equilibrium \cite{hatsopoulos_1976_unifiedb},
\begin{defRKR}\label{def:litsurv_stableEq}
    The state $\mathbf{A}_0$ will be in \textit{stable equilibrium} if and only if for any other state $\mathbf{A}_i$ of the system, the corresponding entropy $\mathit{S}_i$ is less than that of $\mathbf{A}_0$ ($\mathit{S}_0$) for same values of the state parameters, energy, and the number of particles.
\end{defRKR}

In what follows, we will look into the concept of stability in more detail and will try to rephrase the second law of thermodynamics from thereon. This rephrasing is key in the SEA scheme of things, as we can then incorporate the second law of thermodynamics in the quantum paradigm and get going to the interesting stuff.
\section{Stability and the second law of thermodynamics}\label{sec:litsurv_2ndLaw}

\begin{figure}[H]
    \centering
    \includegraphics[height=0.6\textheight]{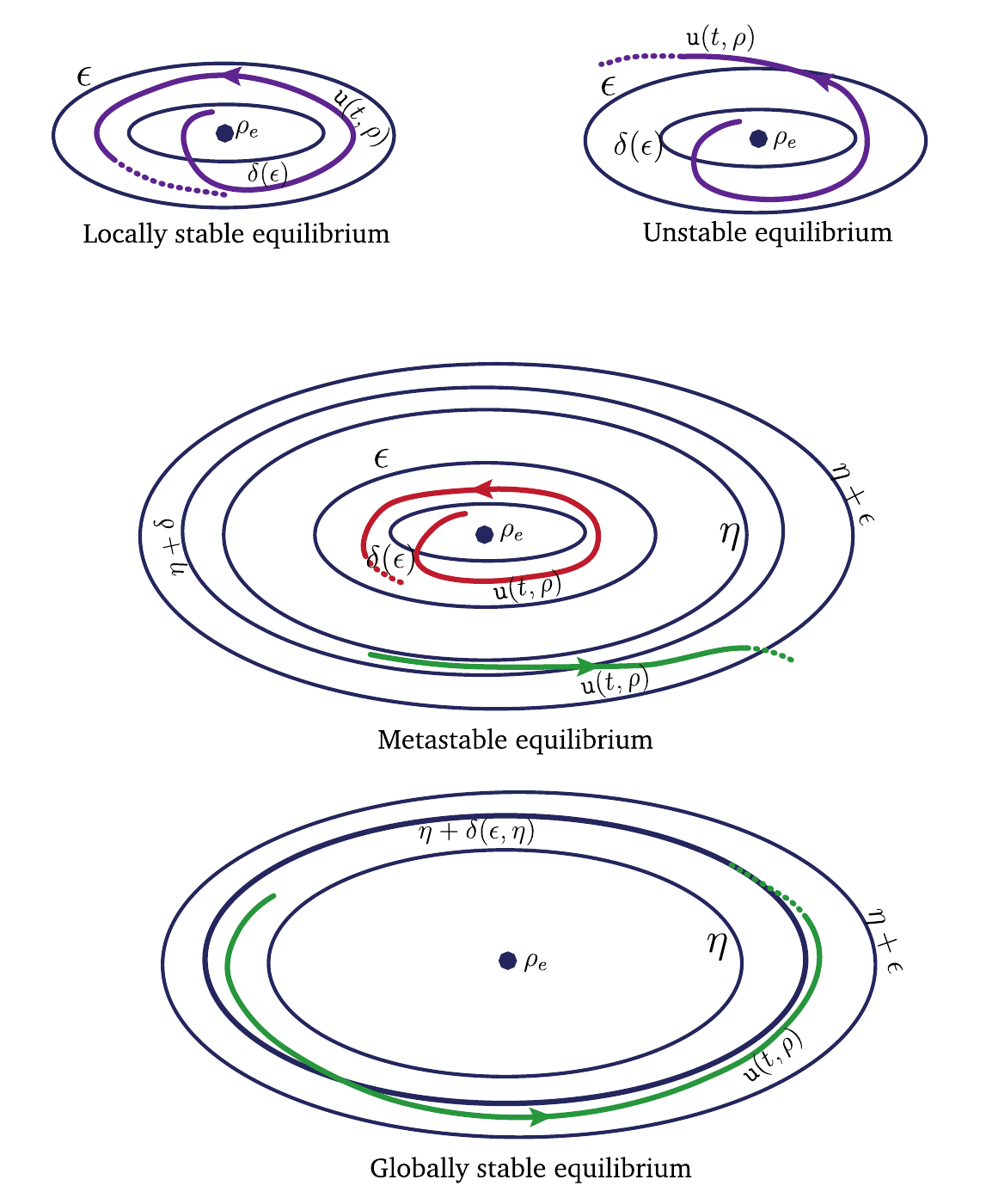}
    \caption[Schematics for various equilibrium trajectories]{\label{fig:litsurv_stability}A schematic description of different types of equilibrium as discussed in the text. The trajectories in purple, red, and green denote various trajectories traced by $\mathtt{u}(t,\rho)$ in the state space. Each orbit is denoted by its distance from the equilibrium state $\rho_e$. In the metastable equilibrium case, we can see how originating from two different distances from the equilibrium state $\rho_e$, two trajectories can end up in two different modes of equilibrium. The red one remains locally stable, while the green one becomes metastable.}
\end{figure}

Stability has a position of utmost importance in thermodynamics. The laws of thermodynamics mostly describe systems that have achieved a `steady state', \textit{i.e.,} are in a stable equilibrium. However, this definition of equilibrium is mostly statistical and thus cannot be directly implemented for systems with fewer degrees of freedom. Hence, we must re-evaluate our understanding of equilibrium that suits the task at hand.\\
Consider the trajectories generated by a given dynamics of the form $\mathtt{u}(t,\rho)$ containing the state $\rho$ at time $t=0$. A state $\rho_e$ is in equilibrium when $\mathtt{u}(t,\rho_e)=\rho_e$ for all $t$, or in other words $\dv{\rho}{t} = 0$. We use a metric to define distance function $d(x,y)$ in the state space as follows
\begin{enumerate}
    \item $d(x,y) = 0$ only when $x=y$,
    \item $d(x,y)\ge 0$ for all $x,~y$,
    \item $d(x,z)+d(z,y)\ge d(x,y)$ for any triplet $x,y,z$ (triangle inequality),
    \item $d(x,y)=d(y,x)$.
\end{enumerate}
Let us begin with the second law of thermodynamics, the following statement is inspired by Hatsopoulos-Keenan statement \cite{hatsopoulos_1965_principles}, and the Kelvin-Planck, Clausius, and Carath\'{e}odory statements follow from this \cite{gyftopoulos_2005_thermodynamics}.
\begin{quotation}
    \textbf{The second law of thermodynamics:} There exists a unique globally stable equilibrium state among all the states of a system for a given value of the energy, number of constituents, and the parameters of the Hamiltonian.
\end{quotation}
It is well understood that thermodynamic equilibrium is the Lyapunov locally stable equilibrium. We begin with Lyapunov equilibrium \cite{beretta_1986_theorem,beretta_2005_nonlinear,beretta_2009_nonlinear,beretta__what}. The Lyapunov local stability condition is given as follows-
\begin{defRKR}\label{def:litsurv_loc_stabi}
    Consider an equilibrium state, $\rho_e$. It is \textit{locally stable} if and only if for every $\epsilon>0$ we can find a $\delta(\epsilon)>0$  such that $d(\rho,\rho_e)<\delta(\epsilon)$ implies $d(\mathtt{u}(t,\rho),\rho_e)<\epsilon$ for all $t>0$ and $\rho$.
\end{defRKR}
To expound on the above, any trajectory that passes through $\rho$ which is at a distance $\delta(\epsilon)$ from the equilibrium state $\rho_e$, cannot go beyond the distance $\epsilon$ for any time $t$. A nice way of visualizing this is presented in Fig. \ref{fig:litsurv_stability}.
Any state that does not follow Def. (\ref{def:litsurv_loc_stabi}) is a \textit{locally unstable} state. It is claimed in some literature, that the entropy functional as described in the previous section is a Lyapunov function \cite{beretta__what}. However, that seems not to be the case, partly because Lyapunov stability excludes metastable states. The other reason is, while maximal entropic states are also locally stable, the converse is not always true \cite{beretta_1986_theorem}.
Consider the case of the unitary evolution of pure states, $\mathtt{u}(t,\rho) = \mathcal{U}_t\rho\mathcal{U}_t^\dagger$ with $\mathcal{U}_t = \exp(-\imi\mathit{H}t)$ for $\hbar=1$. The value of entropy functional is zero, while these states satisfy the local stability criteria as set by the definition above, they are not unique. The function $d(\rho_1,\rho_2) = \tr\abs{\rho_1-\rho_2}$ will give same distance for all states $\rho$ at all times $t$ from the equilibrium state $\rho_e|~\comm{\mathit{H}}{\rho_e}=0$ implying $d(\mathtt{u}(t,\rho),\rho_e)=d(\mathtt{u}(0,\rho),\rho_e)$ \cite{beretta_1986_theorem}. One thus needs to assure that other equilibrium states which might be metastable or limit cycles are not globally stable also. The chief reason behind such a requirement is the second law of thermodynamics. There is also the apparent conflict between mechanics and thermodynamics which needs addressing. The basic statement of a stable equilibrium from the point of view of mechanics based on the minimum energy principle is \cite{beretta_2009_nonlinear}
\begin{quotation}
    Among all the states of a system consistent with a given set of values for the numbers of constituents and the parameters of the Hamiltonian, the state of lowest energy is the only stable equilibrium state.
\end{quotation}
As we can see, while the second law asks for entropy maximization, mechanics asks for energy minimization. This paradoxical situation can be resolved once we consider the pure states to be a subset of all states available to a quantum system. That is states which satisfy the criteria $\rho^2=\rho$ form what is known as limit cycles, which while being locally stable are not maximal entropic states. Hence they satisfy Lyapunov criteria while not being a global equilibrium state. However, by admitting states of the form $\rho^2\ne\rho$ one opens up the possibility of accommodating maximal entropic states, which coincide with the thermodynamic equilibrium state as well.  The Lyapunov definition no longer remains necessary and sufficient for thermodynamic equilibrium by the second law, it only remains necessary. The Lyapunov criteria falling short of accommodating these states must be augmented with the definition of global stability and metastability. Which are given as follows \cite{beretta__what}
\begin{defRKR}\label{def:litsurv_loc_metasble}
    An equilibrium state $\rho_e$ is \textit{metastable} if and only if it is locally stable but there is an $\eta>0$ and an $\epsilon>0$ such that for every $\delta>0$ there is a trajectory $\mathtt{u}(t,\rho)$ passing at $t=0$ between distance $\eta$ and $\eta+\delta$ from $\rho_e$, and reaching at later time $t>0$ some distance further than $\eta+\epsilon$.
\end{defRKR}
To elaborate, consider a state $\rho$ at $t=0$, is at distance between $\eta$ and $\eta+\delta$, implying $\eta<d(\mathtt{u}(0,\rho),\rho_e)<\eta+\delta$. If after some time, it is found that the trajectory lies outside the range $\eta+\epsilon$, that is $d(\mathtt{u}(t,\rho),\rho_e)\ge\eta+\epsilon$, the state has started veering off from its locally stable trajectory as in Fig. \ref{fig:litsurv_stability}. This means, while the states which are already within the distance $\delta$, to begin with, are in the distance $\epsilon$ for all times (locally stable), metastable implies, states that slightly depart from that distance steer away from the stable orbit. This allows us to define global stability in the following way,
\begin{defRKR}\label{def:litsurv_glob_stable}
    An equilibrium state $\rho_e$ is \textit{globally stable} if for every $\eta>0$ and $\epsilon>0$ there exists a $\delta(\eta,\epsilon)>0$ such that every trajectory $\mathtt{u}(t,\rho)$ that starts within the distance $\eta$ and $\eta+\delta$ from $\rho_e$, stays within $\eta+\epsilon$ distance from it at all times.
\end{defRKR}
One can see how this connects to the standard definition of global stability, \textit{an equilibrium state is stable if it can be altered into another state by interactions that leave net effects on the environment.}\\
As we can see from the above discussion, the entropy functional, along with the entropy non-decrease property, provides a reliable way to reach global equilibrium states via the second law of thermodynamics. We must ask ourselves at this point, what do we mean by states in the context of quantum mechanics? Where do the above-mentioned trajectories find their application? Are these states described as the usual states of quantum mechanical formulation? Or do they come with philosophical implications of their own? Let us find that out in the next section.

\section{Ontology of states}\label{sec:litsurv_State}

The search for the true philosophical meaning of a quantum mechanical state is an age-old problem and has an affinity to drag the questioner into a rabbit hole. This thesis will try not to indulge in those difficult problems. Yet, given the nature of the issue under consideration, some mention of the philosophical implications of the assumptions considered in the SEA formalism must be given, lest we are to ignore all the nuances of the formalism. \\
It is to be understood, that the approach to non-equilibrium as envisioned in the SEA formalism is towards a fundamental nature of decoherence in contrast to the coarse-grained nature of the same as discussed in Sec \ref{sec:litsurv_decoherence}. Hence it is also important to observe that the states discussed in the context of SEA are not outcomes of measurement statistics of an ergodic system performed over a long period of time \cite{hatsopoulos_1976_unified}. One must also acknowledge the fact that it is not possible in quantum mechanics to predict which eigenvalue a single measurement will yield. Hence, it may be concluded that the outcomes of some or all measurements performed on a given system are going to be irreducibly dispersed. That is why we deal with averages when talking about the value of a property of a quantum system. This implies that there is an epistemic rule of correspondence \cite{hatsopoulos_1976_unifiedc,park_1968_nature}, by which physical conditions of a state are experimentally known only through repeated measurements performed on replicas of the same prepared in a particular way. This suggests that the preparation procedure of a state needs to be specified and uniquely associated with it before studying any quantal evolution of the same. This requirement is peculiar to quantum mechanics \cite{hatsopoulos_1976_unifiedc}. \\
A density matrix $\rho$ is used to express the probability distribution associated with measurement corresponding to a given preparation. A density matrix thus produced can be pure $\rho^2=\rho$, or mixed $\rho^2\ne\rho$. While there is no ambiguity regarding the ontological status of the pure states, it is the mixed states that are mostly in a pickle. The dominant narrative is that mixed states are a convex combination of pure states and the weighted sum implies epistemic ignorance. However, this idea is contested by Park, Hatsopoulos, Beretta and colleagues \cite{hatsopoulos_1965_principles,park_1968_nature,hatsopoulos_1976_unifiedc,beretta_2009_nonlinear}. In their point of view, if a preparation results in a mixed state, then the mixed density matrix is the true representation of the state. Park also argued \cite{park_1968_nature} that a $\rho$ can be numerically subdivided into infinite many combinations of pure and mixed states, thus one can represent our `ignorance' in so many ways. All of these combinations will have the following format (each $\rho_k$ can be mixed or pure),
\begin{equation}\label{eq:litsurv_onto_convex}
    \text{for } 0\le w_k\le 1, ~ \rho = \sum_k w_k\rho_k, ~ \text{with} ~ \sum_k w_k = 1.
\end{equation}
Such a statement begs another set of questions, can we set up a procedure through which we can operationally distinguish between preparations ($\mathtt{Z}_1$) that are due to epistemic ignorance and result in a `mixture' of pure states and preparations ($\mathtt{Z}_2$) that are genuinely mixed? Hatsopoulos \textit{et. al.} \cite{hatsopoulos_1976_unifiedc} answer this in positive. They provide one of the first hints of contextual and non-contextual preparations \cite{hatsopoulos_1976_unifiedc}. Instead of delving into the details, we can with certain confidence conclude that the preparations of the former type, $\mathtt{Z}_1$ can be treated with information-theoretic tools, while the latter type requires a more fundamental approach. \\
This is where the ontological status of the state in the context of SEA gets cemented. We present the following hypothesis \cite{beretta_2009_nonlinear,ray_2023_nosignalinga} which together with the second law of thermodynamics as stated in Sec. \ref{sec:litsurv_2ndLaw} sets the ground for the inclusion of a fourth postulate in the scheme of quantum evolution as will be discussed later in this chapter.
\begin{quotation}
    \textbf{Ontological state hypothesis:} A density matrix can be used to represent the evolution of a single particle. Moreover, both pure and mixed state density matrices are real ontological objects. Mixed state density matrices are not merely a result of epistemic ignorance represented through a mixture of pure states.
\end{quotation}
Before concluding this section, we must acknowledge in the vein of Ref. \cite{vonspakovsky_2014_trends} that there is more than one way of preparing a mixed density operator. Each of those preparations bears a physically different meaning. However, if no local measurement can distinguish between those preparations, it can be safely said that irrespective of their origin, operationally those density matrices are physically equivalent.
\section{The four postulates}\label{sec:litsurv_fourthPost}
Now that we have discussed the second law of thermodynamics, the role of the entropy functional in identifying the globally stable state, and the meaning of those states, it is about time that we bring all of these together in a coherent fashion. The introduction of the second law in the context of QM takes place through the equilibrium state postulate that we will discuss below. Before, we proceed, for the sake of continuity we will revisit the postulates of the Schr\"{o}dinger-von Neumann theory of unitary dynamics as well.
\subsection*{I. The correspondence postulate}
\textit{Observables are represented by some linear Hermitian operators defined on a Hilbert space where the quantum mechanical states reside. Some of these operators represent physical observables.}\\
A note on the wording. Some operators are not necessarily Hermitian (PT symmetry case with non-Hermitian operators \cite{bender_1998_real}). While most of the operators are physical observables, not all physical observables are operators, for example, temperature \cite{hatsopoulos_1976_unified}.
\subsection*{II. The mean-value postulate}
\textit{We can associate a real linear functional $\expval{\mathit{P}}$ of the Hermitian operators $\mathit{P}$, such that for every ensemble of measurements performed on unambiguously prepared copies of a system, $\expval{\mathit{P}}$ is the average of $\mathit{P}$ operations.}\\
It can be shown \cite{hatsopoulos_1976_unified,hatsopoulos_1976_unifieda,hatsopoulos_1976_unifiedb,hatsopoulos_1976_unifiedc} that using the above two postulates, $\rho$, the density matrix can be used as a representation of state with unit trace and positive semidefinite structure.
\subsection*{III. The dynamical postulate}
\textit{For a reversible process the time evolution of the density operator is given by the Schr\"{o}dinger-von Neumann equation of motion as given below}
\begin{equation}\label{eq:litsurv_SchroEvo}
    \dv{\rho}{t} = -\dfrac{\imi}{\hbar}\comm{\mathit{H}}{\rho}.
\end{equation}
Where $\hbar$ is the reduced Planck's constant, and $\mathit{H}$ is the Hamiltonian operator.
\subsection*{IV. The stable equilibrium postulate}
\textit{Any independent separable system has a unique stable equilibrium state which is fixed by a given value of energy, number of constituents, and other parameters of the Hamiltonian.}\\
The second law is included in the QM picture via this postulate. As we can see, the existence of an accessible globally stable equilibrium state means that now we can theorize and formulate how quantum states will evolve towards that state. This will be the main focus of the steepest entropy ascent ansatz, which will be discussed in the next section.
\section{Steepest entropy ascent}\label{sec:litsurv_sea}
We begin by stating the steepest entropy ascent (SEA) ansatz, which is based on the ontological state hypothesis, the second law of thermodynamics presented as the fourth postulate and is given below \cite{ray_2022_steepestb}-
\begin{quotation}
    \textbf{Steepest entropy ascent ansatz:} for a given system (closed or open), there exists a globally unique stable equilibrium state $\rho_e$ (from the second law). Any other state evolves towards $\rho_e$ under given constraints of motion such that the local entropy production rate is maximum, and it does so in the direction of the gradient of the entropy functional (steepest ascent).
\end{quotation}
In what follows, we will see how the SEA ansatz gives rise to a new equation of motion which in some sense is a generalized version of the KSGL equation discussed above in Eq. \eqref{eq:litsurv_KSGL1}. In the following section, we will discuss the geometric interpretation of the SEA \ac{EoM}.\\
To find the SEA EoM we begin with the state operator $\rho$, we follow Refs. \cite{beretta_2014_steepest,ray_2022_steepestb}. To maintain the positivity of $\rho$ at all times, we consider $\gamma = \sqrt{\rho}$ to construct the EoM. Later we will see that the final EoM includes only $\rho$. This $\gamma$ is an element of the linear manifold $\mathscr{L}$, a state space of the Hilbert space $\mathscr{H}$. One can define a symmetric inner product of the linear operators (not necessarily Hermitian) defined by $\inprc{A}{B} = \tr(A^\dagger B+B^\dagger A)/2$. We have,
\begin{equation}\label{eq:litsurv_gamma}
    \rho = \gamma\gamma^\dagger.
\end{equation}
Consider the generators of motion by elements of the set $\{\mathit{C}_i(\gamma)\}$. Each of these $\mathit{C}_i$ represents various generators, for example, Hamiltonian ($\hamiltonian$), probability ($\mathit{I}$), number operators ($\mathit{N}$), and so on. We use the entropy functional ($\mathit{S}$) as defined in Sec. \ref{sec:litsurv_entropy}. The functional derivative of $\mathit{C}_i,~\mathit{S}$ with respect to the change in $\gamma$ is denoted by
\begin{align}\label{eq:litsurv_func_der}
    \begin{split}
        \Psi_i &=~ \vstSpdv{\mathit{C}_i(\gamma)}{\gamma}\\
        \Phi &=~ \vstSpdv{\mathit{S}}{\gamma}.
    \end{split}
\end{align}
To formulate the steepest entropy path, we need to set the constraints of the motion. They are entropy non-decrease, energy conservation, probability conservation, etc. Denoting $\Pi_\gamma = \dv{\gamma}{t}$, we can use the following equation to express the constraints
\begin{align}\label{eq:litsurv_constr-eq}
    \begin{split}
        \dv{\mathit{S}}{t} = \Pi_S~~ & ~~\text{with~~} \Pi_S = \inprc{\Phi}{\Pi_\gamma}\ge 0\\
        \dv{\mathit{C}_i}{t} = \Pi_{\mathit{C}_i}~~ &~~ \text{with~~} \Pi_{\mathit{C}_i} = \inprc{\Psi_i}{\Pi_\gamma} =0.
    \end{split}
\end{align}
The evolution of $\gamma$ in time ($\dot{\gamma}$) has two principal components, one due to the pure unitary Hamiltonian evolution ($\dot{\gamma}_H$), and the other is due to dissipative motion ($\dot{\gamma}_D$) that can be written as \cite{beretta_2009_nonlinear},
\begin{equation}\label{eq:litsurv_gamma_component}
    \dot{\gamma} = \dot{\gamma}_H + \dot{\gamma}_D.
\end{equation}
The pure unitary Hamiltonian evolution follows Schr\"{o}dinger equation of motion
\begin{equation}\label{eq:litsurv_gammH}
    \dot{\gamma}_H = - \dfrac{\imi}{\hbar}\hamiltonian\gamma.
\end{equation}
To figure out the dissipative part, we need SEA formalism. As we have seen so far, constraints play an important role in determining the trajectory of SEA motion. However, one needs to associate a metric to the manifold $\mathscr{L}$ so that the distance function can be determined. Besides, a metric will allow us to fix the norm of the $\dot{\gamma}$ so that we can focus only on variation in direction. Let us consider the metric given as $\mathit{G}(\gamma)$, this allows us to write a small line segment in the state space as
\begin{equation}\label{eq:litsurv_metric_dl}
    \dd{l} = \sqrt{\seaexp{\Pi_\gamma}{\mathit{G}(\gamma)}{\Pi_\gamma}}.
\end{equation}
Hence, using the norm constraint, and the constraints of Eq. \eqref{eq:litsurv_constr-eq}, we can invoke Lagrange's multiplier method to write down the following Lagrangian type functional
\begin{equation}\label{eq:litsurv_lagrange_func}
    \Upsilon = \Pi_S - \sum_i \itbar{\beta}_i\Pi_{\mathit{C}_i} - \dfrac{\tau}{2}\seaexp{\Pi_\gamma}{\mathit{G}(\gamma)}{\Pi_\gamma}.
\end{equation}
We have used $\itbar{\beta}_i$ and $\tau$ as Lagrange's multipliers independent of $\Pi_\gamma$. Now we will use the calculus of variation to minimize the Lagrangian $\Upsilon$ about the variations with respect to $\Pi_\gamma$ by taking functional derivative as under (to distinguish from QM states, instead of the Dirac ket, we use the $\kae{\cdot}$ for SEA states and functions on states)
\begin{equation}\label{eq:litsurv_lagrange_funcder}
    \vstSpdv{\Upsilon}{\Pi_\gamma} = \kae{\Phi} - \sum_i\itbar{\beta}_i\kae{\Psi_i}-\tau\mathit{G}\kae{\Pi_\gamma}.
\end{equation}
Using the criteria, $\vSpdv{\Upsilon}{\Pi_\gamma} = 0$, we get the following equation for $\Pi_\gamma=\dot{\gamma}_D$,
\begin{equation}\label{eq:litsurv_gamma_eom}
    \kae{\Pi_\gamma} = \frac{1}{\tau}\mathit{G}^{-1}\left(\Phi-\sum_i\tbar{\beta_i}\Psi_i\right).
\end{equation}
Next, we assume a metric of the form $\mathit{G}(A) = \frac{1}{\tau}\mathcal{L}^{-1}A$ where $\mathcal{L}$ is a positive definite hermitian operator, and clearly $\mathit{G}^{-1}(A) = \tau \mathcal{L}A$ and $\mathit{G}^{-1}(A)^\dagger = \tau A^\dagger\mathcal{L}$. Let us look at the expressions for $\kae{\Phi}$ and $\kae{\Psi_i}$ \cite{beretta_2009_nonlinear,beretta_2014_steepest,ray_2022_steepestb}
\begin{align}\label{eq:litsurv_Phi_Psi}
    \begin{split}
        \kae{\Psi_i} &=~ \kae{2\mathit{C}_i\gamma}\\
        \kae{\Phi} &=~ \kae{-2k_\text{B}\left(\ln\gamma\gamma^\dagger+\mathit{I}\right)\gamma}.
    \end{split}
\end{align}
Now looking at the dissipative part of the motion, we get the following relation from Eq. \eqref{eq:litsurv_gamma}, and \eqref{eq:litsurv_gamma_component},
\begin{equation}\label{eq:litsurv_rhoD}
    \dv{\rho_D}{t} = \Pi_\gamma\gamma^\dagger+\gamma\Pi_{\gamma^\dagger}.
\end{equation}
Now if we insert the expressions from Eq. \eqref{eq:litsurv_Phi_Psi} into Eq. \eqref{eq:litsurv_gamma_eom}, we get,
\begin{equation}\label{eq:litsurv_gamma_eomEx}
    \Pi_\gamma = -2\mathcal{L}\left(k_\text{B}(\ln\gamma\gamma^{\dagger}+\mathit{I})\gamma+\sum_{i}\itbar{\beta}_i\mathit{C}_i\gamma\right).
\end{equation}
Using this along with Eq. \eqref{eq:litsurv_rhoD}, we get
\begin{align}\label{eq:litsurv_rhoD_deri}
    \begin{split}
        \dv{\rho_D}{t} = & -2\left[k_\text{B}\mathcal{L}(\ln(\gamma\gamma^{\dagger}))\gamma\gamma^{\dagger} +\sum_{i}\itbar{\beta}_i\mathcal{L}(\mathit{C}_i)\gamma\gamma^{\dagger} + k_\text{B}\mathcal{L}\gamma\gamma^{\dagger} \right.\\
            &~~\left.+k_\text{B}\gamma\gamma^{\dagger}\mathcal{L} +k_\text{B}\gamma\gamma^{\dagger}\ln(\gamma\gamma^{\dagger})\mathcal{L} + \sum_{i}\itbar{\beta}_i\gamma\gamma^{\dagger}\mathit{C}_i\mathcal{L}\right].
    \end{split}
\end{align}
Identifying $\gamma\gamma^{\dagger}=\rho$, we can rearrange the r.h.s of the Eq. \eqref{eq:litsurv_rhoD_deri} to get the following one,
\begin{equation}\label{eq:litsurv_rhoD_fin}
    \dv{\rho_D}{t} = -2\left[k_\text{B}\acomm{\mathcal{L}(\ln(\rho))}{\rho}+k_\text{B}\acomm{\mathcal{L}}{\rho} +\sum_{i}\itbar{\beta}_i(\mathcal{L}\mathit{C}_i\rho+\rho\mathit{C}_i\mathcal{L})\right].
\end{equation}
We now will determine $\itbar{\beta}_i$, which will allow us to express Eq. \eqref{eq:litsurv_rhoD_fin} in a more aesthetic manner.
Using the constraints of Eq. \eqref{eq:litsurv_constr-eq}, with Eq. \eqref{eq:litsurv_gamma_eom} the following equation is found,
\begin{equation}\label{eq:litsurv_beta_constraint}
    \sum_{i}\bae{\Psi_j}\mathcal{L}\kae{\Psi_i}\itbar{\beta}_i = \bae{\Psi_j}\mathcal{L}\kae{\Phi}.
\end{equation}
Let us consider three generators of motion giving rise to three constraints, and correspondingly three $\itbar{\beta}$s. The Eq. \eqref{eq:litsurv_beta_constraint} thus can be solved using Cramer's method, the solution is given below,

\begin{equation}\label{eq:litsurv_betaBarOm}
    \tbar{\Omega} =  \vmqty{\seaexp{\Psi_1}{\mathcal{L}}{\Psi_1} &  \seaexp{\Psi_1}{\mathcal{L}}{\Psi_2} & \seaexp{\Psi_1}{\mathcal{L}}{\Psi_3}\\
        \seaexp{\Psi_2}{\mathcal{L}}{\Psi_1} & \seaexp{\Psi_2}{\mathcal{L}}{\Psi_2} &\seaexp{\Psi_2}{\mathcal{L}}{\Psi_3}\\
        \seaexp{\Psi_3}{\mathcal{L}}{\Psi_1} & \seaexp{\Psi_3}{\mathcal{L}}{\Psi_2} & \seaexp{\Psi_3}{\mathcal{L}}{\Psi_3}},
\end{equation}
which must be non-zero for the solution to exist. Then we can have,
\begin{equation}\label{eq:litsurv_betaBar1}
    \itbar{\beta}_1 =  \dfrac{1}{\tbar{\Omega}} \vmqty{\seaexp{\Psi_1}{\mathcal{L}}{\Phi} &  \seaexp{\Psi_1}{\mathcal{L}}{\Psi_2} & \seaexp{\Psi_1}{\mathcal{L}}{\Psi_3}\\
        \seaexp{\Psi_2}{\mathcal{L}}{\Phi} & \seaexp{\Psi_2}{\mathcal{L}}{\Psi_2} &\seaexp{\Psi_2}{\mathcal{L}}{\Psi_3}\\
        \seaexp{\Psi_3}{\mathcal{L}}{\Phi} & \seaexp{\Psi_3}{\mathcal{L}}{\Psi_2} & \seaexp{\Psi_3}{\mathcal{L}}{\Psi_3}},
\end{equation}
\begin{equation}\label{eq:litsurv_betaBar2}
    \itbar{\beta}_2 =  \dfrac{1}{\tbar{\Omega}}\vmqty{\seaexp{\Psi_1}{\mathcal{L}}{\Psi_1} &  \seaexp{\Psi_1}{\mathcal{L}}{\Phi} & \seaexp{\Psi_1}{\mathcal{L}}{\Psi_3}\\
        \seaexp{\Psi_2}{\mathcal{L}}{\Psi_1} & \seaexp{\Psi_2}{\mathcal{L}}{\Phi} &\seaexp{\Psi_2}{\mathcal{L}}{\Psi_3}\\
        \seaexp{\Psi_3}{\mathcal{L}}{\Psi_1} & \seaexp{\Psi_3}{\mathcal{L}}{\Phi} & \seaexp{\Psi_3}{\mathcal{L}}{\Psi_3}},
\end{equation}
\begin{equation}\label{eq:litsurv_betaBar3}
    \itbar{\beta}_3 =  \dfrac{1}{\tbar{\Omega}}\vmqty{\seaexp{\Psi_1}{\mathcal{L}}{\Psi_1} &  \seaexp{\Psi_1}{\mathcal{L}}{\Psi_2} & \seaexp{\Psi_1}{\mathcal{L}}{\Phi}\\
        \seaexp{\Psi_2}{\mathcal{L}}{\Psi_1} & \seaexp{\Psi_2}{\mathcal{L}}{\Psi_2} &\seaexp{\Psi_2}{\mathcal{L}}{\Phi}\\
        \seaexp{\Psi_3}{\mathcal{L}}{\Psi_1} & \seaexp{\Psi_3}{\mathcal{L}}{\Psi_2} & \seaexp{\Psi_3}{\mathcal{L}}{\Phi}}.
\end{equation}
On column rearrangement, we get:
\begin{equation}\label{eq:litsurv_betaBar2re}
    \itbar{\beta}_2 =  - \dfrac{1}{\tbar{\Omega}}\vmqty{ \seaexp{\Psi_1}{\mathcal{L}}{\Phi} &\seaexp{\Psi_1}{\mathcal{L}}{\Psi_1} &  \seaexp{\Psi_1}{\mathcal{L}}{\Psi_3}\\
        \seaexp{\Psi_2}{\mathcal{L}}{\Phi} &\seaexp{\Psi_2}{\mathcal{L}}{\Psi_1} & \seaexp{\Psi_2}{\mathcal{L}}{\Psi_3}\\
        \seaexp{\Psi_3}{\mathcal{L}}{\Phi} &\seaexp{\Psi_3}{\mathcal{L}}{\Psi_1} & \seaexp{\Psi_3}{\mathcal{L}}{\Psi_3}},
\end{equation}
and,
\begin{equation}\label{eq:litsurv_betaBar3re}
    \itbar{\beta}_3 =  \dfrac{1}{\tbar{\Omega}}\vmqty{\seaexp{\Psi_1}{\mathcal{L}}{\Phi}& \seaexp{\Psi_1}{\mathcal{L}}{\Psi_1} &  \seaexp{\Psi_1}{\mathcal{L}}{\Psi_2} \\
        \seaexp{\Psi_2}{\mathcal{L}}{\Phi}&\seaexp{\Psi_2}{\mathcal{L}}{\Psi_1} & \seaexp{\Psi_2}{\mathcal{L}}{\Psi_2}\\
        \seaexp{\Psi_3}{\mathcal{L}}{\Phi}&\seaexp{\Psi_3}{\mathcal{L}}{\Psi_1} & \seaexp{\Psi_3}{\mathcal{L}}{\Psi_2} }.
\end{equation}
We can assume $\mathcal{L} = \tfrac{1}{4k_\text{B}\tau}\mathit{I}$, $\mathit{I}$ for Fisher-Rao metric\cite{ray_2022_steepestb}, $k_\text{B}$ to handle scaling of the entropy term, and 4 as a constant scaling factor. To find a more explicit solution let us note the following
\begin{align}\label{eq:litsurv_traceRels}
    \begin{split}
        \seaexp{\Psi_i}{\mathcal{L}}{\Phi} =&~ -\dfrac{1}{\tau}\left(\tr(\rho\half\acomm{\mathit{C}_i}{\ln(\rho)})+\tr(\half\acomm{\mathit{C}_i}{\rho})\right)\\
        \seaexp{\Psi_i}{\mathcal{L}}{\Psi_j} =&~ \dfrac{1}{k_\text{B}\tau}\tr(\rho\half\acomm{\mathit{C}_i}{\mathit{C}_j}).
    \end{split}
\end{align}
Using this expression in Eqs. \eqref{eq:litsurv_betaBarOm}, \eqref{eq:litsurv_betaBar1}, \eqref{eq:litsurv_betaBar2re}, and in \eqref{eq:litsurv_betaBar3re} we get,
\begin{equation}\label{eq:litsurv_betaBarOm1}
    \tbar{\Omega} =  \dfrac{1}{(k_\text{B}\tau)^3}\vmqty{\tr(\tfrac{\rho}{2}\acomm{\mathit{C}_1}{\mathit{C}_1}) &  \tr(\tfrac{\rho}{2}\acomm{\mathit{C}_1}{\mathit{C}_2}) & \tr(\tfrac{\rho}{2}\acomm{\mathit{C}_1}{\mathit{C}_3})\\
        \tr(\tfrac{\rho}{2}\acomm{\mathit{C}_2}{\mathit{C}_1}) & \tr(\tfrac{\rho}{2}\acomm{\mathit{C}_2}{\mathit{C}_2}) &\tr(\tfrac{\rho}{2}\acomm{\mathit{C}_2}{\mathit{C}_3})\\
        \tr(\tfrac{\rho}{2}\acomm{\mathit{C}_3}{\mathit{C}_1}) & \tr(\tfrac{\rho}{2}\acomm{\mathit{C}_3}{\mathit{C}_2}) & \tr(\tfrac{\rho}{2}\acomm{\mathit{C}_3}{\mathit{C}_3})},
\end{equation}
\begin{equation}\label{eq:litsurv_betaBar_F1}
    \itbar{\beta}_1 =  -\dfrac{1}{k_\text{B}^2\tau^3\tbar{\Omega}}\vmqty{\tr(\tfrac{\rho}{2}\acomm{\mathit{C}_1}{\ln(\rho)+\mathit{I}}) &  \tr(\tfrac{\rho}{2}\acomm{\mathit{C}_1}{\mathit{C}_2}) & \tr(\tfrac{\rho}{2}\acomm{\mathit{C}_1}{\mathit{C}_3})\\
        \tr(\tfrac{\rho}{2}\acomm{\mathit{C}_2}{\ln(\rho)+\mathit{I}}) & \tr(\tfrac{\rho}{2}\acomm{\mathit{C}_2}{\mathit{C}_2}) &\tr(\tfrac{\rho}{2}\acomm{\mathit{C}_2}{\mathit{C}_3})\\
        \tr(\tfrac{\rho}{2}\acomm{\mathit{C}_3}{\ln(\rho)+\mathit{I}}) & \tr(\tfrac{\rho}{2}\acomm{\mathit{C}_3}{\mathit{C}_2}) & \tr(\tfrac{\rho}{2}\acomm{\mathit{C}_3}{\mathit{C}_3})},
\end{equation}
\begin{equation}\label{eq:litsurv_betaBar_F2}
    \itbar{\beta}_2 =  \dfrac{1}{k_\text{B}^2\tau^3\tbar{\Omega}}\vmqty{\tr(\tfrac{\rho}{2}\acomm{\mathit{C}_1}{\ln(\rho)+\mathit{I}}) &\tr(\tfrac{\rho}{2}\acomm{\mathit{C}_1}{\mathit{C}_1}) &   \tr(\tfrac{\rho}{2}\acomm{\mathit{C}_1}{\mathit{C}_3}) \\
        \tr(\tfrac{\rho}{2}\acomm{\mathit{C}_2}{\ln(\rho)+\mathit{I}}) &\tr(\tfrac{\rho}{2}\acomm{\mathit{C}_2}{\mathit{C}_1}) & \tr(\tfrac{\rho}{2}\acomm{\mathit{C}_2}{\mathit{C}_3})\\ \tr(\tfrac{\rho}{2}\acomm{\mathit{C}_3}{\ln(\rho)+\mathit{I}}) &
        \tr(\tfrac{\rho}{2}\acomm{\mathit{C}_3}{\mathit{C}_1}) &  \tr(\tfrac{\rho}{2}\acomm{\mathit{C}_3}{\mathit{C}_3})},
\end{equation}
\begin{equation}\label{eq:litsurv_betaBar_F3}
    \itbar{\beta}_3 =  -\dfrac{1}{k_\text{B}^2\tau^3\tbar{\Omega}}\vmqty{\tr(\tfrac{\rho}{2}\acomm{\mathit{C}_1}{\ln(\rho)+\mathit{I}}) &\tr(\tfrac{\rho}{2}\acomm{\mathit{C}_1}{\mathit{C}_1}) &  \tr(\tfrac{\rho}{2}\acomm{\mathit{C}_1}{\mathit{C}_2}) \\ \tr(\tfrac{\rho}{2}\acomm{\mathit{C}_2}{\ln(\rho)+\mathit{I}})&
        \tr(\tfrac{\rho}{2}\acomm{\mathit{C}_2}{\mathit{C}_1}) & \tr(\tfrac{\rho}{2}\acomm{\mathit{C}_2}{\mathit{C}_2}) \\ \tr(\tfrac{\rho}{2}\acomm{\mathit{C}_3}{\ln(\rho)+\mathit{I}})&
        \tr(\tfrac{\rho}{2}\acomm{\mathit{C}_3}{\mathit{C}_1}) & \tr(\tfrac{\rho}{2}\acomm{\mathit{C}_3}{\mathit{C}_2}) }.
\end{equation}
One can do elementary operations on the columns of the determinants in Eqs. \eqref{eq:litsurv_betaBar_F1}-\eqref{eq:litsurv_betaBar_F3}, and then consider the scaling $\tbar{\Omega} = \tfrac{1}{(k\tau)^3}\Omega$ to get the following expressions for $\itbar{\beta}$
\begin{align}\label{eq:litsurv_beta1}
    \begin{split}
        \itbar{\beta_1} = & -\dfrac{1}{k_\text{B}^2\tau^3\tbar{\Omega}}\vmqty{\tr(\tfrac{\rho}{2}\acomm{\mathit{C}_1}{\ln(\rho)}) &  \tr(\tfrac{\rho}{2}\acomm{\mathit{C}_1}{\mathit{C}_2}) & \tr(\tfrac{\rho}{2}\acomm{\mathit{C}_1}{\mathit{C}_3})\\
            \tr(\tfrac{\rho}{2}\acomm{\mathit{C}_2}{\ln(\rho)}) & \tr(\tfrac{\rho}{2}\acomm{\mathit{C}_2}{\mathit{C}_2}) &\tr(\tfrac{\rho}{2}\acomm{\mathit{C}_2}{\mathit{C}_3})\\
            \tr(\tfrac{\rho}{2}\acomm{\mathit{C}_3}{\ln(\rho)}) & \tr(\tfrac{\rho}{2}\acomm{\mathit{C}_3}{\mathit{C}_2}) & \tr(\tfrac{\rho}{2}\acomm{\mathit{C}_3}{\mathit{C}_3})}, \\
        & = -k_\text{B}\beta_1,
    \end{split}
\end{align}
\begin{align}\label{eq:litsurv_beta2}
    \begin{split}
        \itbar{\beta_2} = & \dfrac{1}{k_\text{B}^2\tau^3\tbar{\Omega}}\vmqty{\tr(\tfrac{\rho}{2}\acomm{\mathit{C}_1}{\ln(\rho)}) &\tr(\tfrac{\rho}{2}\acomm{\mathit{C}_1}{\mathit{C}_1}) &   \tr(\tfrac{\rho}{2}\acomm{\mathit{C}_1}{\mathit{C}_3}) \\
            \tr(\tfrac{\rho}{2}\acomm{\mathit{C}_2}{\ln(\rho)}) &\tr(\tfrac{\rho}{2}\acomm{\mathit{C}_2}{\mathit{C}_1}) & \tr(\tfrac{\rho}{2}\acomm{\mathit{C}_2}{\mathit{C}_3})\\ \tr(\tfrac{\rho}{2}\acomm{\mathit{C}_3}{\ln(\rho)}) &
            \tr(\tfrac{\rho}{2}\acomm{\mathit{C}_3}{\mathit{C}_1}) &  \tr(\tfrac{\rho}{2}\acomm{\mathit{C}_3}{\mathit{C}_3})}, \\
        & = k_\text{B}\beta_2,
    \end{split}
\end{align}
\begin{align}\label{eq:litsurv_beta3}
    \begin{split}
        \itbar{\beta_3} = & -\dfrac{1}{k_\text{B}^2\tau^3\tbar{\Omega}}\vmqty{\tr(\tfrac{\rho}{2}\acomm{\mathit{C}_1}{\ln(\rho)}) &\tr(\tfrac{\rho}{2}\acomm{\mathit{C}_1}{\mathit{C}_1}) &  \tr(\tfrac{\rho}{2}\acomm{\mathit{C}_1}{\mathit{C}_2}) \\ \tr(\tfrac{\rho}{2}\acomm{\mathit{C}_2}{\ln(\rho)})&
            \tr(\tfrac{\rho}{2}\acomm{\mathit{C}_2}{\mathit{C}_1}) & \tr(\tfrac{\rho}{2}\acomm{\mathit{C}_2}{\mathit{C}_2}) \\ \tr(\tfrac{\rho}{2}\acomm{\mathit{C}_3}{\ln(\rho)})&
            \tr(\tfrac{\rho}{2}\acomm{\mathit{C}_3}{\mathit{C}_1}) & \tr(\tfrac{\rho}{2}\acomm{\mathit{C}_3}{\mathit{C}_2}) }, \\
        & = -k_\text{B}\beta_3.
    \end{split}
\end{align}
We use the scaled $\beta$ to write Eq. \eqref{eq:litsurv_rhoD_fin} in the following way,
\begin{equation}\label{eq:litsurv_SEA_EoM}
    \dv{\rho_D}{t} = -\dfrac{1}{\tau}\left[\rho\ln(\rho)+\half\sum_i(-1)^i\beta_i\acomm{\mathit{C}_i}{\rho}\right].
\end{equation}
Using \ac{beta} thus expressed in Eqs. \eqref{eq:litsurv_beta1}-\eqref{eq:litsurv_beta3}, along with $\Omega$ in Eq. \eqref{eq:litsurv_SEA_EoM}, and little algebra, we get the following compact form ($\hbar,k_\text{B}=1$).
\begin{align}\label{eq:litsurv_SEA_compact}
    \begin{split}
        &\dv{\rho}{t} +\imi\comm{\mathcal{H}}{\rho} =\\
        &  -\dfrac{1}{\tau}\tfrac{\vmqty{\mathit{B}\rho\ln(\rho) & \half\acomm{\mathit{C}_1}{\rho} &  \half\acomm{\mathit{C}_2}{\rho} &  \half \acomm{\mathit{C}_3}{\rho}\\
                \tr(\tfrac{\rho}{2}\acomm{\mathit{C}_1}{\mathit{B}\ln(\rho)}) & \tr(\rho\mathit{C}_1^2) &  \tr(\tfrac{\rho}{2}\acomm{\mathit{C}_1}{\mathit{C}_2}) & \tr(\tfrac{\rho}{2}\acomm{\mathit{C}_1}{\mathit{C}_3})\\ \tr(\tfrac{\rho}{2}\acomm{\mathit{C}_2}{\mathit{B}\ln(\rho)}) &\tr(\tfrac{\rho}{2}\acomm{\mathit{C}_2}{\mathit{C}_1}) & \tr(\rho\mathit{C}_2^2) & \tr(\tfrac{\rho}{2}\acomm{\mathit{C}_1}{\mathit{C}_3})\\
                \tr(\tfrac{\rho}{2}\acomm{\mathit{C}_3}{\mathit{B}\ln(\rho)}) &	 \tr(\tfrac{\rho}{2}\acomm{\mathit{C}_3}{\mathit{C}_1}) & \tr(\tfrac{\rho}{2}\acomm{\mathit{C}_3}{\mathit{C}_2}) & \tr(\rho\mathit{C}_3^2)}}{\vmqty{\tr(\tfrac{\rho}{2}\acomm{\mathit{C}_1}{\mathit{C}_1}) &  \tr(\tfrac{\rho}{2}\acomm{\mathit{C}_1}{\mathit{C}_2}) & \tr(\tfrac{\rho}{2}\acomm{\mathit{C}_1}{\mathit{C}_3})\\
                \tr(\tfrac{\rho}{2}\acomm{\mathit{C}_2}{\mathit{C}_1}) & \tr(\tfrac{\rho}{2}\acomm{\mathit{C}_2}{\mathit{C}_2}) &\tr(\tfrac{\rho}{2}\acomm{\mathit{C}_2}{\mathit{C}_3})\\
                \tr(\tfrac{\rho}{2}\acomm{\mathit{C}_3}{\mathit{C}_1}) & \tr(\tfrac{\rho}{2}\acomm{\mathit{C}_3}{\mathit{C}_2}) & \tr(\tfrac{\rho}{2}\acomm{\mathit{C}_3}{\mathit{C}_3})}}.
    \end{split}
\end{align}
Where we have used the Eq. \eqref{eq:litsurv_SchroEvo} to write the Hamiltonian evolution part. This equation was first introduced by Beretta in Ref. \cite{beretta_1984_quantuma}. Hence, we will call it \ac{BSEA}. On the other hand, by not explicitly expressing $\beta$s and using Eq. \eqref{eq:litsurv_SEA_compact}, we write the BSEA evolution equation of motion in the nice form as under,
\begin{equation}\label{eq:litsurv_SEA_nice}
    \dv{\rho}{t} =-\imi\comm{\hamiltonian}{\rho}-\dfrac{1}{\tau}\left[\mathit{B}\rho\ln(\rho)+\half\sum_i(-1)^i\beta_i\acomm{\mathit{C}_i}{\rho}\right]
\end{equation}
If we compare Eq. \eqref{eq:litsurv_KSGL1} and \eqref{eq:litsurv_SEA_nice}, we can see the latter does not contain terms of the form $V_j^\dagger\rho V_j$, which populates zero eigenvalued states. And instead, the BSEA equation contains terms involving $\ln(\rho)$ which considers the contribution due to the entropy non-decrease principle. To restrict the evolution to non-zero eigenvalued subspace of $\rho$, an operator $B$ is introduced which is diagonal in the eigenbasis of $\rho$ and is constructed by substituting each non-zero eigenvalues of $\rho$ by one. Thus, one avoids a major issue (preservation of rank space of $\rho$) with KSGL formalism while maintaining a nonlinear evolution of $\rho$. A very convenient way of writing the above equation is through the introduction of the following operator
\begin{equation}\label{eq:litsurv_D}
    \mathcal{D} = \dfrac{1}{2\tau}\left(\mathit{B}\ln(\rho)+\sum_i(-1)^i\beta_i\mathit{C}_i\right),
\end{equation}
$\mathit{B}$ is a diagonal operator formed by substituting the nonzero eigenvalues of $\rho$ with ones. We get
\begin{equation}\label{eq:litsurv_SEA_withD}
    \dv{\rho}{t} = -\imi\comm{\hamiltonian}{\rho} - \acomm{\mathcal{D}}{\rho}.
\end{equation}
Thus endeth our derivation of the SEA EoM. In what follows, we will discuss the geometric interpretation of the above formalism.
\section{A Geometric construction}\label{sec:litsurv_Geometry}
\begin{figure}[H]
    \centering
    \includegraphics[width=\textwidth]{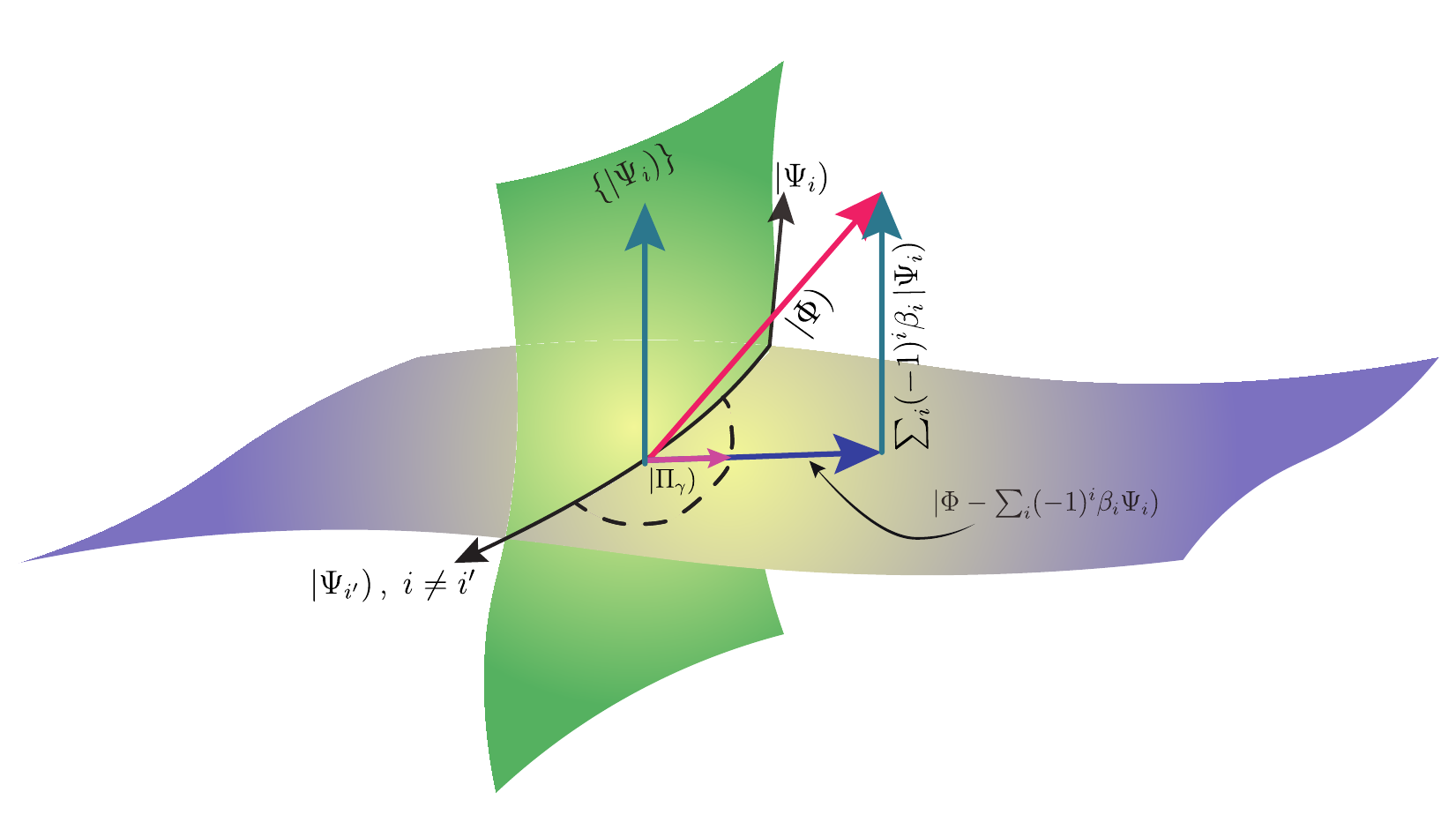}
    \caption[]{\label{fig:litsurv_geometry} A geometric interpretation of the SEA EoM.}
\end{figure}

An attempt at deriving a nonlinear version of Schr\"{o}dinger equation which approximates the Lindbladian evolution equation was taken up by Gisin \textit{et. al.} on geometric grounds \cite{gisin_1995_relevant}. There the flow generated by $\delta\rho_t$ was projected onto the subspace of the Bloch sphere to find the two components of motion, one dissipative and one due to the Hamiltonian evolution. However, this approach was more of a heuristic way to find relevant nonlinear extensions of QM and was not rigorously pursued further. Whereas, SEA formalism is mostly based on geometric grounds, making it more appealing. But before proceeding with the interpretation, we must lay down some logic and terminology required to appreciate the picture as in Fig. \ref{fig:litsurv_geometry}. \\
Consider a set of vectors $\{\bm{v}_0,\bm{v}_1,\cdots,\bm{v}_n\}$ and the linear manifold spanned by their real linear combinations as $\mathsf{L}(\bm{v}_0,\bm{v}_1,\cdots,\bm{v}_n)$. Given some other vector $\bm{u}$ not in $\mathsf{L}$, let $\bm{u}_\mathsf{L}$ denote the orthogonal projection onto $\mathsf{L}$, such that for any vector $\bm{v}$ in $\mathsf{L}$, we have \cite{beretta_2009_nonlinear}
\begin{equation}\label{eq:litsurv_orthogonal_p}
    \bm{v}\vdot\bm{u}_\mathsf{L} = \bm{v}\vdot\bm{u}.
\end{equation}
The projection of $\bm{u}$ onto $\mathsf{L}$ can be expressed using Gram matrix as
\begin{equation}\label{eq:litsurv_orthogonal_g}
    \bm{u}_\mathsf{L} = \sum_{i,j}^r (\bm{u}\vdot\bm{h}_i)[\mathsf{G}]_{ij}^{-1}\bm{h}_j,
\end{equation}
where, $\{\bm{h}_1,\cdots,\bm{h}_r\}$, for $r\le n$ span $\mathsf{L}$, and $\mathsf{G}$ is the Gram matrix given as under
\begin{equation}\label{eq:litsurv_gram}
    \mathsf{G} = \mqty[\bm{h}_1\vdot\bm{h}_1 & \cdots & \bm{h}_r\vdot\bm{h}_1\\
        \vdots & \ddots & \vdots \\
        \bm{h}_1\vdot\bm{h}_r & \cdots & \bm{h}_r\vdot\bm{h}_r].
\end{equation}
Because the set of vectors, $\{\bm{h}_r\}$ is composed of linearly independent vectors, the determinant, $\det(\mathsf{G})$ is strictly positive. We can find the component of $\bm{u}$ orthogonal to $\mathsf{L}$ via the relation
\begin{equation}\label{eq:litsurv_u_orthogonal}
    \bm{u}_{\perp\mathsf{L}} = \bm{u}-\bm{u}_\mathsf{L}.
\end{equation}
As mentioned in Ref. \cite{beretta_2009_nonlinear}, there exists a second method of finding the projection, given as the ratio of two determinants,
\begin{equation}\label{eq:litsurv_orthogonal_d}
    \bm{u}_\mathsf{L} = -\dfrac{\vmqty{0 & \bm{h}_1 & \cdots & \bm{h}_r\\
            \bm{u}\vdot\bm{h}_1 & \bm{h}_1\vdot\bm{h}_1 & \cdots & \bm{h}_r\vdot\bm{h}_1\\
            \vdots & \vdots & \ddots & \vdots\\
            \bm{u}\vdot\bm{h}_r & \bm{h}_1\vdot\bm{h}_r & \cdots& \bm{h}_r\vdot\bm{h}_r}}{\vmqty{\bm{h}_1\vdot\bm{h}_1 & \cdots & \bm{h}_r\vdot\bm{h}_1\\
            \vdots & \ddots & \vdots \\
            \bm{h}_1\vdot\bm{h}_r & \cdots & \bm{h}_r\vdot\bm{h}_r}}.
\end{equation}
Using Eq. \eqref{eq:litsurv_u_orthogonal} and \eqref{eq:litsurv_orthogonal_d}, we get,
\begin{equation}\label{eq:litsurv_gram_final}
    \bm{u}_{\perp\mathsf{L}} = \dfrac{\vmqty{\bm{u} & \bm{h}_1 & \cdots & \bm{h}_r\\
            \bm{u}\vdot\bm{h}_1 & \bm{h}_1\vdot\bm{h}_1 & \cdots & \bm{h}_r\vdot\bm{h}_1\\
            \vdots & \vdots & \ddots & \vdots\\
            \bm{u}\vdot\bm{h}_r & \bm{h}_1\vdot\bm{h}_r & \cdots &\bm{h}_r\vdot\bm{h}_r}}{\vmqty{\bm{h}_1\vdot\bm{h}_1 & \cdots & \bm{h}_r\vdot\bm{h}_1\\
            \vdots & \ddots & \vdots \\
            \bm{h}_1\vdot\bm{h}_r & \cdots & \bm{h}_r\vdot\bm{h}_r}}.
\end{equation}
A comparative study of Eq. \eqref{eq:litsurv_SEA_compact} with \eqref{eq:litsurv_gram_final} reveals the underlying geometric structure clearly. We can consider the linear manifold spanned by the vectors $\kae{\Psi_i}$ as $\mathsf{L}$. The functional $\kae{\Phi} = \vSpdv{\mathit{S}}{\gamma}$ acts as $\bm{u}$ whose component perpendicular to the manifold $\mathsf{L}$ is the one that drives the diffusion rate equation. In Fig. \ref{fig:litsurv_geometry}, we see a short arrow lying in the direction of the $\kae{\Phi}_{\perp\mathsf{L}}$. This is due to the norm fixing of $\Pi_\gamma$ as given by Eq. \eqref{eq:litsurv_metric_dl}. The Lagrange's multipliers $\beta_i$ turn out to be the components of the projection of $\kae{\Phi}$ onto each of the vectors $\kae{\Psi_i}$. The sum of these projections form the green vector on the greenish manifold in  Fig. \ref{fig:litsurv_geometry}. The vector difference between $\kae{\Phi}$, and this sum is the perpendicular vector lying in the purplish manifold orthogonal to the greenish one. Thus we can construct a vector that seeks out the direction of the gradient that is orthogonal to the manifold spanned by the constraints of motion \cite{beretta_2014_steepest,ray_2022_steepestb}.  This construction is at the heart of the SEA approach.
\section{The relaxation time}\label{sec:litsurv_tau}
The \ac{tau} is the most difficult of the Lagrange multipliers to address in the SEA EoM. The relaxation time associated with the system is represented by $\tau$. It is related to the pace of evolution of the state operator, as stated in the literature \cite{beretta_2014_steepest,beretta_2009_nonlinear,li_2016_generalized,li_2018_steepest,ray_2022_steepestb}. Using a Fisher-Rao metric, which becomes a uniform metric in probability space, one may get the following expression from Eq. \eqref{eq:litsurv_metric_dl}.
\begin{equation}\label{eq:litsurv_dot_epsilon}
    \dv{l}{t} = 2\sqrt{\dot{\gamma}_D\cdot\dot{\gamma}_D} = \dot{\epsilon}.
\end{equation}
Here $\dot{\epsilon}$ is a small positive number, which fixes the norm of $\Pi_{\gamma}$ and maximizes the direction as a consequence \cite{beretta_2014_steepest}. From the evolution equation of state operator $\gamma_D$, we have,
\begin{equation}\label{eq:litsurv_gammaD_tau}
    \dot{\gamma}_D = \kae{\Pi_{\gamma}} = \frac{1}{\tau}\kae{\Phi-\sum_{i}\beta_i\mathbf{C_i}}.
\end{equation}
Using these two and defining $\kae{\Lambda}$ as an affinity vector that draws the motion towards SEA evolution, one can write the following expressions involving $\tau$ \cite{ray_2022_steepestb}
\begin{align}\label{eq:litsurv_tau}
    \begin{split}
        \tau & = \frac{\sqrt{\inprc{\Lambda}{\Lambda}}}{\dot{\epsilon}},                                                  \\
        & = \frac{\seaexp{\Phi-\sum_{i}\beta_i\mathbf{C_i}}{\hat{G}^{-1}}{\Phi-\sum_{i}\beta_i\mathbf{C_i}}}{\Pi_S}.
    \end{split}
\end{align}
$\tau$ is also inversely related to the rate of entropy formation. As a result, with greater $\tau$, we will witness less entropy creation and dissipation; the system does not relax rapidly since the speed is high. In the event of low $\tau$ values, the system will relax faster and entropy formation will be increased. Both these $\tau$ characteristics are examined in this thesis in chapters \ref{ch:qubitSEA} and \ref{ch:qwalker}. As can be seen, $\tau$ is reliant on $\rho$, but as the literature suggests, constant non-zero $\tau$ can also be useful in revealing the characteristics of the general motion.

\section*{Summary}
We have begun this chapter with a brief overview of the phenomenological modeling of decoherence via the application of a Lindblad-type master equation. We commented on some of the shortcomings of this formalism which acted as motivation for using the SEA formalism. Thereafter, we introduce and derive the form of the entropy functional from the principle of reversible and irreversible processes as a fundamental property of such evolution. We then discuss the concept of stability as understood in the context of SEA, and rephrase the second law of thermodynamics. Furthermore, we present a discussion on the ontic status of the mixed states on a level similar to that of pure states in the von Neumann formalism. This provides us the necessary background to state the postulates of quantum mechanics and state the steepest entropy ascent ansatz. We then derive the Beretta SEA equation of motion for a single constituent of matter. We have also provided a geometric backdrop to embed the concept of SEA evolution. Finally, we present our remarks on the nature and role of the relaxation time parameter.

\conditionalBlankPage

\chapter{SEA in a Two-level System}\label{ch:qubitSEA}
\blankpage
\newpage
\begin{figure}[H]
    \centering
    \captionsetup{justification=centering}
    \includegraphics[height=0.75\textheight, width=0.75\textwidth]{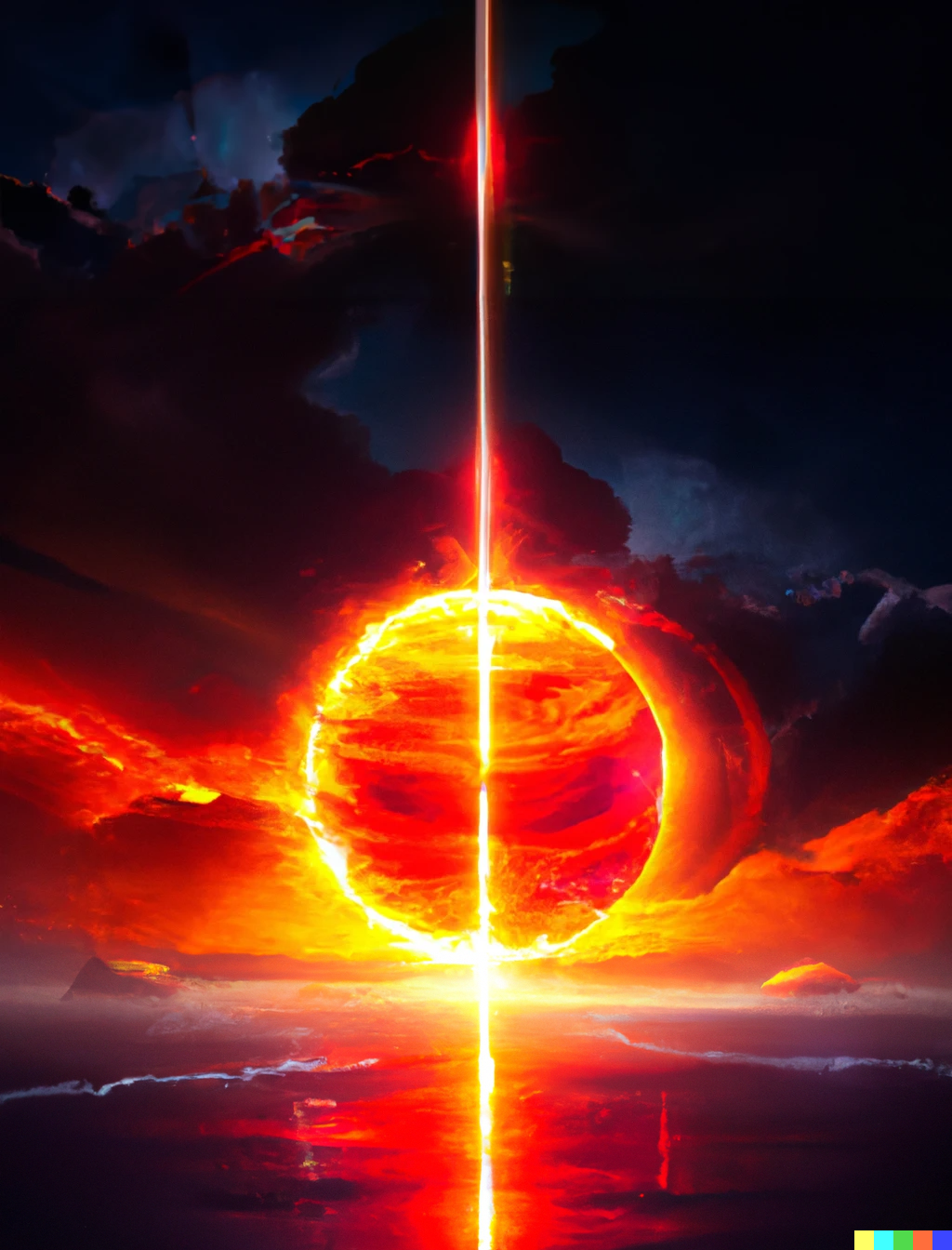}
    \caption*{\emph{The rising sun above the sea,}\\
        \emph{smothered the duality in me}\\
        \emph{while its shimmering rays of hope,}\\
        \emph{sent my doubts flying.}
    }
\end{figure}
\thispagestyle{empty}
\newpage
\blankpage
\lettrine[lines=3]{O}{ur} discussion in the previous chapter demonstrated that the steepest entropy ascent (SEA) formalism maximizes local entropy production in an isolated system, thus producing `spontaneous decoherence' adhering to the second law of thermodynamics. However, unless we consider some simple physical systems and apply the SEA principle, we won't be able to appreciate the robustness of the approach, therefore will miss out on the physical implications of the same. The simplest possible case to study such a nonlinear evolution is the two-level system, a qubit. In this chapter, we will consider the qubit and its evolution under SEA. We will first present the exact analytical solution, followed by a scalable yet approximated analytical solution. We will conclude the chapter by comparing these results with full numerical results.
\section{The Bloch sphere}\label{sec:qubit_bloch_sphere}
A general two-level system can be represented by a parametrized form of the density matrix $\rho$. In this form, a vector $\bm{r}$ is associated with $\rho$, such that $\bm{r}$ is the radial vector of a Riemann sphere. In literature, this representation is also known as the Bloch sphere representation \cite{nielsen_2000_quantum}. Each point of this sphere is a valid state, the points on the surface represent idempotent states, $\rho^2 = \rho$, while those on the inside represent mixed-density matrices. As the general two-level system can be represented by the generators of SU$(2)$ algebra, we get the following representation
\begin{equation}\label{eq:qubit_bloch}
    \rho = \half\left(\mathit{I}+ \bm{r}\vdot\bm{\sigma}\right).
\end{equation}
Where $\bm{\sigma}$ is the Pauli vector comprised of three traceless $2\times 2$ matrices given as follows
\begin{equation}\label{eq:qubit_pauli}
    \sigma_1 = \mqty[0 & 1\\ 1 & 0], ~~ \sigma_2 = \mqty[0 & -\imi\\ \imi & 0], ~~ \sigma_3 = \mqty[1 & 0\\0& -1].
\end{equation}
\begin{figure}[t]
    \centering
    \includegraphics[height=0.25\textheight]{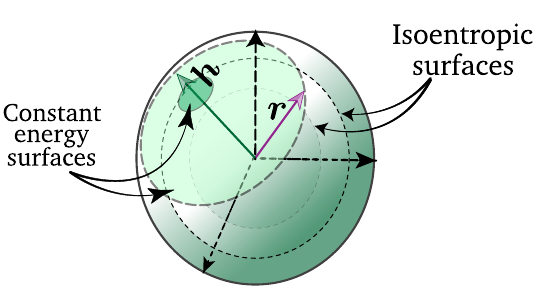}
    \caption[Two-level Bloch sphere]{\label{fig:qubit_bloch}Bloch sphere representation of a two-level system. The purple arrow denotes the Bloch vector $\bm{r}$, while the vector along the precession axis for Hamiltonian is denoted by $\bm{h}$ using a green arrow. Isoentropic and constant energy surfaces are appropriately labeled. Image is taken from \cite{ray_2022_steepestb}.}
\end{figure}
One can see from Eq. \eqref{eq:qubit_bloch}, if $\bm{r} = \bm{0}$, we get $\rho = \half\mathit{I}$, which is the maximally mixed state. Thus at the center of the Bloch sphere lies the state with maximum entropy, the value of which is given by, using Eq. \eqref{eq:litsurv_ent_form}, $\mathit{S} = k\ln(2)$.
In this form, the eigenvalues of $\rho$ are
\begin{equation}\label{eq:qubit_bloch_eigen}
    \lambda_\pm = \dfrac{1\pm r}{2}.
\end{equation}
The entropy thus computed has the following dependence on the magnitude $r$ of $\bm{r}$,
\begin{align}\label{eq:qubit_entropy_onr}
    \begin{split}
        \mathit{S}=& -k\left[\lambda_+\ln(\lambda_+)+\lambda_-\ln(\lambda_-)\right],\\
        =&-k\left[\frac{1+r}{2}\ln(\frac{1+r}{2})+\frac{1-r}{2}\ln(\frac{1-r}{2})\right].
    \end{split}
\end{align}
From Eq. \eqref{eq:qubit_entropy_onr}, it is evident that the entropy function is zero on the surface $r=1$. It is also clear that the constant entropic surfaces are formed by concentric balls within the Bloch sphere as shown in Fig. \ref{fig:qubit_bloch}. We also have $\tr\rho = 1$, and denoting $\tr\rho^2 = R$, one can write
\begin{equation}\label{eq:qubit_R}
    r = \sqrt{2R-1},
\end{equation}
implying $0\le r\le 1$, for $\half\le R\le 1$.\\
Similarly, the most general Hamiltonian acting on a point on the sphere can be written as
\begin{equation}\label{eq:qubit_hamiltonian}
    \hamiltonian = \left(\omega_0\mathit{I}+\omega\bm{h}\vdot\bm{\sigma}\right).
\end{equation}
We identify $\bm{h}$ as the unit vector in the direction of the axis of rotation induced by $\hamiltonian$. We also note the eigenvalues of $\hamiltonian$ are as under
\begin{equation}\label{eq:qubit_energy_eigen}
    h_\pm = \omega_0\pm\omega, ~~ \text{with }~~ \omega_0 = \half\tr\hamiltonian.
\end{equation}
The precession frequency is $2\omega$. As Fig. \ref{fig:qubit_bloch} shows, for a fixed entropy (fixed $r$), the radius vector traces out path along circles perpendicular to $\bm{h}$ (constant energy surfaces).
\section{Exact analytical solution}\label{sec:qubit_GPB}
\begin{figure}[t]
    \centering
    \includegraphics[]{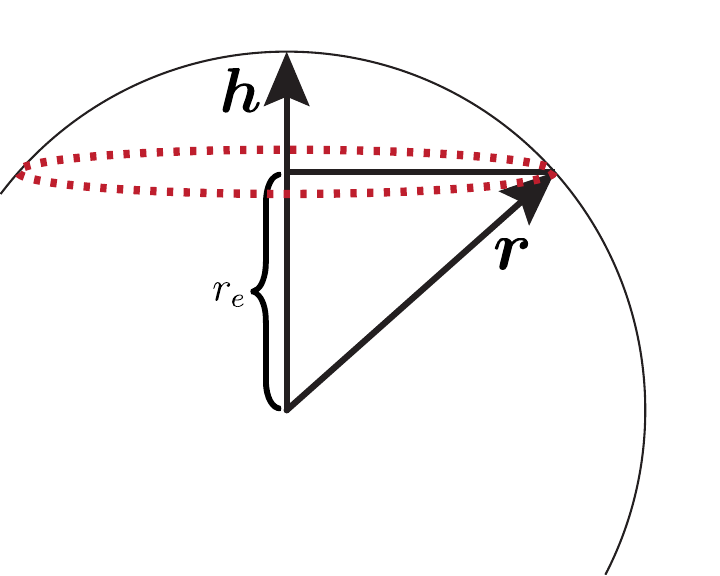}
    \caption[Schematic for $r_e$.]{\label{fig:qubit_r_e}A schematic for showing $r_e = \bm{h}\vdot\bm{r}$.}
\end{figure}

The exact analytical solution to the two-level system was given by Gian Paolo Beretta \cite{beretta_1985_entropy}. Before we discuss that here, let us note that for a single particle, only two constraints \textit{i.e.,} that of energy and probability conservation are required. This reduces the BSEA equation of motion Eq. \eqref{eq:litsurv_SEA_compact} in the following expression,
\begin{equation}\label{eq:qubit_sea_2const}
    \dv{\rho}{t} +\imi\comm{\mathit{H}}{\rho} =  -\dfrac{1}{\tau}\tfrac{\vmqty{\rho\ln(\rho) & \rho &  \half\acomm{\rho}{\mathit{H}} \\
            \tr(\rho\ln(\rho)) & 1 &  \tr(\rho\mathit{H}) \\ \tr(\rho\mathit{H}\ln(\rho)) &		\tr(\rho\mathit{H}) & \tr(\rho\mathit{H}^2) \\
        }}{\vmqty{1 &  \tr(\rho\mathit{H}) \\
            \tr(\rho\mathit{H}) & \tr(\rho\mathit{H}^2) }}.
\end{equation}
Because of the constant energy criteria, the Bloch vector $\bm{r}$ is constrained to rotate in a plane at a distance $r_e = \bm{h}\vdot\bm{r}$ from the center about $\bm{h}$, see the red dotted circled trajectory in Fig. \ref{fig:qubit_r_e}. The overall SEA motion will drag $\bm{r}$ towards $\bm{h}$ such that $r_e$ remains unchanged. Given the definition of $\rho$, and $\hamiltonian$ above, we compute the trace relations below \cite{ray_2022_steepestb}
\begin{align}\label{eq:qubit_trace_rel}
    \begin{split}
        \tr(\rho) =& 1,\\
        \tr(\rho\hamiltonian) =& \left(\omega_0+\omega r_e\right),\\
        \tr(\rho\hamiltonian^2) =& \left(\left(\omega^2+\omega_0^2\right)+2\omega_0\omega r_e\right),\\
        \tr(\rho\ln(\rho)) =& \half\left(\ln(\dfrac{1-r^2}{4})+r\ln(\dfrac{1+r}{1-r})\right),\\
        \tr(\rho\hamiltonian\ln(\rho)) =& \dfrac{\omega_0}{2}\left(\ln(\dfrac{1-r^2}{4})+r\ln(\dfrac{1+r}{1-r})\right)\\
        &+ \dfrac{\omega}{2}\left(\ln(\dfrac{1-r^2}{4})+\dfrac{1}{r}\ln(\dfrac{1+r}{1-r})\right)r_e.
    \end{split}
\end{align}
Using these traces, we can find the expressions for Lagrange's multipliers using Eq. \eqref{eq:litsurv_beta1}, and \eqref{eq:litsurv_beta2} with two constraints. Note, $\Omega = \tr(\rho\hamiltonian^2)-(\tr(\rho\hamiltonian))^2$. We have
\begin{align}\label{eq:qubit_beta_exact}
    \begin{split}
        \beta_1  &=~ \dfrac{1}{\Omega}\left[\tr(\rho\ln\rho)\tr(\rho\hamiltonian^2)-\tr(\rho\hamiltonian\ln\rho)\tr(\rho\hamiltonian)\right],\\
        \beta_2  &=~ \dfrac{1}{\Omega}\left[\tr(\rho\ln\rho)\tr(\rho\hamiltonian)-\tr(\rho\hamiltonian\ln\rho)\right].
    \end{split}
\end{align}
Let us denote the following,
\begin{align}\label{eq:qubit_AB}
    \begin{split}
        \mathbf{A} &=~ \left(\ln(\dfrac{1-r^2}{4})+r\ln(\dfrac{1+r}{1-r})\right),\\
        \mathbf{B}&=~\left(\ln(\dfrac{1-r^2}{4})+\dfrac{1}{r}\ln(\dfrac{1+r}{1-r})\right).
    \end{split}
\end{align}
Using Eqs. \eqref{eq:qubit_trace_rel}, \eqref{eq:qubit_beta_exact}, and \eqref{eq:qubit_AB}, we get
\begin{align}
    \beta_I \equiv \beta_1 & = \dfrac{1}{2\omega^2\left(1-r_e^2\right)}\Bigl[\omega^2\left[\mathbf{A}- r_e^2\mathbf{B}\right] + \omega\omega_0 \left[\mathbf{A}-\mathbf{B}\right] \Bigr], \label{eq:qubit_betai} \\
    \beta_H \equiv \beta_2 & = \frac{r_e}{2\omega\left(1-r_e^2\right)}\left[\mathbf{A}-\mathbf{B}\right]. \label{eq:qubit_betah}
\end{align}
Next, we find the expressions for the commutation and anti-commutations given as under
\begin{align}\label{eq:qubit_comm_rel}
    \begin{split}
        \comm{\hamiltonian}{\rho} = &~ \imi\omega\left(\bm{h}\cross\bm{r}\right)\cdot\bm{\sigma},\\
        \acomm{\hamiltonian}{\rho} = &~ \left(\omega_0+\omega r_e\right)\mathit{I}+\left(\omega_0\bm{r}+\omega\bm{h}\right)\cdot\bm{\sigma},\\
        \acomm{\ln(\rho)}{\rho} = &~ \half\mathbf{A}\mathit{I} + \half \mathbf{B}\bm{r}\cdot\bm{\sigma}.
    \end{split}
\end{align}
Now that we have gathered all the necessary expressions, let us recall Eq. \eqref{eq:litsurv_SEA_nice}, and employ them for the dissipative part as follows,
\begin{equation}\label{eq:qubit_analytic_diff_re}
    \dv{(\bm{r}-\bm{r}_e)}{t} = -\dfrac{1-r^2}{2\tau(1-r_e^2)}\ln(\dfrac{1+r}{1-r})\dfrac{(\bm{r}-\bm{r}_e)}{r}.
\end{equation}
In writing the above differential equation, we have used $\dv{\rho}{t} = \half\dv{\bm{r}}{t}\vdot\bm{\sigma}$, and $\bm{r}_e= r_e\bm{h}$. If we confine the motion on the equatorial plane perpendicular to the vector $\bm{h}$ on the Bloch sphere, we get $r_e = 0$, see Fig. \ref{fig:qubit_r_e}. Setting this, we have the following equation \cite{beretta_1985_entropy,ray_2022_steepestb},
\begin{equation}\label{eq:qubit_analytic_diff}
    \dv{r}{t} = -\dfrac{1}{2\tau}(1-r^2)\ln(\dfrac{1+r}{1-r}).
\end{equation}
The solution to the above equation is given as ($r^0 \equiv r(0) = \varepsilon$),
\begin{equation}\label{eq:qubit_analytic_sol}
    r_t = \tanh\left[\half\exp(-\dfrac{t}{\tau})\ln(\dfrac{1+\varepsilon}{1-\varepsilon})\right].
\end{equation}
Since this solution was first given by Beretta, we will refer to this solution as the \ac{GPB} solution.
\section{Fixed Lagrange's multiplier method}\label{sec:qubit_FLM}
The GPB solution, without any doubt, is elegant and exact. However, it cannot be scaled to systems with higher dimensionality (more than two-level systems). Besides, the GPB solution is not available for $r_e \ne 0$ states, which leaves most of the Bloch sphere unexplored. As a consequence one resorts to numerical techniques \cite{cano-andrade_2015_steepestentropyascent}. While numerical solutions are exact and using current computing resources doesn't require much time to be solved, they suffer from two major setbacks. Firstly, the numerical solutions obscure the interesting features and clues of the underlying physical implications of the solutions presented. Secondly, as we scale up in dimensions, and include more complex studies to undertake, these solutions require a lot of computational resources, which may not be readily available. To address and resolve this, in a recent work \cite{ray_2022_steepestb}, this author developed an approximate analytical method to solve the SEA EoM for a single particle. The method relies on fixing the $\beta_i$'s as used in Eq. \eqref{eq:litsurv_SEA_nice}. This means fixing the component of the projections on the constraint manifold as discussed in Sec. \ref{sec:litsurv_Geometry}. As a result, we do not get the exact solution, but it has been shown in Ref. \cite{ray_2022_steepestb}, in higher dimensions we get a good agreement with the numerical results. Since one fixes the $\beta_i$s, \textit{i.e.,} the Lagrange's multipliers, this solution method is known as the fixed Lagrange's multiplier (FLM) scheme. Depending on the region of interest, such as near equilibrium or far-off equilibrium, one fixes the $\beta_i$s accordingly. Below we outline the solution via FLM, and then in the next section, we compare FLM with GPB and numerical results for the two-level system under discussion.\\
We note that $\rho_t$, the density matrix at time $t$ is a unit trace, semi-positive definite matrix for all $t$ during the evolution of the state, suggesting it can be diagonalized throughout the process. We can use this property and the following definition using similarity transformation to go to the diagonal basis of $\rho_t$ at all times.
\begin{defRKR}\label{def:qubit_u_t}
    $\mathsf{U}_t$ exist at each time instance $t$ which takes $\rho_t$ to a diagonal matrix $\rho_t^d$ as given,
    \begin{equation}\label{eq:qubit_defnU_t}
        \rho_t = \mathsf{U}_t\rho_t^d\mathsf{U}_t^{-1}.
    \end{equation}
\end{defRKR}
Next, we state and prove a theorem on the first-order time derivative of time-dependent square matrices,
\begin{theorem}\label{th:qubit_matrix_derivative}
    For a time-dependent square matrix $\mathrm{A(t)}$, with left and right eigenvectors as $\mathrm{y}_i$ and $\mathrm{x}_i$, respectively, and corresponding eigenvalue $\lambda_i$, we have,
    \begin{equation}\label{eq:qubit_matrix_derivative}
        \dv{\lambda_i}{t} = \mathrm{y}_i^{\text{T}}\dv{\mathrm{A}}{t}\mathrm{x}_i.
    \end{equation}
\end{theorem}
\begin{proof}
    We note the following facts,
    \begin{align}
        \mathrm{Ax}_i                       & = ~ \lambda_i\mathrm{x}_i,                      \\
        \mathrm{y}_i^{\text{T}}\mathrm{A}   & = ~ \lambda_i\mathrm{y}_i^{\text{T}},           \\
        \mathrm{y}_i^{\text{T}}\mathrm{x}_j & = ~ \delta_{ij}. \label{eq:qubit_eigvect_ortho}
    \end{align}

    Combining all of the above, we get the following equation,
    \begin{equation}\label{eq:qubit_matrix_eigen}
        \mathrm{y}_i^{\text{T}}\mathrm{Ax}_i = \lambda_i.
    \end{equation}
    Taking derivatives on both sides, we get (suppressing the index $i$),
    \begin{align}\label{eq:qubit_matrix_steps}
        \begin{split}
            \dv{\mathrm{y}^{\text{T}}}{t}\mathrm{Ax} + \mathrm{y}^{\text{T}}\dv{\mathrm{A}}{t}\mathrm{x} + \mathrm{y}^{\text{T}}\mathrm{A}\dv{\mathrm{x}}{t} & = ~ \dv{\lambda}{t}, \\
            \dv{\mathrm{y}^{\text{T}}}{t}\lambda\mathrm{x} + \mathrm{y}^{\text{T}}\dv{\mathrm{A}}{t}\mathrm{x} + \lambda\mathrm{y}^{\text{T}}\dv{\mathrm{x}}{t} & = ~ \dv{\lambda}{t}, \\
            \lambda\left[\dv{(\mathrm{y}^{\text{T}}\mathrm{x})}{t}\right] + \mathrm{y}^{\text{T}}\dv{\mathrm{A}}{t}\mathrm{x} & = ~ \dv{\lambda}{t}.
        \end{split}
    \end{align}
    From the last line, and using Eq. (\ref{eq:qubit_eigvect_ortho}), we get the main result,
    \begin{equation}\label{eq:qubit_matrix_theorem_pr}
        \dv{\lambda_i}{t} = \mathrm{y}_i^{\text{T}}\dv{\mathrm{A}}{t}\mathrm{x}_i.
    \end{equation}
\end{proof}
A direct consequence of the above theorem is the following result,
\begin{equation}\label{eq:qubit_matrix_der_complete}
    \dv{\mathrm{A}}{t} = \mathrm{Y}\dv{\Lambda}{t}\mathrm{X}^{\text{T}},
\end{equation}
where $\mathrm{Y}$ and $\mathrm{X}$ are created by column-wise stacking the respective eigenvectors, while $\Lambda = \textbf{diag}\left[\lambda_i\right]$.
We note that, in our case of $\rho_t$, The \ref{th:qubit_matrix_derivative} suggests that, $\mathsf{U}_t = \mathrm{Y}$, and $\mathsf{U}_t^{-1} = \mathrm{X}$. Having established this identification, we consider Eq. \eqref{eq:litsurv_SEA_nice} and multiply it from left and right with $\mathsf{U}_t^{-1}$ and $\mathsf{U}_t$ respectively. This implies we get the following equation,
\begin{equation}\label{eq:qubit_eqdiag}
    \dv{\rho_t^d}{t} = -\dfrac{1}{2\tau}\left[\acomm{\ln(\rho_t^d)}{\rho_t^d}+\sum_{i}(-1)^i\beta_i\acomm{\mathit{C_i}^d}{\rho_t^d}\right] -i\comm{\hamiltonian^d}{\rho_t^d}.
\end{equation}
In the above equation, we have used $\hamiltonian^d = \mathsf{U}_t^{-1}\hamiltonian\mathsf{U}_t$, and $\mathit{C}_i^d = \mathsf{U}_t^{-1}\mathit{C}_i\mathsf{U}_t$. In the restricted class of problems, for which the eigenbasis of $\rho$ changes solely due to the Hamiltonian, we can have $\mathsf{U}_t$ as constant \cite{ray_2022_steepestb}. In essence, this means that $\rho$ remains diagonal in the energy basis throughout the evolution. Since most of the problems encountered in this thesis belong to this special class of problems, we will focus our attention here.\\
$\rho$ has a spectral decomposition in its eigenbasis $\{\ket{\lambda_i}\}$, and $\rho^d$ is diagonal in the standard basis $\{\ket{i}\}$ ($\ket{i}$ is an $N$-dimensional unit vector),
\begin{align}\label{eq:qubit_spectral}
    \rho^d = \sum_{ij}\lambda_i\delta_{ij}\dyad{i}{j}; & \quad\text{ and }
    \begin{split}
        \rho =& \sum_{i}\lambda_i\dyad{\lambda_i}{\lambda_i},\\
        =&\sum_{ij}\lambda_i\delta_{ij}\dyad{\lambda_i}{\lambda_j}.
    \end{split}
\end{align}
$\delta_{ij}$ is the Kronecker delta. Using these expansions in equation (\ref{eq:qubit_eqdiag}), and by choosing $\left[\hamiltonian^d\right]_{ij}=H_{ij}^d$, and $\left[\mathit{C_s}^d\right]_{ij}=C_{ij}^s$, the r.h.s of equation (\ref{eq:qubit_eqdiag}) can be modified as,
\begin{align}\label{eq:qubit_mod_eqdiag_rhs}
    \begin{split}
        &-\frac{1}{2\tau}\sum_{ijm}\delta_{im} \Big[2\lambda_m\ln(\lambda_m)\delta_{mj}+\sum_{s}(-1)^s\beta_s\left[\lambda_m+\lambda_j\right]C_{mj}^s\Big]\dyad{i}{j}\\ &-i\sum_{ij}\left[\lambda_j-\lambda_i\right]H_{ij}^d\dyad{i}{j}.
    \end{split}
\end{align}

The complete expression upon considering a Euclidean metric reads as
\begin{align}
    \begin{split}
        \sum_{ij}\dot{\lambda}_i\delta_{ij}\dyad{i}{j} = & -i\sum_{ij}\left[\lambda_j-\lambda_i\right]H_{ij}^d\dyad{i}{j}\\
        &-\frac{1}{2\tau}\sum_{ij}\Big[2\lambda_i\ln(\lambda_i)\delta_{ij}- 2\beta_I\lambda_i\delta_{ij}+\beta_H\left[\lambda_j+\lambda_i\right]H_{ij}^d \Big]\dyad{i}{j}.
    \end{split}
\end{align}
According to our construction l.h.s of equation (\ref{eq:qubit_eqdiag}) is diagonal,
\begin{equation}\label{eq:qubit_eigenequation}
    \dv{\lambda_i}{t} = -\frac{1}{\tau}\left[\lambda_i\ln(\lambda_i)-\beta_I\lambda_i+\beta_H\lambda_i H^d_{ii}\right].
\end{equation}
For diagonal density matrices, similarity transformation is identity, so we get,
\begin{equation}\label{eq:qubit_p_iequation}
    \dv{p_i}{t} = -\frac{1}{\tau}\left[ p_i\ln(p_i)-\beta_I p_i+\beta_H p_i H_{ii}\right],
\end{equation}
where $p_i=\left[\rho_t\right]_{ii}$. Both the equations (\ref{eq:qubit_eigenequation}), and (\ref{eq:qubit_p_iequation}) have a similar type of solution, namely that of almost identical non-linear ODE. Using standard techniques and FLM approximation, we arrive at the following expression,
\begin{equation}\label{eq:qubit_p_isol}
    p_i(t) = \exp(\exp(w_i-\frac{t}{\tau})+\beta_I-\beta_H H_{ii}).
\end{equation}
We have $w_i=\ln(\ln(p^0_i)-\beta_I+\beta_H H_{ii})$, where $p^0_i$ is the $i^{\text{th}}$ diagonal entry of initial $\rho$. The solution produced above can be written in a straightforward form, identifying $\mu^c_i = \beta_H H_{ii}-\beta_I$, or for general cases as $\sum_{s}(-1)^s\beta_s C^s_{ii}$; $\tilde{v}_i = e^{w_i}$, and $\mu^t = \exp(-t/\tau)$, we get,
\begin{equation}\label{eq:qubit_p_isimple}
    p_i(t) \equiv p^t_i = \exp(\tilde{v}_i\mu^t-\mu^c_i).
\end{equation}
And we find $\tilde{v}_i$ as,
\begin{align}
             & p_i(0) =  p_i^0,                                        \\
             & \exp(\tilde{v}_i\mu^0-\mu^c_i) =  p_i^0,                \\
    \implies & \tilde{v}_i =                       \ln(p_i^0)+\mu^c_i.
\end{align}
Hence, we can write $\rho^d_t = \sum_{i}p_i^t\dyad{i}{i}$. For a general initial $\rho$ with off-diagonal terms can be written as - $\rho = \mathsf{U}_t\rho^d_t\mathsf{U}_t^{-1}$, whereas if we consider only diagonal $\rho$'s, we get $\rho = \rho^d_t$. Including the Hamiltonian evolution, we get the following equation for uniform metric ($\mathcal{U}_t \equiv\exp(-i\mathit{H}t)$, and projections $\mathbb{P}_m = \dyad{m}{m}$),
\begin{equation}\label{eq:qubit_fullevolution}
    \rho_t =\mathcal{U}_t\mathsf{U}_t\left(\sum_{m}\exp(\mu^t_m-\mu^c_m)\mathbb{P}_m\right)\mathsf{U}_t^{-1}\mathcal{U}_t^{\dagger},
\end{equation}
where, $\mu^t_m = \left(\ln(p_m^0)+\mu^c_m\right)e^{-t/\tau} = \tilde{v}_m\mu^t$, and $\mu^c_m = \sum_{s}(-1)^s\beta_sC_{mm}^s$.
So far, as we can see, equation (\ref{eq:qubit_fullevolution}) represents the evolution of $\rho_t$s diagonal in the energy basis. Thus we have a general FLM solution for $N$-level systems. In the next section, we will apply this to the case of qubits and analyze with respect to numerical solutions of the Eq. \eqref{eq:litsurv_SEA_nice}.
\section{FLM on a qubit}\label{sec:qubit_FLM_qubit}
Since pure states form the limit cycles of the SEA motion, they follow Schr\"{o}dinger dynamics. To study the SEA dynamics, one, therefore, quenches the system from a pure state and prepares it in a mixed state \cite{cano-andrade_2015_steepestentropyascent}.
\begin{align}\label{eq:qubit_quench}
    \begin{split}
        \rho = &\varepsilon \rho_0 + \left(1-\varepsilon\right)\rho_u\\
        \sum_{i}\lambda_i\dyad{\lambda_i}{\lambda_i}=& \sum_{i}\varepsilon\lambda_i^0\dyad{\lambda_i^0}{\lambda_i^0}+\left(1-\varepsilon\right)\frac{\mathrm{I}}{\tr(\mathrm{I})}.
    \end{split}
\end{align}
If $\rho$ is diagonal in the $\rho_0$ basis, then
\begin{equation}\label{eq:qubit_initial}
    \lambda'_i = \varepsilon\lambda_i^0+(1-\varepsilon)\frac{1}{N},
\end{equation}
where $N=\tr(\mathrm{I})$, which for qubit is 2. $\varepsilon$ is a variable parameter $\in\left[0,1\right]$, with zero value denoting the completely mixed state. Armed with all these and a Euclidean metric, we consider Eqs. (\ref{eq:qubit_eigenequation}), \eqref{eq:qubit_bloch_eigen},
\begin{align}\label{eq:qubit_qlambda_diff}
    \begin{split}
        \dv{\lambda_{\pm}}{t} = & -\frac{1}{\tau}\left[\lambda_{\pm}\ln(\lambda_{\pm})+(\beta_HH_{\pm}^d-\beta_I)\lambda_{\pm}\right]\\
        \implies\pm\dot{r} = & \mp\frac{1}{\tau}\Big[\left(1\pm r\right)\ln(\frac{1\pm r}{2}) +(\beta_H H_{\pm}^d-\beta_I)\left(1\pm r\right)\Big].
    \end{split}
\end{align}
After that, we can write for the dissipative part of the motion as before,
\begin{align}\label{eq:qubit_r_sol_pm}
    \begin{split}
        r_t = & -1 + 2\exp(\left(\mu^t_+-\mu^c_+\right)), \text{ for $\lambda_+$, and}\\
        r_t = & 1 - 2\exp(\left(\mu^t_--\mu^c_-\right)), \text{ for $\lambda_-$}.
    \end{split}
\end{align}
Thence,
\begin{equation}\label{eq:qubit_rtime}
    r_t = \left(\exp(\mu^t_+-\mu^c_+)-\exp(\mu^t_--\mu^c_-)\right),
\end{equation}\\
and also,
\begin{equation}\label{eq:qubit_lambda_pm}
    \lambda_{\pm}(t) = \exp(\mu^t_{\pm}-\mu^c_{\pm}).
\end{equation}
Where, $\mu^t_{\pm} = \left(\ln(\lambda_{\pm}')+(\beta_H H_{\pm}-\beta_I)\right)e^{-\frac{t}{\tau}}$, $ \lambda' $ as in equation (\ref{eq:qubit_initial}). The full evolution is given by,
\begin{equation}\label{eq:qubit_qubitevol}
    \mathcal{U}_t\mqty[\dmat{\exp(\mu^t_+-\mu^c_+),\exp(\mu^t_--\mu^c_-)}]\mathcal{U}_t^{\dagger}.
\end{equation}
This solution above in Eq. (\ref{eq:qubit_qubitevol}) works when we have Lagrange multipliers fixed using initial conditions \textit{i.e.,} FLM method. Otherwise, in general $\beta_i$'s depend on time-dependent $r$ and on constant $r_e=\hat{h}\cdot\vec{r}$.\\

\begin{figure}[t]
    \centering
    \includegraphics[height=0.35\textheight]{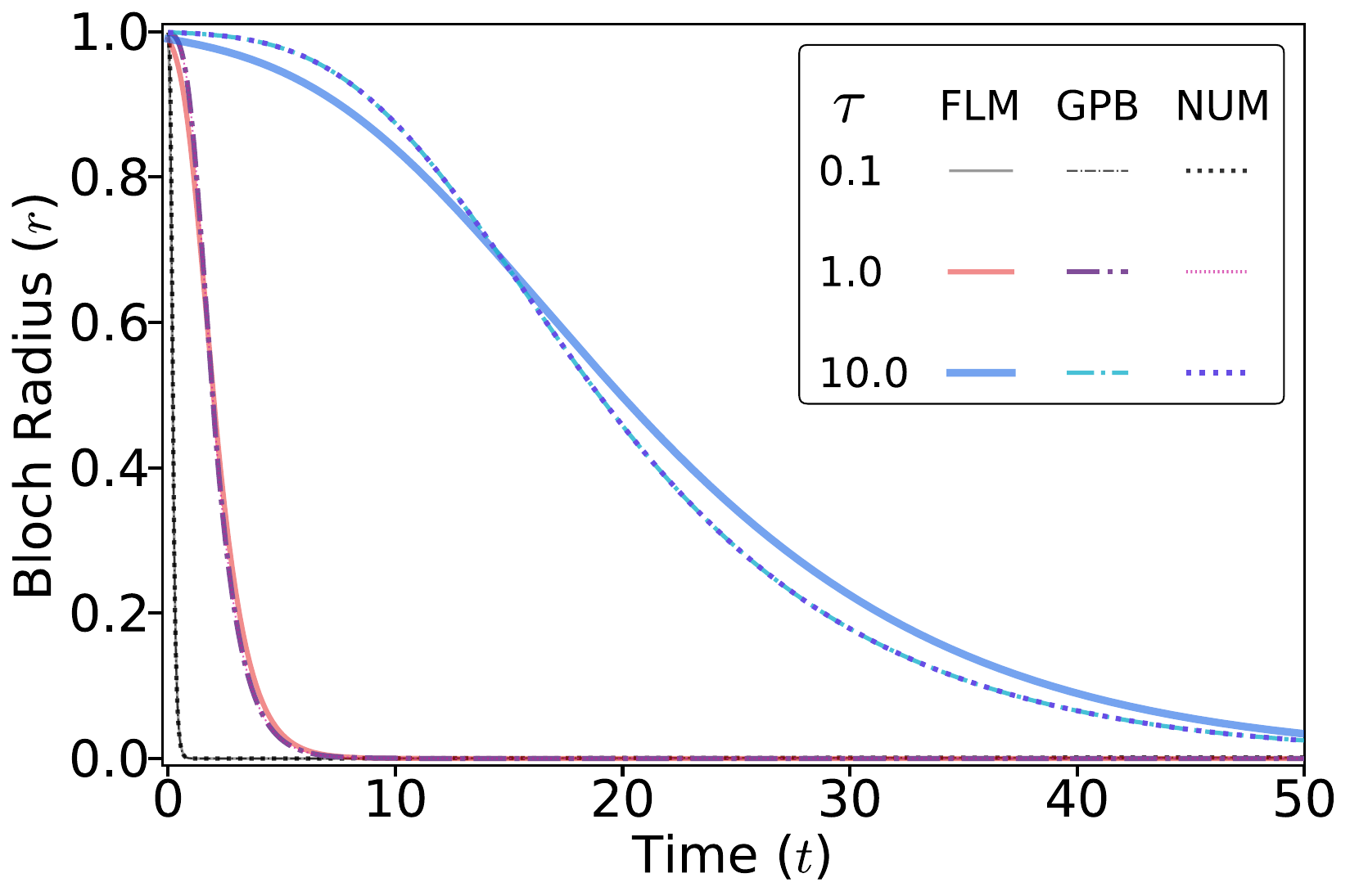}
    \caption[Magnitude decay of the Bloch radius over time, using out-of-equilibrium $\rho$.]{\label{fig:qubit_radial} Relaxation time dependence decay of the magnitude of the Bloch radius $r_t$ over time. FLM computations using initial $\rho$ to fix the $\beta_i$s are denoted by solid lines,  GPB Eq. (\ref{eq:qubit_analytic_sol}) are plotted using dot-dash, and the direct numerical simulation (NUM) of Eq. (\ref{eq:litsurv_SEA_nice}) are shown in dotted lines. Image cited from Ref. \cite{ray_2022_steepestb}.}
\end{figure}
Let us now understand the SEA approach through simple well-known, and well-studied physical conditions. We take $\hat{h} = \hat{z}$, and focus on the states lying on the equatorial plane of the Bloch sphere, $r_e=0$. $\hamiltonian$ in this scenario becomes $\omega\sigma_z$, which is diagonal in the standard basis. Using the expression for $ \lambda' $ provided in equation (\ref{eq:qubit_initial}), we write down the $\beta_i$'s as (Eqs. (\ref{eq:qubit_betai}), and (\ref{eq:qubit_betah})),
\begin{align}\label{eq:qubit_betass}
    \begin{split}
        \beta_I = & \dfrac{k}{2}\left[\ln(\dfrac{1-\varepsilon^2}{4})+\varepsilon\ln(\dfrac{1+\varepsilon}{1-\varepsilon})\right],\\
        \beta_H = &\quad 0.
    \end{split}
\end{align}
\begin{figure}[t]
    \centering
    \includegraphics[height=0.35\textheight]{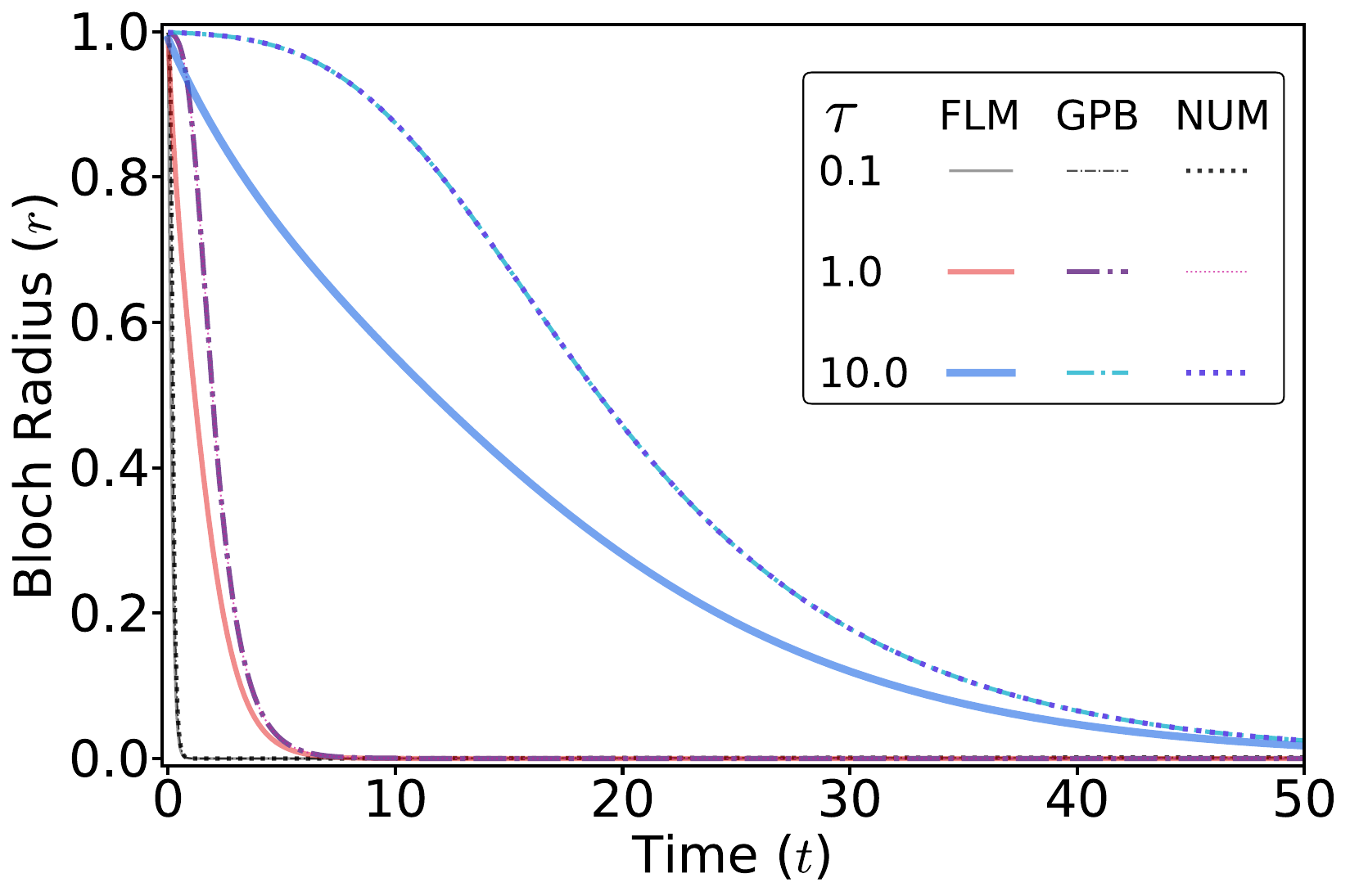}
    \caption[Magnitude decay of the Bloch radius over time, using equilibrium $\rho$.]{\label{fig:qubit_radial_w} Relaxation time dependence decay of the magnitude of the Bloch radius $r_t$ over time. FLM computations using final $\rho$ to fix the $\beta_i$s are denoted by solid lines,  GPB Eq. (\ref{eq:qubit_analytic_sol}) are plotted using dot-dash, and the direct numerical simulation (NUM) of Eq. (\ref{eq:litsurv_SEA_nice}) are shown in dotted lines. Image cited from Ref. \cite{ray_2022_steepestb}.}
\end{figure}
Let the initial $\rho$ be $a\dyad{0}{0} + b\dyad{1}{1}$. Then radius $ r' $ after quenching is $r' = \varepsilon r = \varepsilon\abs{a-b}$. We get, $\mu^t_{\pm} = \left(\ln(\frac{1\pm\varepsilon}{2})-\beta_I\right)e^{-\frac{t}{\tau}}$,
and $\mu^c_{\pm}= \beta_I$. And finally, the evolution equation becomes,
\begin{equation}\label{eq:qubit_revolution}
    r_t = \left(e^{\theta_+^t}-e^{\theta_-^t}\right),
\end{equation}
where, $\theta_{\pm}^t = \mu^t_{\pm}-\mu^c_{\pm}$.\\
Using the Eq. \eqref{eq:qubit_revolution} for $\omega = 5$, and $\varepsilon  = 0.999$ (just an arbitrary number close to one), one can plot the evolution of $r_t$ vs time as in Fig. \ref{fig:qubit_radial}. In this figure, we have plotted $r_t$ for various values of the relaxation time of the system. In the same figure, we have also included the GPB solution (Eq. \eqref{eq:qubit_analytic_sol}), and the numerical solution (NUM) resulting from solving Eq. \eqref{eq:litsurv_SEA_nice}. As we can see from the figure, FLM results fit nicely within the neighborhood of GPB/NUM solutions.
However, it depends on what kind of initial condition are we basing our FLM computation on. In Fig. \ref{fig:qubit_radial}, we have considered FLM values from the far-off equilibrium region. Meaning, we have computed $\beta_i$s from initial density matrices.
\begin{figure}[H]
    \centering
    \subfigure[\label{fig:qubit_2dspiral}]{\includegraphics[height=0.23\textheight]{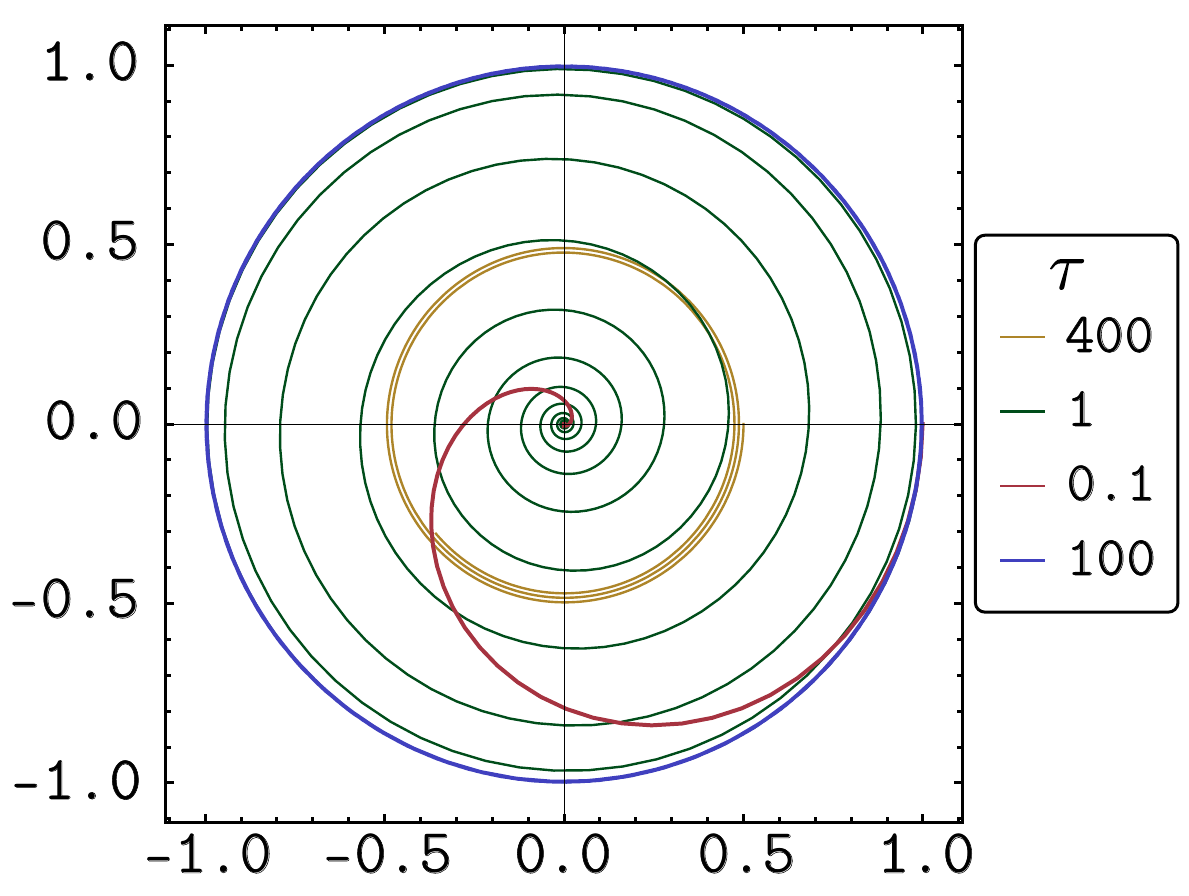}}
    \subfigure[\label{fig:qubit_entropyspiral}]{\includegraphics[height=0.23\textheight]{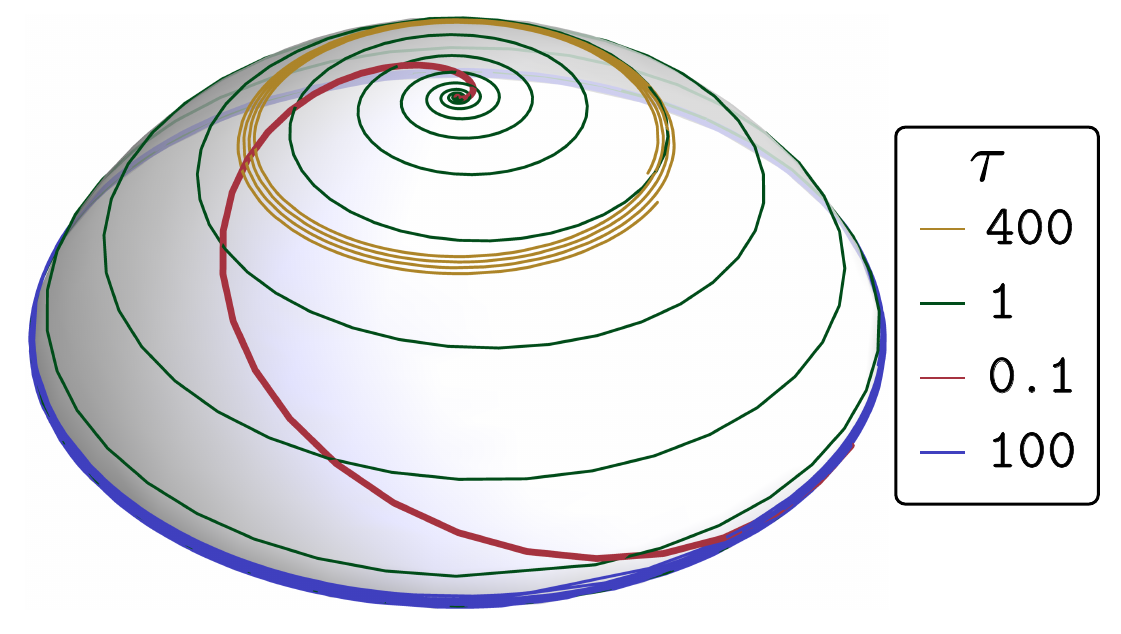}}
    \caption[Plot of spiral trajectories in the Bloch sphere]{\label{fig:qubit_spiralmain} (a) $\tau$ dependent spirals formed by tracing the state operator in the equatorial plane of the Bloch sphere as viewed towards Nadir from the North pole. The evolution has been studied till $t=30$. (b) The same spirals are projected onto the surface of revolution generated from the entropy function in Eq. \eqref{eq:qubit_entropy_onr}. The case with $\tau = 400$ corresponds to the case with $\varepsilon = 0.5$, while the rest of the cases have $\varepsilon = 0.999$. Lower $\tau$ states rise faster, and the lowest has the steepest ascent. Image cited from Ref. \cite{ray_2022_steepestb}.}
\end{figure}
However, if we computed the same using $\rho_u$ instead, the FLM fit might not have given such a nice overall behavior, see Fig. \ref{fig:qubit_radial_w}.
This suggests the following, whilst the FLM scheme works nice enough to approximate the exact analytical solution for the two-level case, and is not limited by special conditions such as being restricted to the equatorial plane of the Bloch sphere, care must be taken as to what values we are considering for computing the $\beta_i$s in FLM. For example, if our interest is in observing equilibrium behavior better, we should go with FLM computed using $\rho_u$ as in Fig. \ref{fig:qubit_radial_w}. Although, for a better approximation of far from equilibrium behavior, FLM at some initial $\rho$ seems to do a better job, see Fig. \ref{fig:qubit_radial}. \\
As evident from the form of the general dynamical Eq. (\ref{eq:litsurv_SEA_withD}), $\tau$ acts as modulating factor, where high $\tau$ results in a smaller dissipation contribution. Besides, system relaxation time $\tau$ is inversely proportional to the rate of entropy production as in Eq. \eqref{eq:litsurv_tau}. Following a quenching process, the system relaxes and could either thermalize or localize. This behavior is dependent mainly on the speed at which this happens. Higher $\tau$ implies slower relaxation, while as commented in literature, lower positive values of $\tau$ result in the steepest ascent of entropy, as we see in Fig. \ref{fig:qubit_spiralmain} which is due to faster relaxation.
In the expression of $\mu_t$ Eq. (\ref{eq:qubit_p_isol}), the exponent has ($\frac{-t}{\tau}$) dependence, which implies in $\frac{t}{\tau}<<1$ we will have a non-dissipative feature, and at $\frac{t}{\tau}\sim 1$, we will have the desired dissipation.

\begin{figure}[t]
    \centering
    \includegraphics[height=0.4\textheight]{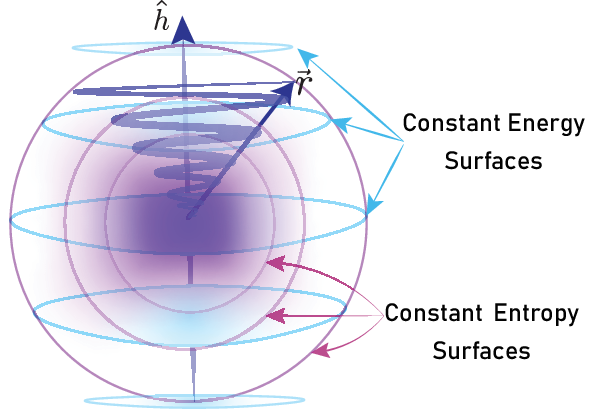}
    \caption[Schematic for general decay of Bloch radius amplitude]{\label{fig:qubit_spirality} Schematic representation of non-energy conserving SEA evolution of the Bloch vector for a qubit. The purple trajectory denotes the said dynamics. Image cited from Ref. \cite{ray_2022_steepestb}.}
\end{figure}
From Fig. \ref{fig:qubit_radial}, we see higher $\tau$ valued states will have more delayed and gradual relaxation.
We show the spiraling motion to the center of the Bloch sphere on the equatorial plane in Fig. \ref{fig:qubit_2dspiral}. Here, we see that high $\tau$ states remain near the pure states for a longer time than low-valued ones, as they almost instantaneously mix to the maximum entropic state. These low values of $\tau$ trajectories represent the steepest entropy ascent solution. This steep \textit{ascent} can be better visualized when we consider the surface of revolution generated from the entropy functional in Eq.  (\ref{eq:qubit_entropy_onr}) and plot these spiral trajectories onto that surface as shown in Fig. \ref{fig:qubit_entropyspiral}.
We can see that as time passes, each trajectory strives to reach the top of the surface, where the point with the highest entropy is located. We can also see that high $\tau$ states maintain a limiting characteristic at the bottom of the surface of revolution, taking nearly forever to reach the top (unitary type behavior). Fig. \ref{fig:qubit_spirality} depicts a schematic development of the generic non-energy conserving motion for a qubit under SEA.
As the value of energy falls, so does the value of $r_e$. The state arrives at the global equilibrium when it reaches zero at the center.

\section*{Summary}
In brief, in this chapter, we have seen a first-hand application of SEA. We first presented the GPB solution, which is exact. we introduced our approximation method namely, the fixed Lagrange's multiplier method. We compared the FLM solution for qubit computed at different conditions with the exact GPB and full numerical results of BSEA. We then commented on the nature of trajectories under this scheme onto the Bloch sphere and the entropy surface.\conditionalBlankPage

\chapter{SEA in a N-level System}\label{ch:qwalker}
\blankpage
\newpage
\thispagestyle{empty}
\begin{figure}[H]
    \centering
    \captionsetup{justification=centering}

    \includegraphics[height=0.75\textheight, width=0.75\textwidth]{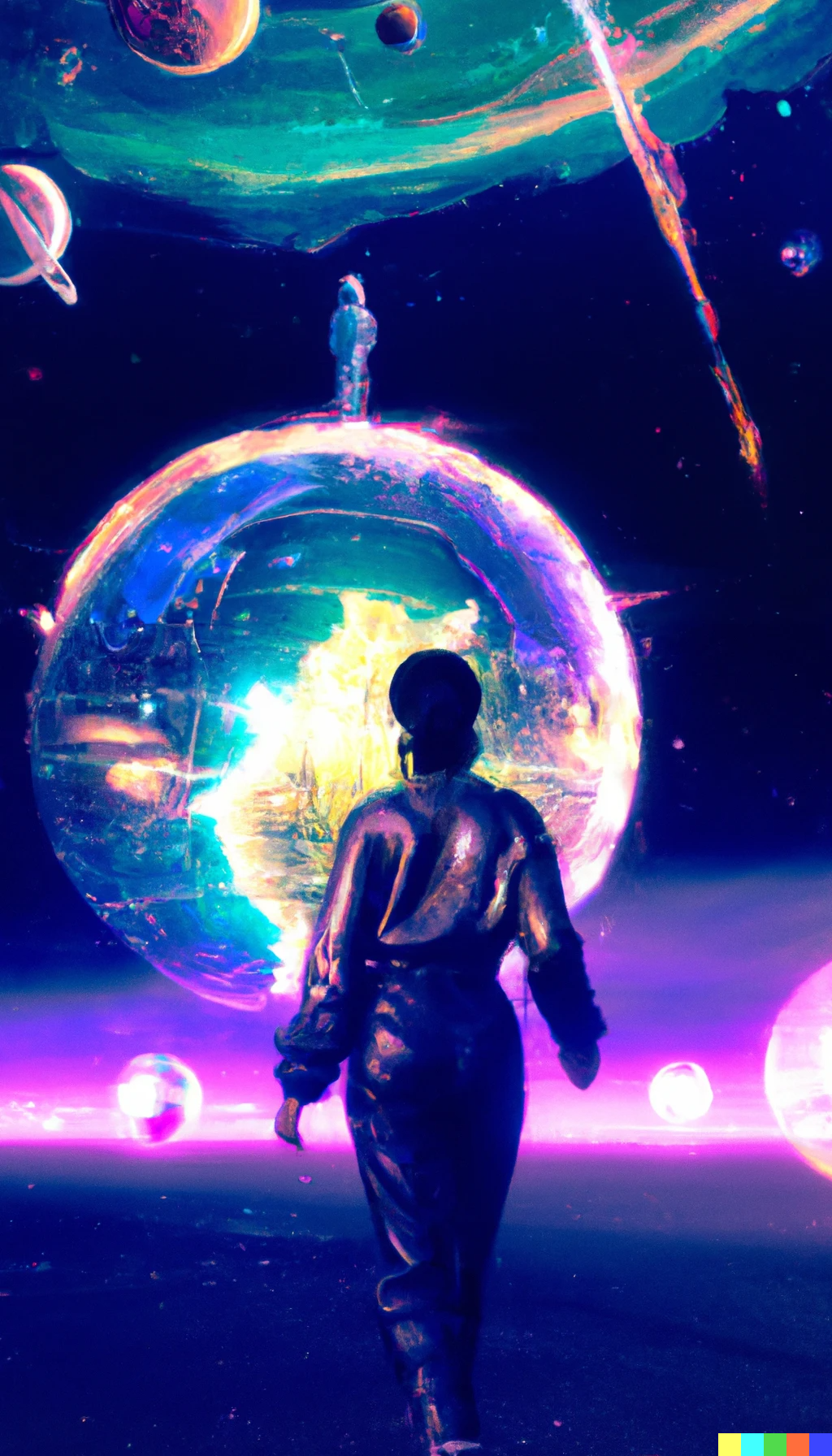}
    \caption*{\emph{I entered the uncertain realm,}\\
        \emph{with my tools and my hopes.}\\
        \emph{I paved new scopes,}\\
        \emph{and kept on traversing.}}
\end{figure}
\newpage
\blankpage
\lettrine[lines=3]{F}{ollowing} the discussion of the steepest entropy ascent formalism for a qubit, \textit{i.e.,} a two-level system, it is only natural that we present the extension of the same to a $N$-level system. As the Bloch sphere picture was quite intuitive for the case of a qubit, it is not so when it comes to more than two-level systems. One of the major reasons being the increase in the dimensionality of the sphere, for a $N$-level system, our Bloch sphere will reside in a space of $N^2-1$ dimensions, which is not at all intuitive \cite{park_1971_general}. Therefore, we look for different modeling of the same. We
arrive at the doorstep of the continuous-time quantum walker model, which easily allows us to understand the behavior of a $N$-level quantum system through some applications of graph theory and related topics. In this chapter, we will study SEA on a single walker performing a continuous-time quantum walk. We will begin with a short review of the theoretical background required by the said study, and will then follow up with the application of the FLM scheme developed in the previous chapter. Our subsequent analysis will include a comparison with numerical results, the dependence of Lagrange's multipliers $\beta_i$s upon the dimensionality of the system among other parameters, and a closer look at the entropy production rate for this system.
\section{Theoretical minimum}\label{sec:qwalker_background}
As discussed in chapter \ref{ch:intro}, the continuous-time quantum walker (CTQW) is a quantum analog of the classical walker. One considers a walker (a resident of the quantum realm), beginning the journey on a grid of some fashion. A graph is usually employed to describe the said network. In the classical case, the walker would toss a coin, and depending on the outcome will choose one path from the set of multiple options available to it. However, in the quantum case, and especially in the case of continuous-time motion, the walker has the opportunity of exploiting the linear superposition of all the available paths as long as there is no measurement of position or any other walk-dependent parameters involved. In the CTQW, there is a hopping probability associated with each neighborhood point for the walker, and the transition can happen at any time. Hence the name continuous-time \cite{manouchehri_2007_continuoustime,manouchehri_2014_physical}. \\
A traditional way of modeling a random walk process involves a graph. A graph $\mathbb{G}$ is a set of nodes or vertices $\mathbb{V}=\{v_i\}$, and the connections between those vertices are called edges $\mathbb{E}=\{e_{ij}|v_i,v_j ~\text{are connected}\}$. There exists an extensive literature on various types of graphs and the consequences of performing walks on them. However, for this thesis, we are only interested in CTQW as a model. Hence we consider a quantum walker walking on some undirected graph $ \mathbb{G} $. $\mathbb{G}$ has no double edge or self-loops, and the number of vertices $N=\abs{\mathbb{V}}$. The adjacency matrix $\mathbf{A}$ of $\mathbb{G}$ can be defined as follows,
\begin{equation}\label{eq:qwalker_adjacency}
    \mathbf{A}: a_{ij} = \begin{cases}
        1 & \text{if } e_{ij} \in \mathbb{E} \\
        0 & \text{otherwise}.
    \end{cases}
\end{equation}
Thereafter, we can define the Laplacian $\mathbf{L}$ of $\mathbb{G}$ as \cite{childs_2004_spatial,farhi_1998_quantum},
\begin{equation}\label{eq:qwalker_lapla}
    \mathbf{L} = \mathbf{D}-\mathbf{A},
\end{equation}
$\mathbf{D}$ is diagonal and has an entry as the degree of the $i^{\text{th}}$ vertex, $d_i$. We associate a hopping probability $\mu_{ij}$ with the probability of transition between two adjacent vertices ($v_i$, $v_j$ ) per unit time. Considering uniform transition rates $\mu_{ij}=\mu$, for the unitary continuous-time quantum walker we can write \cite{manouchehri_2014_physical,kendon_2003_decoherence,childs_2004_spatial},
\begin{equation}\label{eq:qwalker_qweom}
    \derivative{\ket{\Psi}}{t} = -\frac{\imi}{\hbar}\mu\mathbf{L} \ket{\Psi}.
\end{equation}
In Eq. (\ref{eq:qwalker_qweom}), the quantity $\mu \mathbf{L}$ is identified as the Hamiltonian $ \hamiltonian $ of the CTQW. The solution to Eq. (\ref{eq:qwalker_qweom}) is read as,
\begin{equation}\label{eq:qwalker_qwsol}
    \ket{\Psi(t)} = \exp(-\imi\mu \mathbf{L} t)\ket{\Psi(0)}\equiv \mathcal{U}(t)\ket{\Psi(0)}\equiv\mathcal{U}_t\ket{\Psi(0)},
\end{equation}
where, $\ket{\Psi(0)}$ is the initial state of the walk, and $\hbar=1$. In terms of density matrix $\rho$, we can write Eq. (\ref{eq:qwalker_qwsol}) as,
\begin{equation}\label{eq::qwalker_qwrho}
    \rho_t= \mathcal{U}_t\rho^0\mathcal{U}_t^{\dagger},
\end{equation}
implying that the quantum state of the walker undergoes unitary rotation in the state space as the walker exhibits CTQW.

\section{FLM on a CTQW}\label{sec:qwalker_FLM}

\begin{figure}[h]
    \centering
    \includegraphics[height=0.3\textheight]{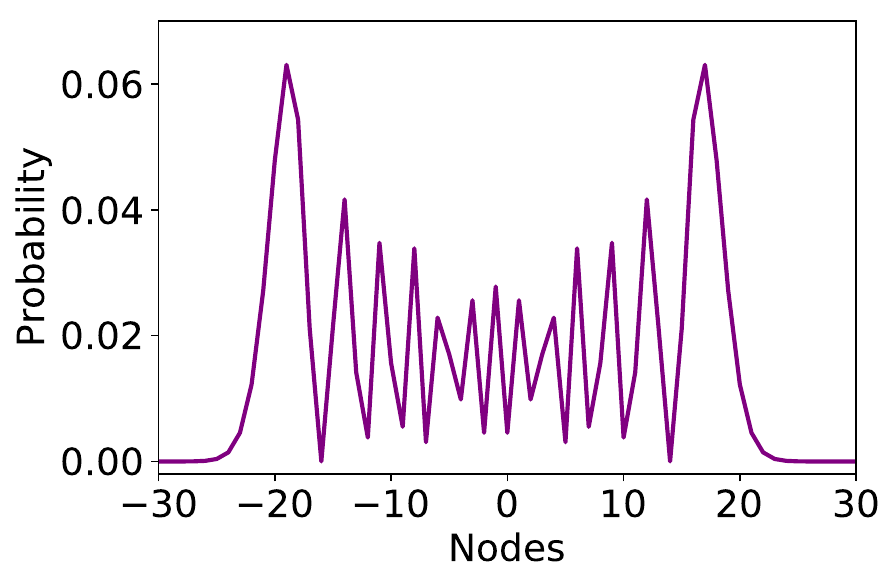}
    \caption[A probability distribution of a continuous-time quantum walker or a cycle graph]{\label{fig:qwalker_unitaryctqw} Plot of the probability vs nodes for a single continuous-time quantum walker on a cycle graph of 100 vertices (nodes) after time $t=10$. The walker was initiated at node 50, which is shifted here to zero for symmetry. Image cited from Ref. \cite{ray_2022_steepestb}.}
\end{figure}

The usual probability distribution computed from Eq. (\ref{eq::qwalker_qwrho}) for a walker performing CTQW on a ring of $N = 100$ nodes after some time $t=10$ is shown in Fig. \ref{fig:qwalker_unitaryctqw}, which is similar to CTQW walk distribution on a line for a short time and large $N$. It is to be noted while we find probability distributions such as this at some time $t$, they are not the same for the discrete and continuous cases. In the continuous case, we have transition amplitudes per unit time ($\mu$), and we consider probabilities at some instance post-initiation. These instances can be recorded as \textit{steps}. Wherein for the discrete case, a coin operation followed by a swap operation constitutes a \textit{step}. For simplicity, we have considered $\mu=1$ \cite{manouchehri_2014_physical}, which means an unbiased transition to any adjacent vertex in an undirected graph $\mathbb{G}$ with no loops \footnote{In the Fig. \ref{fig:qwalker_unitaryctqw}, we have only shown the nodes up to which the walker has spread after t=10; it does not show all nodes.}.
We use the standard basis to describe the density matrix of the walker at some given time as
\begin{equation}
    \rho_t = \sum_{i}p^t_i\dyad{i}{i},
\end{equation}
$p^t_i$ is the probability of finding the walker on a node (vertex) $i$ after some time $t$ since the initiation of the walk. Laplacian of $\mathbb{G}$ can be expressed as
\begin{align}\label{eq:qwalker_laplacianexpand}
    \begin{split}
        \mathbf{L} = & \mathbf{D} - \mathbf{A} \\
        = & \sum_{ij}\left(d_i\delta_{ij}\dyad{i}{j}-\mathbb{E}_{ij}\left(\dyad{i}{j}+\dyad{j}{i}\right)\right),
    \end{split}
\end{align}
$\mathbb{E}_{ij} = 1$ when there is an edge element $e_{ij}\in\mathbb{E}$, and zero otherwise. For the Hamiltonian in the standard basis; the diagonal elements are simply entries of $\mathbf{D}$, $H^d_{ii} = \mu d_{i}$, $d_i$ is the degree at the vertex $v_i$. So in the case of these walks, using the computation of $\beta_i$ carried out below, the Eq. (\ref{eq:qubit_fullevolution}) becomes,
\begin{equation}\label{eq:qwalker_ctqwevolution}
    \rho_t = \exp(-\imi\hamiltonian t)\left(\sum_{m}\exp(\mu^t_m-\mu^c_m)\mathbb{P}_m\right)\exp(\imi\hamiltonian t),
\end{equation}
where, $\mu^t_m = \left(\ln(p_m^0)+\mu^c_m\right)e^{-t/\tau} = \tilde{v}_i\mu_t$, and $\mu^c_m = \mu\beta_H d_i-\beta_I$, and $\mathbb{P}_m$ is the projection operator. $p_m^0$ is found using Eq. (\ref{eq:qubit_initial}).
\begin{figure}[h]
    \centering
    \includegraphics[height=0.4\textheight]{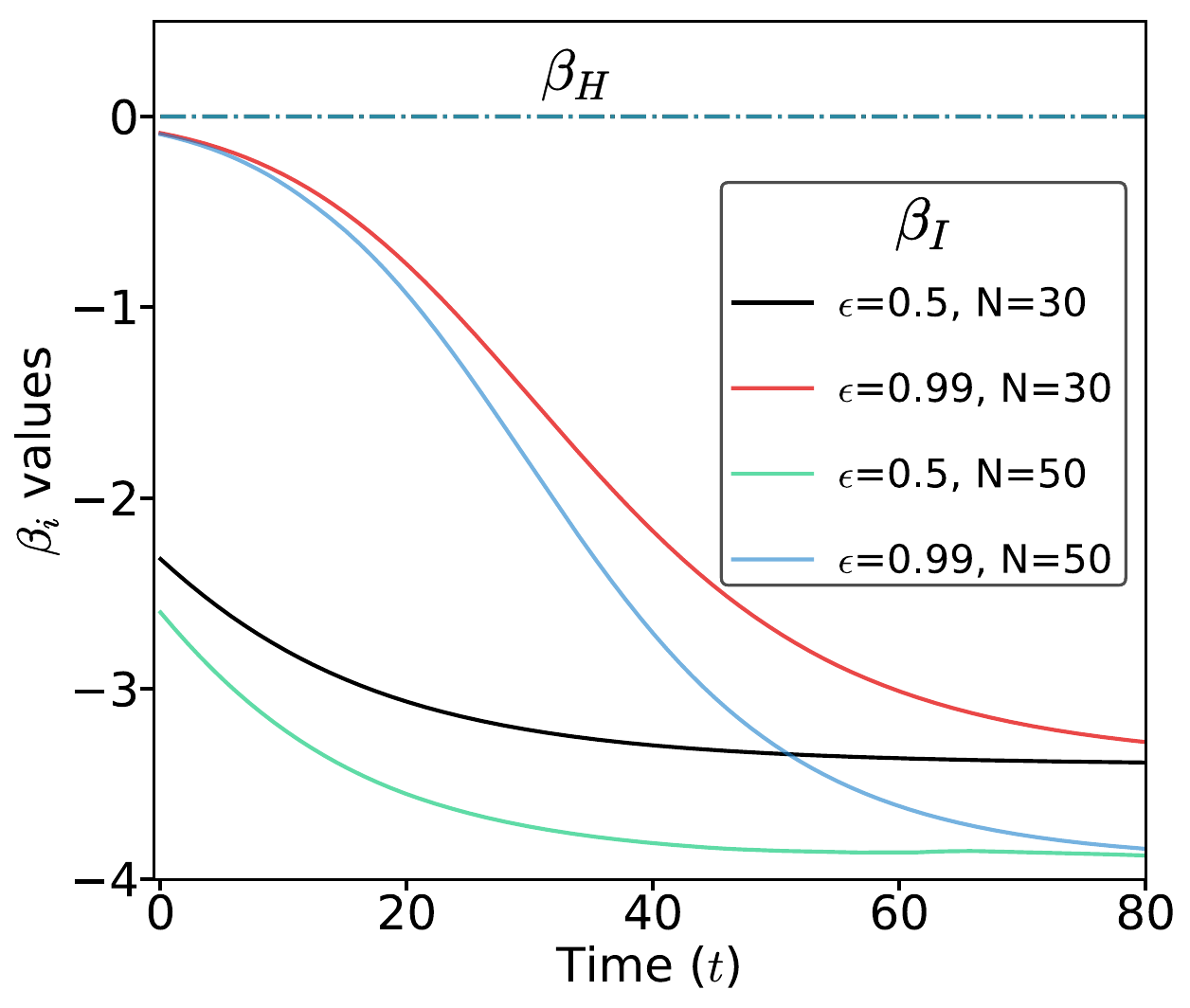}
    \caption[Variation of Lagrange multipliers over time]{\label{fig:qwalker_betai} Plot of ariation of Lagrange multipliers $\beta_i$ over time. In the legend, $ \beta_I $ is tagged, as it varies for different $ N $ and $ \varepsilon $ values. $\beta_H$ is the constant line $ y=0 $. Image cited from Ref. \cite{ray_2022_steepestb}.}
\end{figure}
We model the walk using fixed parameters defined beforehand so that we can characterize and comprehend the solutions to Eq. \eqref{eq:qwalker_ctqwevolution}. A cycle graph $\mathcal{C}_N$ being 2-regular, has the following Hamiltonian in the standard basis,
\begin{equation}\label{eq:qwalker_hamilctqw}
    \hamiltonian = \sum_{i}\left(2\dyad{i}{i} - \mathbb{E}_{i,\tilde{i}}\left(\dyad{i}{\tilde{i}}+\dyad{\tilde{i}}{i}\right)\right),
\end{equation}
where $\tilde{i} = i \pmod{N} + 1$. Using this $\hamiltonian$, and the equilibrium distribution $ \rho_u $, $ \rho_u = \dfrac{1}{N}\mathbf{I} $, and $ \hamiltonian $ as given in Eq. (\ref{eq:qwalker_hamilctqw}), we begin by computing trace function as required in the Eqs. (\ref{eq:qubit_sea_2const}) which are as given below:
\begin{align}\label{eq:qwalker_trace_beta}
    \begin{split}
        \tr(\rho) =& 1,\\
        \tr(\rho\hamiltonian) =& d,\\
        \tr(\rho\hamiltonian^2) =&N\left(d^2+2\right),\\
        \tr(\rho\ln(\rho)) =& -\ln(N),\\
        \tr(\rho\hamiltonian\ln(\rho)) =& -d\ln(N).
    \end{split}
\end{align}
Using these traces and noticing that $\Omega$ is given by
\begin{equation}\label{eq:qwalker_omega}
    \tr(\rho\hamiltonian^2)\tr(\rho) - \left(\tr(\rho\hamiltonian)\right)^2 = \left(N-1\right)d^2+2N,
\end{equation}
we can write the expressions below for $\beta_i$s,
\begin{align}\label{eq:qwalker_cqwbetanumerical}
    \begin{split}
        \beta_H = &~ 0,\\
        \beta_I = & -\ln(N),
    \end{split}
\end{align}
with $k=1$. In Fig. \ref{fig:qwalker_betai}, we plot the variation of $\beta_I$ for two different $N$ values of 50 and 30, respectively. We numerically solve the Eq. (\ref{eq:qubit_sea_2const}) and use the $\rho$ thus produced at each iteration to compute $ \beta_i $'s as defined in Eqs. \eqref{eq:litsurv_beta1}, \eqref{eq:litsurv_beta2}. We can see from the plot that the final value of $\beta_I$ is dependent on $N$ and mean energy. Consider the red and black lines, for instance. As time passes, we observe that they merge towards a consistent value provided by Eq. \eqref{eq:qwalker_cqwbetanumerical}, implying that although there is an initial dependency on $\varepsilon$, as equilibrium approaches, all of the $\beta_I$ take the same value. If one desires to evaluate the behavior far from equilibrium, it is prudent to consider the initial $\beta_I$ in FLM. Otherwise, using an equilibrium distribution to fix the multipliers for FLM will correctly describe equilibrium behavior. The $\beta_H$ plot is shown by the $ y=0 $ line in the graph, which remains constant in this case. The variation in $ \beta_I $ lies within a single order of magnitude and does not reflect a strong difference in probability amplitudes, as seen in Fig. \ref{fig:qwalker_ctqw}.
\section{Analysis of SEA evolution}\label{sec:qwalker_Analysis}
\subsection{Probability amplitudes}\label{secsub:qwalker_Analysis_prob}
\begin{figure}[h]
    \centering
    \includegraphics[height=0.38\textheight]{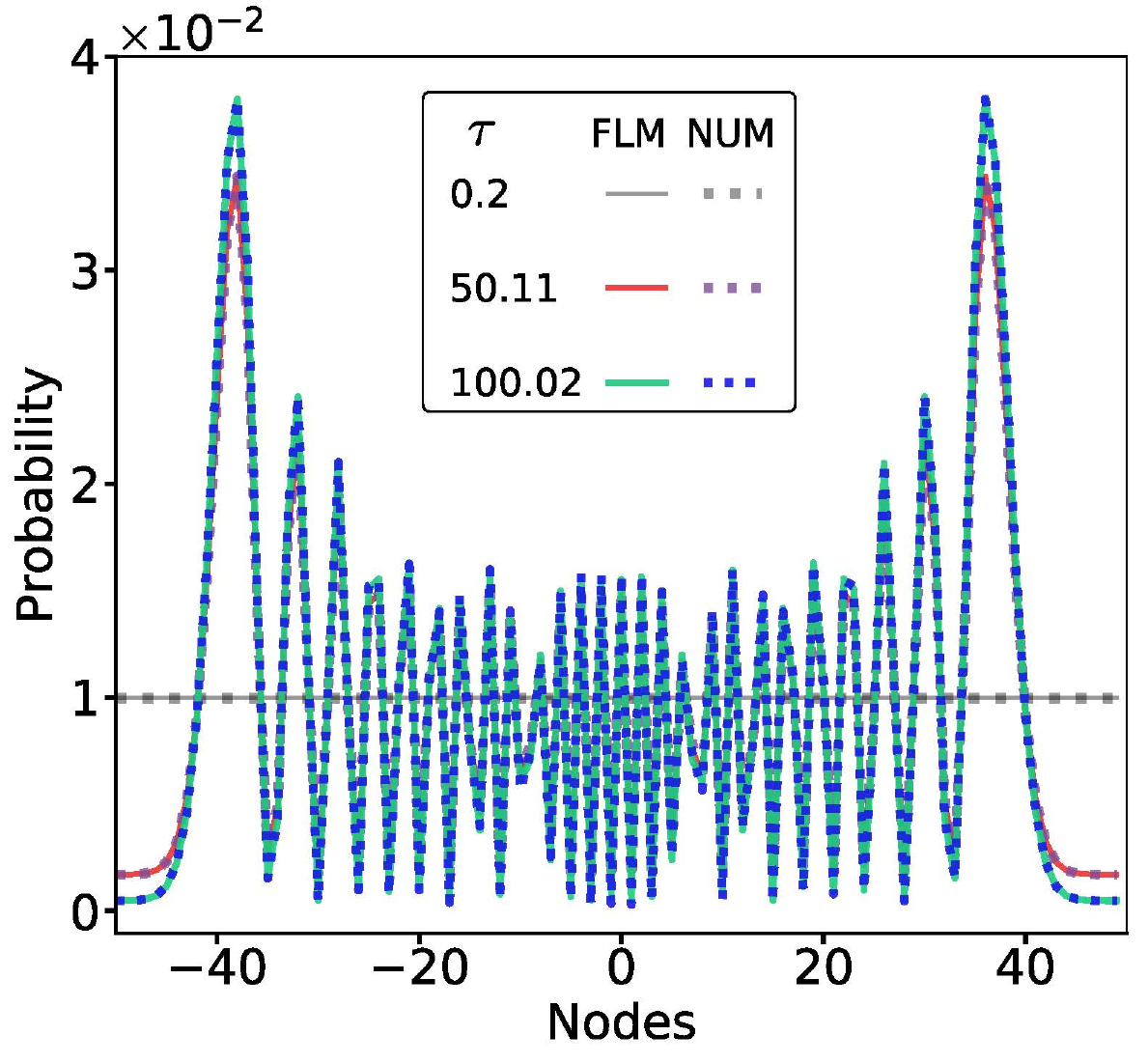}
    \caption[FLM vs numerical comparison of probability amplitudes for CTQW]{\label{fig:qwalker_ctqw}Probability amplitudes for a single CTQW under SEA on a cycle of 100 nodes under SEA conditions after $t=20$, initiated at node 50 (shifted to node 0 for symmetry), after $t=20$. The amplitudes are plotted for various relaxation times $\tau$ as given in the legend. $\varepsilon$ is $0.99$. The plot compares the analytic solution found using the FLM method (solid lines in the plot) and those (dotted lines in the plot above) from the numerical solution to the Eq. (\ref{eq:qubit_sea_2const}) using CTQW Hamiltonian and other relevant substitutions. Image cited from Ref. \cite{ray_2022_steepestb}.}
\end{figure}

To analyze the solution got from FLM, we begin by plotting the probability amplitudes. Using appropriate $\tau$ values, we get Fig. \ref{fig:qwalker_ctqw}.
we find from the plot, that the FLM (solid lines) and NUM (dotted lines, numerical solution of Eq. \eqref{eq:qubit_sea_2const}) are not distinguishable from each other through visual observation alone. As seen in the plot, the probability amplitudes are of the order of $10^{-2}$. From our numerical computation, we've estimated the difference of NUM with FLM results, which is of the order $10^{-4}$ for low $\tau$, and of the order $10^{-3}$ for high $\tau$s. We see, for the distributions considered after $t=20$, a similarity in the behavior of probability emerges as in Figs \ref{fig:qubit_radial}, \ref{fig:qubit_2dspiral}. Higher $\tau$ or states closer to unitary states tend to relax slower. For low enough $\tau$, the rapid relaxation of the system is observed in Fig. \ref{fig:qwalker_ctqw}, and all initial information is lost. On the other hand, high $\tau$ states having lesser entropy generation rates drive the system toward unitary-like behavior. This can also be understood in terms of the localization and delocalization of the walker. The probability distribution for the case of $\tau = 0.2$ in Fig. \ref{fig:qwalker_ctqw} shows strong delocalization. While in the same figure, because of $\tau=50.11, 100.02$, and $t=20 < \tau$, we can say decoherence is yet to set in. That is, it displays linear behavior. As understood so far, low $\tau$ results in more non-linear behavior. But how low? Unfortunately, the answer to such a question remains elusive \cite{beretta_1987_steepest}. In the following, we try to figure that out by using entropy and the rate of entropy production.

\subsection{Entropy and energy}\label{secsub:qwalker_analy_ent}

\begin{figure}[t]
    \centering
    \subfigure[\label{fig:qwalker_AvEn}]{\includegraphics[width=0.49\textwidth,height=0.3\textheight]{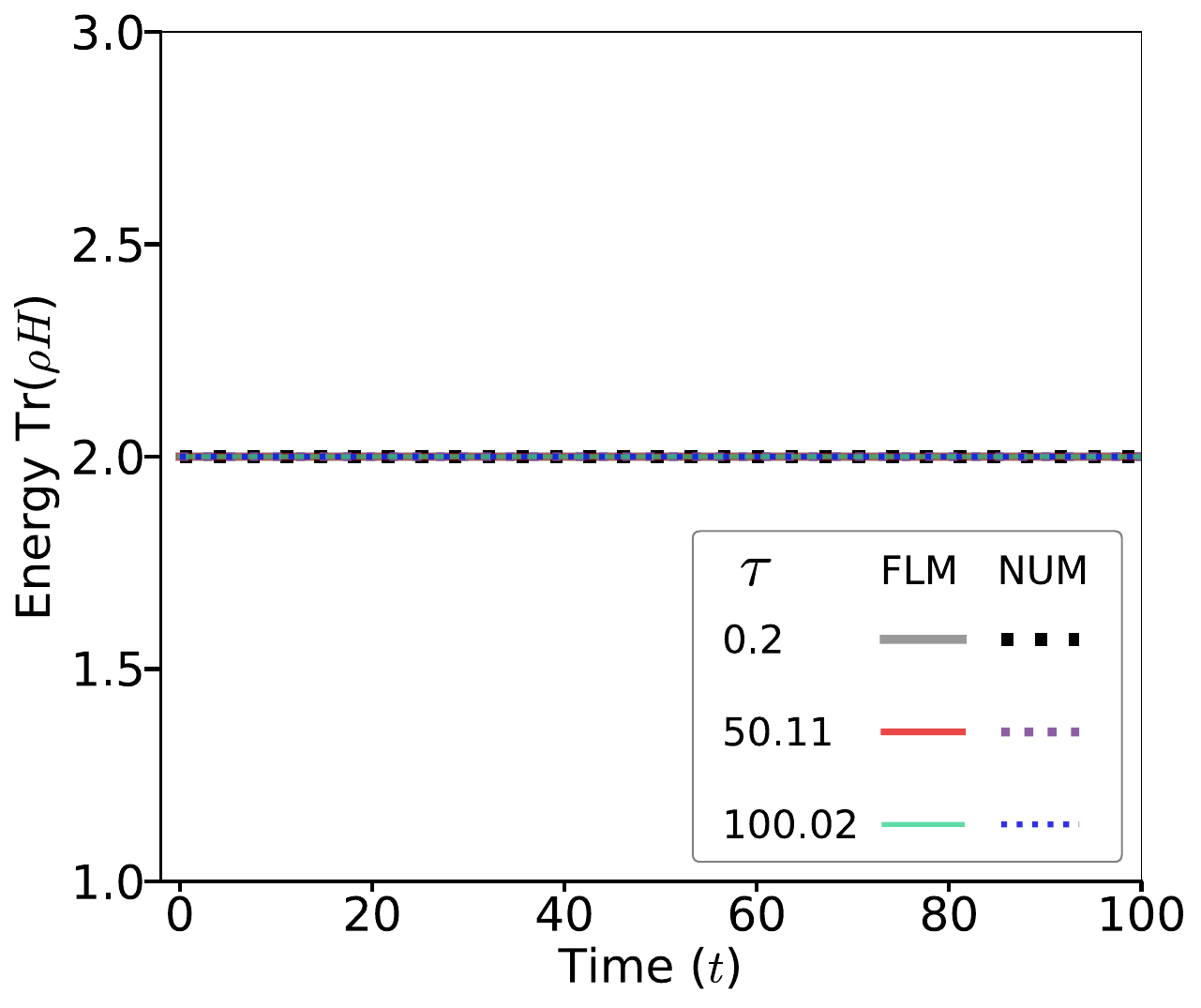}}
    \subfigure[\label{fig:qwalker_S}]{\includegraphics[width=0.49\textwidth,height=0.3\textheight]{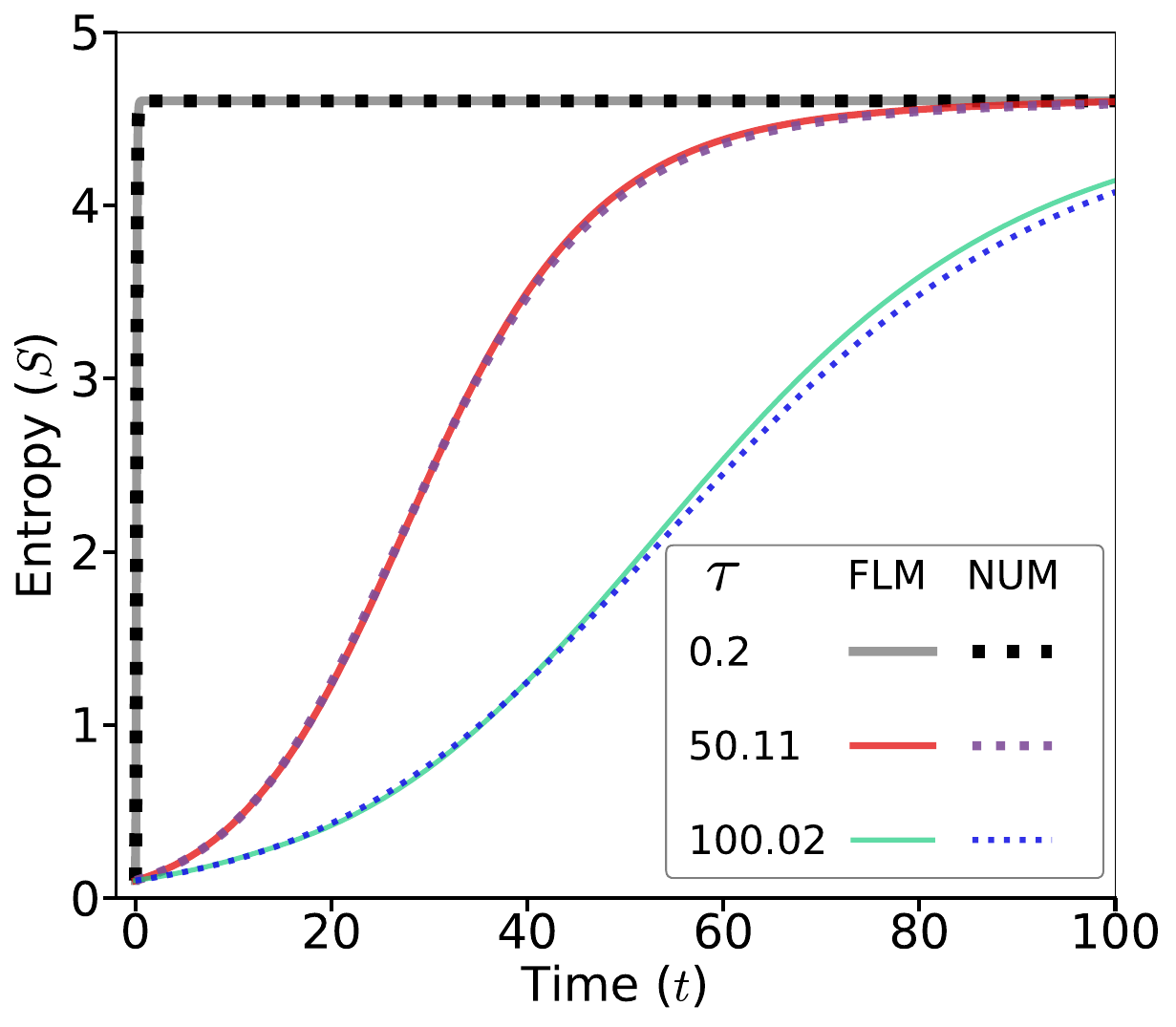}}
    \caption[Energy and entropy plots for CTQW]{\label{fig:qwalker_E_andS}(a) Plot of average energy vs time, and (b) entropy vs time for a CTQW on a cycle graph of 100 nodes for various $ \tau $ (first column in the legend) values and $ \varepsilon = 0.99 $. The walk was performed up to $ t=100 $. FLM (solid lines) denotes the analytically computed results, while NUM (dotted lines) denotes the numerical results. Image cited from Ref. \cite{ray_2022_steepestb}.}
\end{figure}

We first begin by checking whether, despite all the nonlinear evolution, energy indeed remains conserved. To do so, we plot the function $\tr(\rho\hamiltonian)$ for different $\tau$ values in Fig. \ref{fig:qwalker_AvEn}. To check for consistency we plot both the FLM and NUM results. We see energy is constant throughout the dynamics. This implies that the FLM solution respects the primary constraints of motion, and agrees with the numerical results. To further check the efficacy of FLM, we plot the entropy functional ($-\tr(\rho\ln\rho)$) against the same set of $\tau$s as in Fig. \ref{fig:qwalker_AvEn} in Fig. \ref{fig:qwalker_S} for both FLM and NUM results. We find a good agreement again, except for the high $\tau$ case. Here, we see FLM surpassing NUM values, which is to be expected because FLM is an approximation after all. Yet the agreement between the two is reassuring for our scheme. We note that in Fig. \ref{fig:qwalker_S}, except for the low $\tau$ case, where entropy directly shoots up to the maximum value, there is a monotonous increment. This validates the construction of SEA EoM, which is based on the entropy non-decrease principle. We proceed then to plot the rate of entropy production for similar data. $\Pi_S$, the entropy generation rate functional from Eqs. \eqref{eq:litsurv_constr-eq}, \eqref{eq:litsurv_Phi_Psi} is given by,
\begin{align}\label{eq:qwalker_piS}
    \begin{split}
        \Pi_S =& -k\tr(\left(\ln(\rho)+1\right)\dv{\rho}{t})\\
        =& 2k^2\tr(\left(\ln(\rho)+1\right)\mathcal{L}\acomm{\ln(\rho)}{\rho})+ 2k^2\sum_{i}(-1)^i\beta_i\tr(\left(\ln(\rho)+1\right)(\mathcal{L}\mathit{C}_i\rho+\rho\mathit{C}_i\mathcal{L})).
    \end{split}
\end{align}
\begin{figure}[h]
    \centering
    \subfigure[\label{fig:qwalker_PiS1}]{\includegraphics[width=0.49\textwidth,height=0.3\textheight]{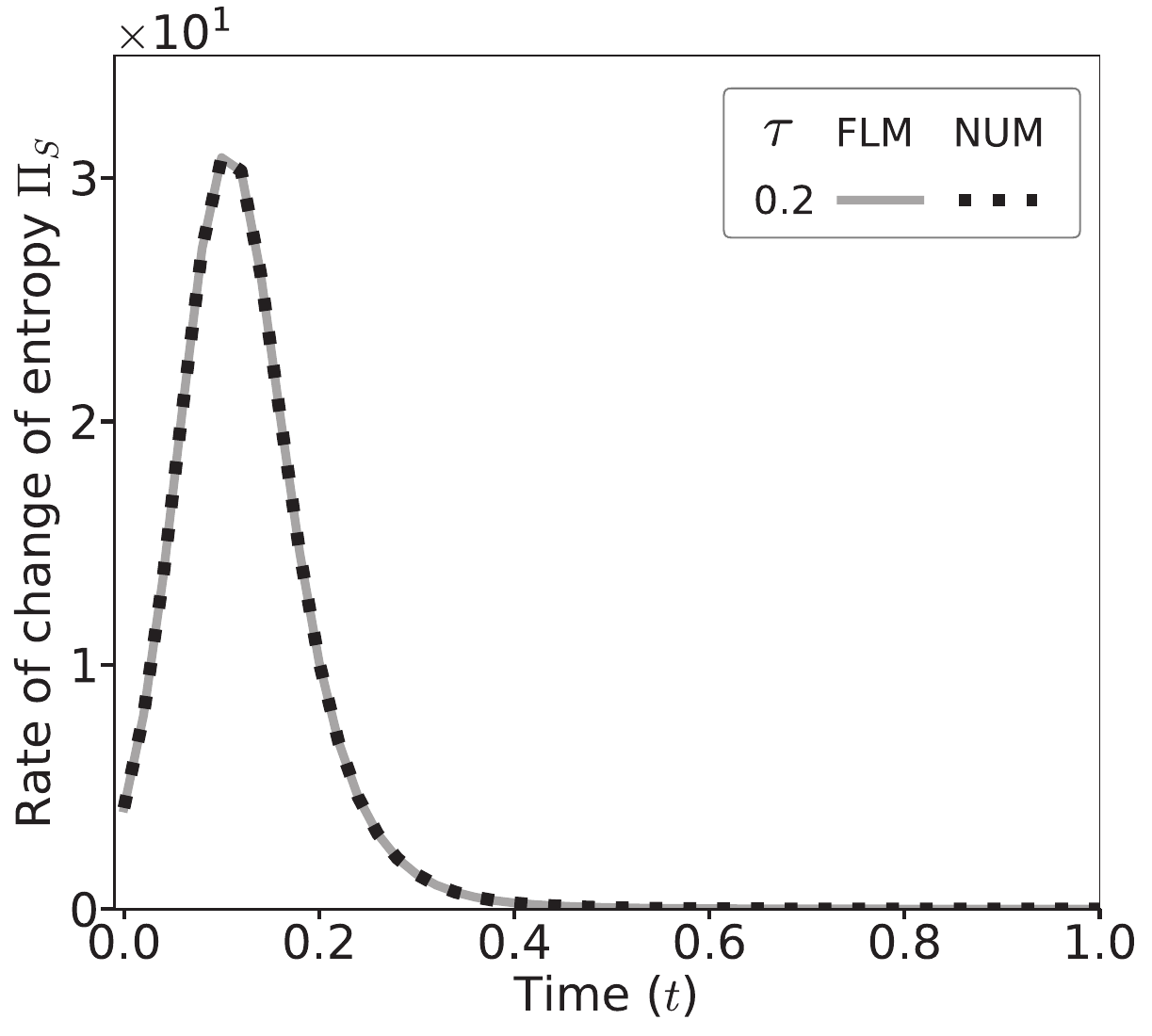}}
    \subfigure[\label{fig:qwalker_PiS2}]{\includegraphics[width=0.49\textwidth,height=0.3\textheight]{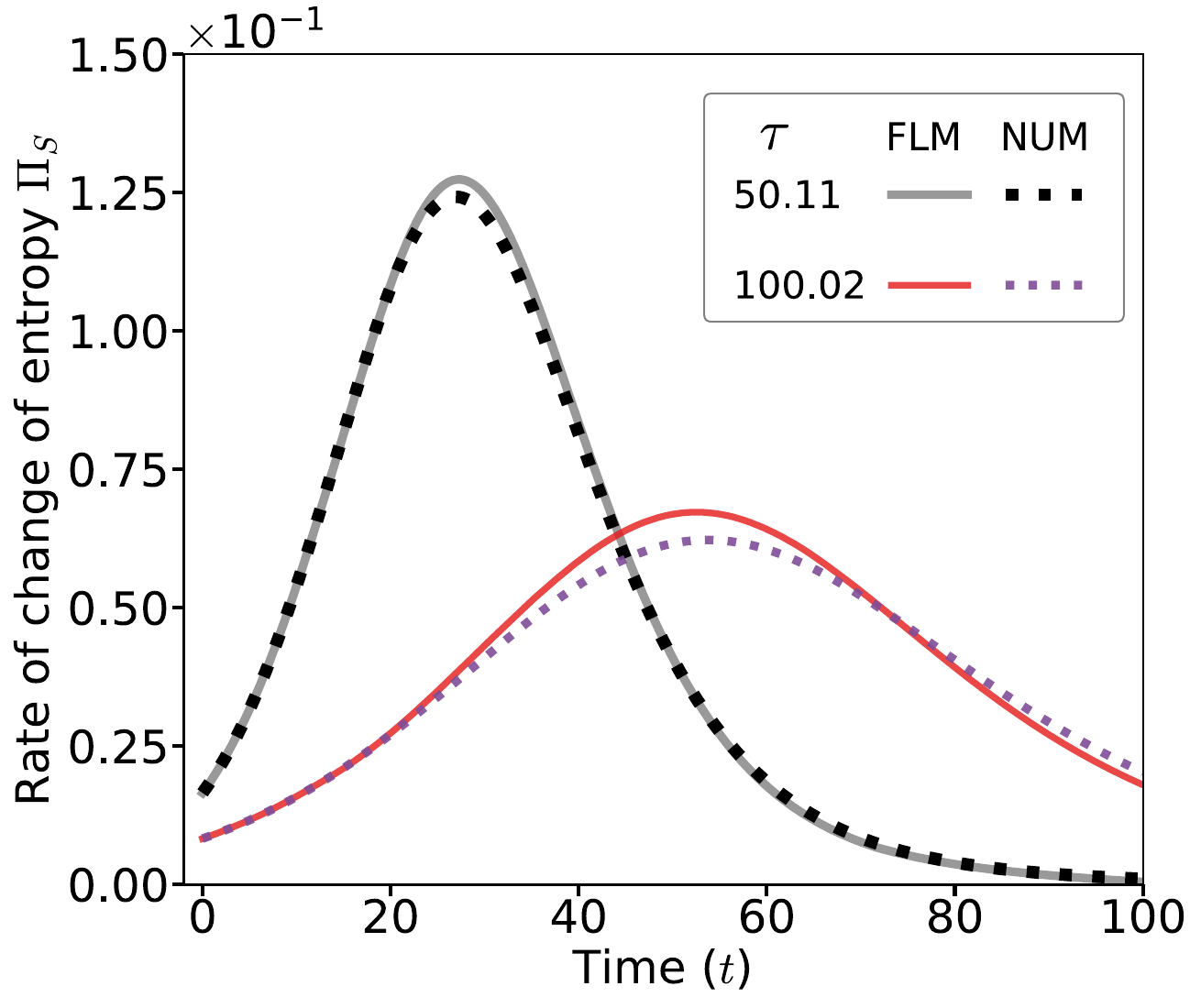}}
    \caption[Rate of change of Entropy vs time for CTQW]{\label{fig:qwalker_Pis}Plot of the rate of change of entropy vs time for a CTQW on a cycle graph of 100 nodes for (a) $ \tau = 0.2 $, and (b) $ \tau = (50.11,100.02) $ with $ \varepsilon = 0.99 $. FLM (solid lines) denotes the analytically computed results, while NUM (dotted lines) denotes the numerical results. Image cited from Ref. \cite{ray_2022_steepestb}.}
\end{figure}
Using $\mathcal{L}= \frac{\mathrm{I}}{4k\tau}$, we get,
\begin{equation}\label{eq:qwalker_entropygeneration}
    \Pi_S = \dfrac{k}{2\tau}\Big(\tr((\ln(\rho)+1)\acomm{\ln(\rho)}{\rho})+ \sum_i(-1)^i\beta_i\tr((\ln(\rho)+1)\acomm{\mathit{C_i}}{\rho}) \Big).
\end{equation}
As before, we see a good agreement between the numerical and FLM results in Fig. \ref{fig:qwalker_Pis}. We begin with the panel of Fig. \ref{fig:qwalker_PiS1}, where we see within $ t<0.1 $ the graph peaks around the value 32, which is twice the order of magnitude higher in other high $ \tau $ cases Fig. \ref{fig:qwalker_PiS2}. Hence, we see a visual confirmation of the SEA ansatz that the steepest entropic path is also the one with the maximum entropy production rate. And this happens at low $ \tau $ values. Also, as noticed in Fig. \ref{fig:qubit_radial}, as $\tau$ increases, we see differences between FLM and NUM results. Despite this difference, there is a strong agreement at initial and later $t$s. This suggests FLM can be relied on to faithfully study the `overall' nature of the dynamics. It goes without saying, for very precise results at high $\tau$, one should rely on NUM results instead of FLM. As we can see, none of the plots discussed so far have been able to give us an indication of how low $\tau$ could be, only suggesting the lower the $\tau$ the steepest the ascent is.
\begin{figure}[t!]
    \centering
    \subfigure[\label{fig:qwalker_cont1}]{\includegraphics[width=0.49\textwidth]{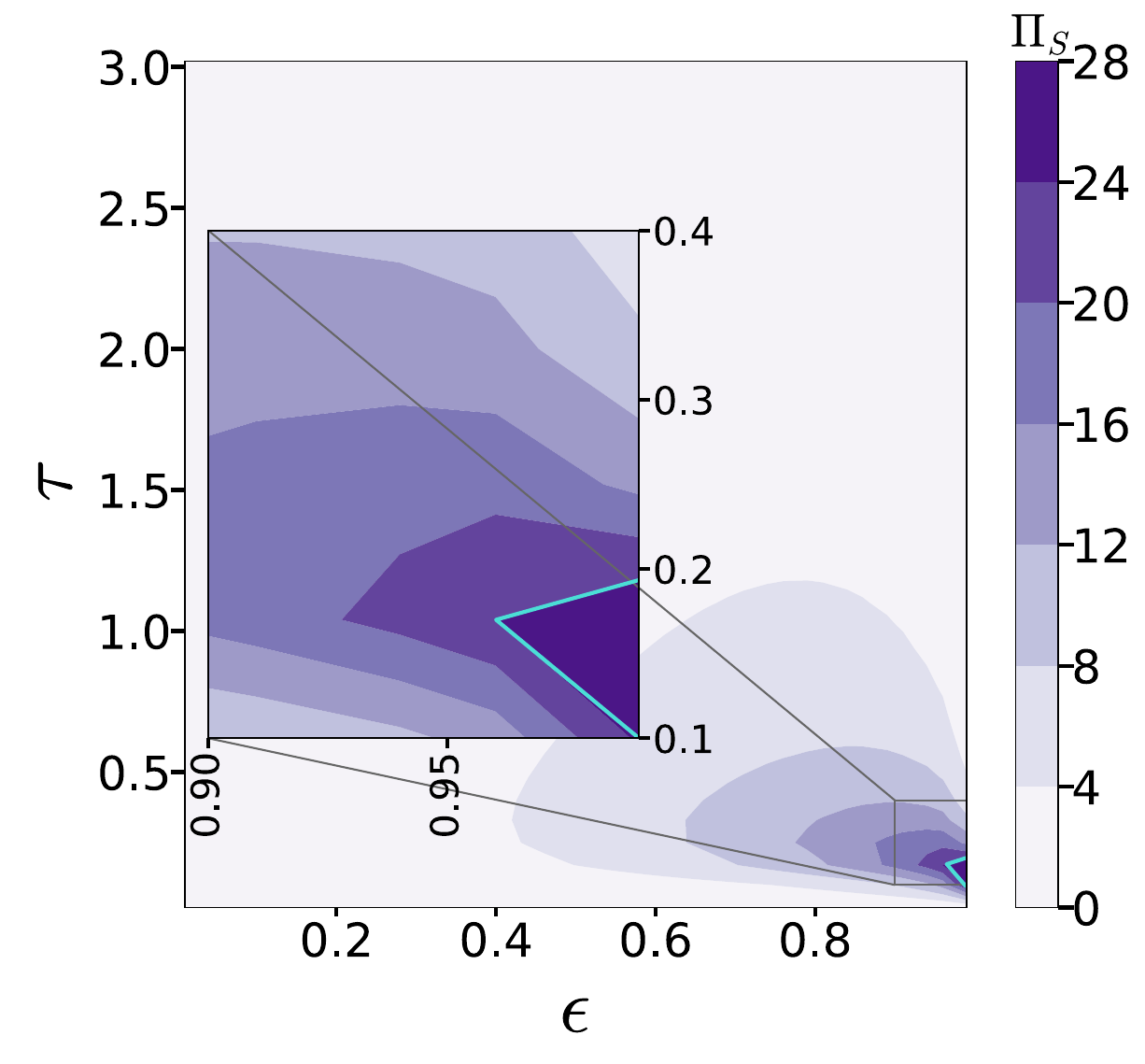}}
    \subfigure[\label{fig:qwalker_cont2}]{\includegraphics[width=0.49\textwidth]{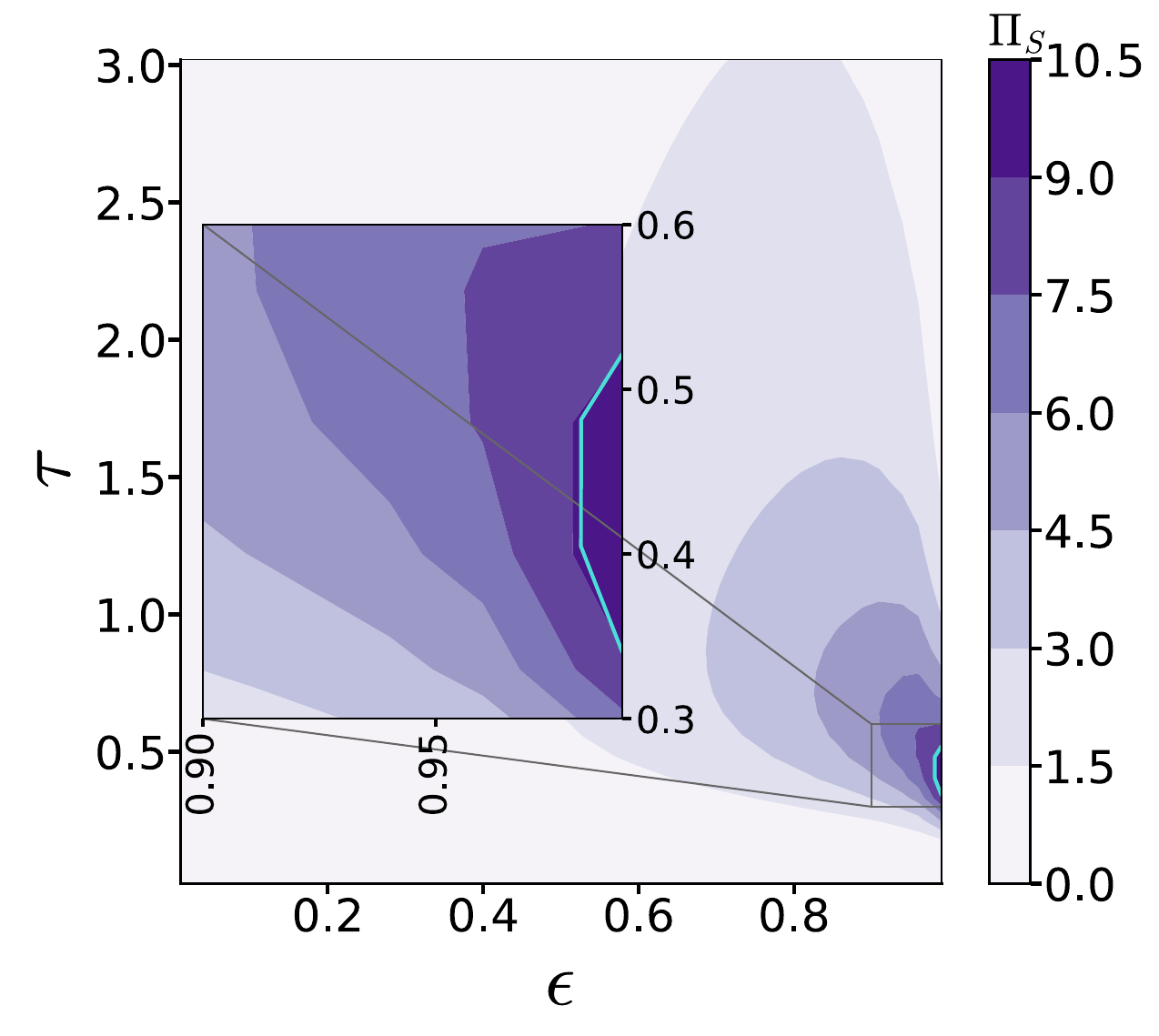}}
    \caption[Entropy generation contour plot vs $\tau$ and $\varepsilon$ for a CTQW]{\label{fig:qwalker_contours}Contour containing $\Pi_S$ values for a CTQW of $N=50$ vs variation in $\tau$ and $\varepsilon$ after time $t=1$ in (a), and $t=3$ in (b). The color bars provide the range and contrast of $\Pi_S$ values. As discussed in this chapter, the high $\Pi_S$ valued zones are concentrated around high $\varepsilon$ and low $\tau$ values. These deep purple areas bounded in cyan represent the maximum entropy generation area. In the insets of panels (a) and (b): zoomed-in view of the bounded region displaying max $\Pi_S$. Image cited from Ref. \cite{ray_2022_steepestb}.}
\end{figure}
To get an approximate picture of this behavior we consider instead `areas' of the high rate of entropy production $\Pi_S$ (\ref{eq:qwalker_entropygeneration}) against $\tau$ and $\varepsilon$ values as given in Fig. \ref{fig:qwalker_contours} for a CTQW with $N=50$. $\tau$ has the greatest contribution to defining $\Pi_S$; as seen in Fig. \ref{fig:qwalker_contours}, higher relaxation times result in essentially negligible entropy formation, which is consistent with our prior findings. As seen in Fig. \ref{fig:qwalker_cont1}, lower $\tau$ states yield larger $\Pi_S$ values early on, which is typical SEA behavior. Furthermore, as time passes, the highest $\Pi_S$ states (bounded by the cyan line in the plots) begin to move higher along the right side of the figures, as seen in Fig. \ref{fig:qwalker_cont2}. Only large $\tau$ valued states remain yet to equilibrate, causing a change in entropy. It is worth noting that the amount $\varepsilon$, by definition, reflects how pure the original state is. As a result, a state with low $\varepsilon$ is projected to attain equilibrium (that is, a noisy state grows louder) rather quickly. However, as shown in Fig. \ref{fig:qwalker_contours}, the rate of entropy production may not be at its peak in those cases. Under SEA evolution, states nearest to pure states have the highest entropy generation rate (the deep purple shaded contours in the diagram). This pattern may be explained by the fact that under the Bloch sphere representation, low entropic states located away from equilibrium must undergo a greater change in entropy while equilibrating. As a result, despite noisy channels growing noisier, their low information content accounts for a low entropy production rate.

\section*{Summary}
To summarize, we have studied the continuous-time quantum walker under the steepest entropy ascent formalism on a simple cycle graph of $N$ vertices (nodes). We have applied the fixed Lagrange's multiplier method developed earlier on the walker evolution. We have demonstrated the justification for using FLM by showing the slow variation of the multipliers over time. We have established the efficacy of FLM by performing various analyses while simultaneously comparing those results with exact numerical values.

\conditionalBlankPage

\chapter{The Bloch Vector for N>2}\label{ch:bloch_SEA}


\blankpage
\newpage
\thispagestyle{empty}
\begin{figure}[H]
    \centering
    \captionsetup{justification=centering}

    \includegraphics[height=0.75\textheight, width=0.75\textwidth]{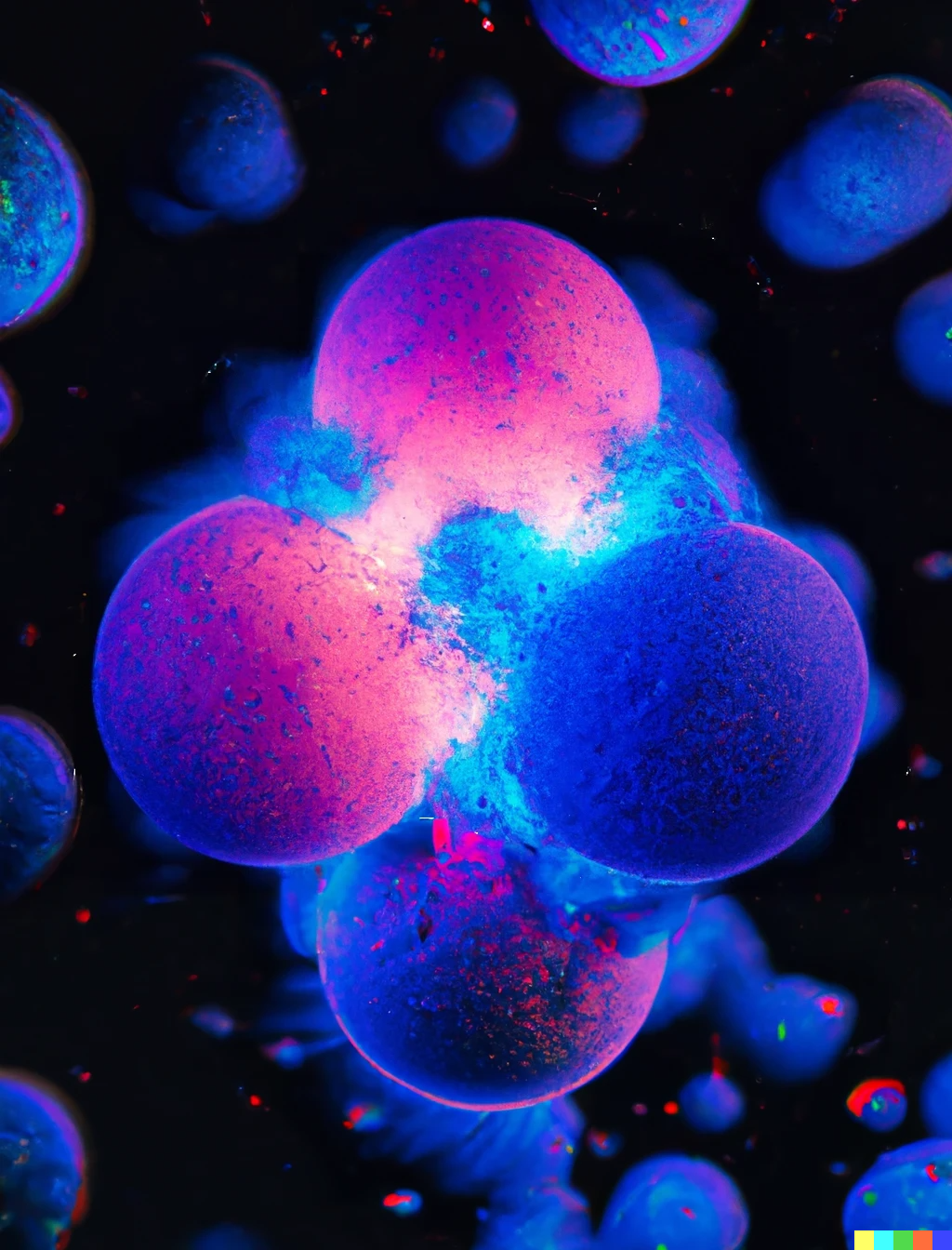}
    \caption*{\emph{Complexities grew and surrounded,}\\
        \emph{my path became quiet and bounded,}\\
        \emph{as the limitations became physical,}\\
        \emph{and difficulties astounding.}}
\end{figure}
\newpage
\blankpage
\lettrine[lines=3]{B}{efore} we begin to contemplate the nature of SEA evolution of a composite system analytically, we must digress. It is imperative that we set the necessary foundation for the mathematical background here. As we have noted earlier, in general, Bloch vector formalism does not provide a bijective map between the space containing all the Bloch vectors of unit radius and the density matrices that represent a physical system \cite{bruning_2012_parametrizations}. Consequently, we deal with nontrivial parametrizations, and the results are not always as convenient and satisfying as that of the two-level systems. As seen in the previous chapter, similar arguments led us to use the continuous-time quantum walker model to understand $N-$level systems. However, one can wish to study SEA evolution analytically for a lower dimensional composite, for example, a two-qubit system. Besides, a proper parametrization can also provide a testing ground for further research involving two qubits analytically. While some concepts such as the Werner state \cite{werner_1989_quantum} and EPR pairs \cite{einstein_1935_can} have been well studied and understood along with the separable cases, the case for mixed states are not so well researched. Here, in this chapter, we will discuss Bloch parametrization. We will introduce some new results regarding the eigenvalues of Bloch vectors for three and four-level systems using some general parametrizations as found in the literature.
\section{The Bloch parametrization}\label{sec:bloch_blochp}

The case of extending the concept of Bloch vectors from two-level systems to $N-$level system has garnered considerable attention over the years. Basically, one needs to generalize the traceless Pauli matrices to higher dimensions to have a mutually orthogonal basis, and expand the general $N\times N$ Hermitian positive semidefinite matrix (\emph{i.e.,} $\rho$) in such suitable basis involving $N^2$ terms. One of the most prominent methods is the multipole expansion method by Park \textit{et. al.} \cite{park_1971_general,band_1971_general}. In this approach the basis used is generated using concepts derived analogous to the algebra involving spins, hence Clebsch-Gordan coefficients are used. On the other hand, using generators of SU$(N)$ one can use the `generalized Gell-Mann' (GGM) matrices to compute the required set of orthogonal basis \cite{kimura_2003_blocha,bertlmann_2008_bloch,boya_2008_geometry,bruning_2012_parametrizations}. The benefit of the latter approach lies in the non-requirement of Clebsch-Gordan coefficient and an iterable method of getting the complete basis. In this thesis, we have followed the method of GGM basis for the sake of all computations.

\subsection{The generalized Gell-Mann matrix representation}\label{secsub:bloch_GGM}
The general Bloch vector formalism when applied to a $N-$level system becomes,
\begin{equation}\label{eq:bloch_bloch_Nd}
    \rho = \dfrac{1}{N}\left(\mathit{I}_N+\sqrt{\dfrac{N(N-1)}{2}}\bm{r}\cdot\bm{\Gamma}\right).
\end{equation}
$\bm{r}$ is a $N^2-1$-dimensional vector with the components given as $\lbrace r_1,r_2,\cdots,r_{N^2-1} \rbrace$ and $\sqrt{\sum_{i=1}^{N^2-1} r_i^2} = r$. $\mathit{I}_N$ is an $N\times N$ identity matrix. Defining dyadic operators or dyads as $\mathbb{P}(i,j) = \dyad{i}{j}$, we can define the $N^2-1$ generators of SU$(N)$ as components of the vector $\bm{\Gamma}$ as \cite{bruning_2012_parametrizations} follows. Firstly, there are $N-1$ diagonal operators,
\begin{equation}\label{eq:bloch_diagonal_gens_Nd}
    \Gamma_\ell = \sqrt{\dfrac{2}{\ell(\ell+1)}}\left(\sum_{j=1}^l \mathbb{P}(j,j)-\ell\cdot\mathbb{P}(\ell+1,\ell+1)\right),
\end{equation}
for $1\le \ell \le N-1$. Then we have the following $N(N-1)$ operators of the form
\begin{align}
    \Gamma_s & = \mathbb{P}(j,i)+\mathbb{P}(i,j),                  & ~ \text{for }~ 1\le i<j \le N  \\
    \Gamma_a & = \imi\left(\mathbb{P}(j,i)-\mathbb{P}(i,j)\right), & ~ \text{for }~ 1\le i<j \le N,
\end{align}
additionally, we have $N\le s \le \dfrac{N(N+1)}{2}-1$, and $\dfrac{N(N+1)}{2}\le a \le N^2-1$, which implies $\bm{\Gamma} = \lbrace \Gamma_\ell\rbrace\cup\lbrace \Gamma_s\rbrace\cup\lbrace \Gamma_a\rbrace$.

Although the above expansion of basis is easy to follow, it does not help in ensuring that $\rho$ is positive semidefinite. We can state this more formally in the following way. Consider the diagonalization of $\rho$ using the similarity transformation given as below
\begin{equation}\label{eq:bloch_similarity}
    \mathit{D}_N = \mathcal{U}\rho\mathcal{U}^\dagger,
\end{equation}
where $\mathcal{U}$ is unitary, and $\rho$ is such that eigenvalue $\lambda_i = \lambda_i(\rho)\ge 0$, and $\tr(\rho) = \sum_i \lambda_i = 1$. Since $\rho$ is unitarily similar to $\mathit{D}_N$, a parametrization of $\mathit{D}_N$ will imply the same for $\rho$. Now, let us consider the parameter set $\mathcal{Q}_N \in \mathbb{R}^M$ where $M=M(N)$ \cite{bruning_2012_parametrizations}. As we have seen with the eigenvalues of $\rho$ for $N=2$ (see Chapter \ref{ch:qubitSEA}), we see for $r\le 1$ the following
\begin{equation}\label{eq:bloch_parameter_2d}
    \mathcal{Q}_2 = \lbrace \bm{r}\in \mathbb{R}^3: r\le 1 \rbrace = \mathcal{B}(\mathbb{R}^3),
\end{equation}
where $\mathcal{B}(\mathbb{R}^3)$ denotes a closed unit ball in $\mathbb{R}^3$ centered at the origin. A map ($\mathrm{F}_2(\bm{r})$) exists that takes the elements from $\mathcal{Q}_2$ to the matrices $\mathit{D}_2$ as in Eq. \eqref{eq:bloch_similarity} as below \cite{kimura_2003_blocha,bruning_2012_parametrizations}
\begin{equation}\label{eq:bloch_map_qubit}
    \mathrm{F}_2(\bm{r}) = \half\mqty(1+r_3 & r_1-\imi r_2\\ r_1+\imi r_2 & 1-r_3), ~~ \bm{r}\in \mathcal{Q}_2.
\end{equation}
This map is onto $\mathit{D}_2$ and one-to-one. Thus $\left(\mathcal{Q}_2,~\mathrm{F}_2\right)$ is a parametrization of $\mathit{D}_2$ with $M=N^2-1=3$, and the parameter set $\mathcal{Q}_2$ forms the Bloch ball. Having established this notation, now we can formally state the problem of parametrization when it comes to $N>2$.

\subsection{The general parametrization}\label{secsub:bloch_gen_parametr}
Firstly, we need to identify $\mathcal{Q}_N$. To do that, a straightforward approach would require considering the characteristic polynomial of Eq. \eqref{eq:bloch_bloch_Nd}, which can be given by \cite{kimura_2003_bloch,bruning_2012_parametrizations}
\begin{equation}\label{eq:bloch_charecter_pol}
    det\left(\lambda\mathit{I}_N-\rho\right) = \sum_{i=0}^{N}(-1)^ia_i\lambda^{N-i}, ~~ a_0 =1.
\end{equation}
The coefficients $a_i$ are determined by the generators $\Gamma_i$ and the parameters $\bm{r}=(r_1,\cdots,r_{N^2-1})\in \mathbb{R}^{N^2-1}$. Let the roots of the polynomial in Eq. \eqref{eq:bloch_charecter_pol} be $\lambda_1,\cdots\lambda_N$. Then we can write
\begin{equation}\label{eq:bloch_poly_root_character}
    \sum_{i=0}^{N}(-1)^ia_i\lambda^{N-1} = \prod_{i=1}^{N}\left(\lambda-\lambda_i\right) =0.
\end{equation}
Before we proceed to find the relations between the roots and the coefficients $a_i$, we notice the following relations between the generators \cite{kimura_2003_bloch},
\begin{align}
    \Gamma_i^\dagger = \Gamma_i, ~~ \tr(\Gamma_i) = 0,                  & ~~\tr(\Gamma_i\Gamma_j) = 2\delta_{ij}, \label{eq:bloch_properties_generators_trace}                                                                 \\
    \comm{\Gamma_i}{\Gamma_j} = 2\imi\sum_{k=1}^{N^2-1}f_{ijk}\Gamma_k, & ~~~~\acomm{\Gamma_i}{\Gamma_j} = \dfrac{4}{N}\delta_{ij}\mathit{I}_N+2\sum_{k=1}^{N^2-1}g_{ijk}\Gamma_k.\label{eq:bloch_properties_generators_comms}
\end{align}
Here, $f_{ijk}$ is a completely antisymmetric tensor, and $g_{ijk}$ is a completely symmetric one. These are the structure constants of the Lie algebra of su($N$).

Now we focus on the relation between the coefficients and the roots. First, we note that the necessary and sufficient condition for the roots to be positive semidefinite is that the coefficients be positive semidefinite and vice versa. Formally stated as below \cite{kimura_2003_bloch,bruning_2012_parametrizations}
\begin{equation}\label{eq:bloch_positive_semid}
    \lambda_i\ge 0 ~(i=1,\cdots,N) ~~ \iff a_i\ge 0 ~(i=1,\cdots,N).
\end{equation}
The Eq. \eqref{eq:bloch_positive_semid} can be proved using Vieta's formula given as
\begin{equation}\label{eq:bloch_vieta}
    a_i = \sum_{1\le i_1<i_2<\cdots i_j}^{N}\lambda_{i_1}\lambda_{i_2}\cdots\lambda_{i_j}.
\end{equation}
These coefficients can be computed explicitly as follows:
\begin{align}\label{eq:bloch_coeffs_complete}
    \begin{split}
        1!a_1 &=~ 1,\\
        2!a_2 &=~ \left(\dfrac{N-1}{N}-\half r^2\right),\\
        3!a_3 &=~ \left(\dfrac{(N-1)(N-2)}{N^2}-\dfrac{3(N-2)}{2N}r^2+\half\sum_{i,j,k=1}^{N^2-1}g_{ijk}r_ir_jr_k\right),\\
        4!a_4 &=~ \left(\dfrac{(N-1)(N-2)(N-3)}{N^3}-\dfrac{3(N-2)(N-3)}{N^2}r^2+\dfrac{3(N-2)}{4N}r^4\right.\\
        &~~\left.+ \dfrac{2(N-3)}{N}\sum_{i,j,k=1}^{N^2-1}g_{ijk}r_ir_jr_k - \dfrac{3}{4}\sum_{i,j,k,l,p=1}^{N^2-1}g_{ijk}g_{klp}r_ir_jr_lr_p\right).
    \end{split}
\end{align}
However, this cumbersome-looking result can be made more convenient by using trace invariants \cite{bruning_2012_parametrizations}. But before that, we need to use Newton's formula involving sums of powers of roots and the coefficients of the polynomials as under \cite{kimura_2003_bloch}. Let us use Eq. \eqref{eq:bloch_poly_root_character} and get the following using Newton's formulas
\begin{equation}\label{eq:bloch_newton}
    ka_k = \sum_{i=1}^{k}(-1)^{i-1}C_{N,i}a_{k-i} ~~~ (1\le k\le N).
\end{equation}
We have defined the power of roots as the collective $C_{N,i} \equiv \sum_{j=1}^{N}\lambda_j^i $. This allows us to write the following
\begin{align}
    \begin{split}\label{eq:bloch_newton_ci}
        1!a_1 &=~ C_{N,1},\\
        2!a_2 &=~ \left(C_{N,1}^2-C_{N,2}\right),\\
        3!a_3 &=~ \left(C_{N,1}^3-3C_{N,1}C_{N,2}+2C_{N,3}\right),\\
        4!a_4 &=~ \left(C_{N,1}^4-6C_{N,1}^2C_{N,2}+8C_{N,1}C_{N,3}+3C_{N,2}^2-6C_{N,4}\right),\\
        \cdots &=~\cdots
    \end{split}
\end{align}
To find the $C_{N,i}$ let us use the following trace relationship using Eq. \eqref{eq:bloch_properties_generators_trace} and Eq. \eqref{eq:bloch_properties_generators_comms},
\begin{align}
    \begin{split}\label{eq:bloch_newton_trace}
        \tr(\Gamma_i\Gamma_j\Gamma_k) &=~ 2z_{ijk},\\
        \tr(\Gamma_i\Gamma_j\Gamma_k\Gamma_l) &=~ \dfrac{4}{N}\delta_{ij}\delta_{kl}+2z_{ijm}z_{mkl},\\
        \tr(\Gamma_i\Gamma_j\Gamma_k\Gamma_l\Gamma_m) &=~ \dfrac{4}{N}\delta_{ij}z_{klm}+\dfrac{4}{N}\delta_{lm}z_{ijk}+2z_{ijn}z_{nko}z_{olm},\\
        \cdots &=~ \cdots,
    \end{split}
\end{align}
while denoting $z_{ijk}=g_{ijk}+\imi f_{ijk}$. Using Eq. \eqref{eq:bloch_bloch_Nd}, we note $C_{N,i} = \tr(\rho^i)$. This allows us to use Eq. \eqref{eq:bloch_newton_trace} to write the following expressions
\begin{align}\label{eq:bloch_ci_trace}
    \begin{split}
        C_{N,1} &=~ 1,\\
        C_{N,2} &=~ \left(4N+2N^2r^2\right)\dfrac{1}{\left(2N\right)^2},\\
        C_{N,3} &=~ \left(8N+12N^2r^2+2N^3r_ir_jr_kg_{ijk}\right)\dfrac{1}{\left(2N\right)^3},\\
        C_{N,4} &=~ \left(16N+48N^2r^2+16N^3r_ir_jr_kg_{ijk}+4N^3r^4+2N^4r_ir_jr_kr_lg_{ijm}g_{mkl}\right)\dfrac{1}{\left(2N\right)^4}.
    \end{split}
\end{align}
When we substitute Eq. \eqref{eq:bloch_ci_trace} in Eq. \eqref{eq:bloch_newton_ci}, we get the following trace relations, which are an improvement over Eq. \eqref{eq:bloch_coeffs_complete},
\begin{align}\label{eq:bloch_coeffs_trace}
    \begin{split}
        1!a_1 &=~ \tr(\rho),\\
        2!a_2 &=~ 1-\tr(\rho^2),\\
        3!a_3 &=~ 1-3\tr(\rho^2)+2\tr(\rho^3),\\
        4!a_4 &=~ 1-6\tr(\rho^2)+8\tr(\rho^3)-6\tr(\rho^4)+3\left(\tr(\rho^2)\right)^2.
    \end{split}
\end{align}

This concludes our introduction to Bloch parametrization. However, our job does not end here. This parametrization provides us a set $\mathcal{Q}_N$ from which we can have a map to $\mathit{D}_N$, but, which $\bm{r}\in\mathbb{R}^{N^2-1}$ actually lies in $\mathcal{Q}_N$ is still a difficult problem to tackle. Most of the literature contain a convenient set of non-zero components of $\bm{r}$ and correspondingly study different `cuts' in the hyper ball $\mathcal{B}(\mathbb{R}^{N^2-1})$ of which $\mathcal{Q}_N$ is a proper subset. Different types of such cuts exist depending on the components of interest \cite{jakobczyk_2001_geometry,kimura_2003_bloch,boya_2008_geometry}. For the purpose of this thesis, and the upcoming computations, it is imperative that we seek out some analytical form of the eigenvalues as in the two-level case, \emph{viz.,} $\dfrac{1\pm r}{2}$, which seems to be lacking in the literature. So in the next sections, we develop the required analytical form of some of the eigenvalues for the cases of $N=3$ and $N=4$ respectively.
\section{The case with $N=3$}\label{sec:bloch_qutrit}
The density matrix $\rho$ for $N=3$ can be written as
\begin{equation}\label{eq:bloch_bloch_3d}
    \rho_3 = \dfrac{1}{3}\left(\mathit{I}_3+\sqrt{3}\bm{r}\cdot\Gamma\right).
\end{equation}
The Bloch vector $\bm{r}$ is eight-dimensional, and the corresponding eight Generalized Gell-Mann matrices can be expressed in the following way. The two diagonal matrices are,
\begin{align}\label{eq:bloch_qutrit_GGM_d}
    \Gamma_1 = \mqty[1 & 0 & 0 \\ 0 & -1 & 0\\ 0 & 0 & 0], ~~&~~ \Gamma_2 = \dfrac{1}{\sqrt{3}}\mqty[1 & 0 & 0\\ 0 & 1 & 0\\ 0 & 0 & -2].
\end{align}
And the six off-diagonal ones are given as
\begin{align}
    \Gamma_3 = \mqty[0 & 1    & 0 \\ 1 & 0 & 0\\0 & 0 & 0], ~~&~~~~ \Gamma_4 = \mqty[0 & 0 & 1\\ 0 & 0 & 0\\ 1 & 0 & 0], & \Gamma_5 = \mqty[0 & 0 & 0\\0 & 0 & 1\\ 0 & 1 & 0],\notag\\
    \Gamma_6 = \mqty[0 & \imi & 0 \\ -\imi & 0 & 0\\0 & 0 & 0], ~~&~~~~ \Gamma_7 = \mqty[0 & 0 & \imi\\ 0 & 0 & 0\\ -\imi & 0 & 0], & \Gamma_8 = \mqty[0 & 0 & 0\\0 & 0 & \imi\\ 0 & -\imi & 0].\label{eq:bloch_qutrit_GGM_od}
\end{align}
Now that we have declared all the necessary components, we are equipped to write down the cubic characteristic polynomial whose roots $\lambda$ will give us the eigenvalues. A straightforward computation yields, with $r = \abs{\sqrt{\sum_{i=1}^{8}r_i^2}}$,
\begin{align}\label{eq:bloch_qutrit_qubic_full}
    \begin{split}
        & \lambda^3-\lambda^2+\frac{1}{3} \lambda  \left(1-r^2\right)+\frac{1}{3} \left(\frac{1}{3} \left(1-2 r_2\right) \left(r^2-\frac{1}{3} \left(4 r_2^2+2 r_2+1\right)\right)\right. \\
        &\left.+\left(r_2-\frac{r_1}{\sqrt{3}}\right) \left(r_4^2+r_7^2\right)
        -\frac{2}{\sqrt{3}} \left(r_3 r_4 r_5+r_6 r_7 r_5-r_4 r_6 r_8+r_3 r_7 r_8\right)\right.\\
        &\left.+\left(\frac{r_1}{\sqrt{3}}+r_2\right) \left(r_5^2+r_8^2\right)\right)=0
    \end{split}
\end{align}
Comparing with Eq. \eqref{eq:bloch_charecter_pol} we get,
\begin{align}\label{eq:bloch_qutrit_a3}
    \begin{split}
        a_3 &= -\frac{1}{3}\left(\frac{1}{3} \left(1-2 r_2\right) \left(r^2-\frac{1}{3} \left(4 r_2^2+2 r_2+1\right)\right)+\left(r_2-\frac{r_1}{\sqrt{3}}\right) \left(r_4^2+r_7^2\right)\right.\\
        &\left.-\frac{2}{\sqrt{3}} \left(r_3 r_4 r_5+r_6 r_7 r_5-r_4 r_6 r_8+r_3 r_7 r_8\right)+\left(\frac{r_1}{\sqrt{3}}+r_2\right) \left(r_5^2+r_8^2\right)\right).
    \end{split}
\end{align}
We re-write Eq. \eqref{eq:bloch_qutrit_qubic_full} with Eq. \eqref{eq:bloch_qutrit_a3} the following way,
\begin{equation}\label{eq:bloch_qutrit_qubic_compact}
    \lambda^3-\lambda^2+\dfrac{1-r^2}{3}\lambda-a_3=0.
\end{equation}
Reading with the following substitutions
\begin{align}\label{eq:bloch_qutrit_rot_subs}
    \begin{split}
        -1+27a_3+3r^2 &=~ \omega,\\
        \left(\omega+\sqrt{-4r^6+\omega^2}\right)^{\sfrac{1}{3}} &=~ \theta,
    \end{split}
\end{align}
Eq. \eqref{eq:bloch_qutrit_qubic_compact} has the following solutions
\begin{align}\label{eq:bloch_qutrit_root_full}
    \begin{split}
        \lambda_1 &=~ \dfrac{1}{3} + \dfrac{2^{\sfrac{1}{3}}r^2}{3\theta}+\dfrac{\theta}{3\cdot2^{\sfrac{1}{3}}},\\
        \lambda_2 &=~ \dfrac{1}{3} - \dfrac{\left(1+\imi\sqrt{3}\right)r^2}{3\cdot2^{\sfrac{1}{3}}\theta}+\dfrac{\left(1-\imi\sqrt{3}\right)\theta}{6\cdot2^{\sfrac{1}{3}}},\\
        \lambda_3 &=~ \dfrac{1}{3} - \dfrac{\left(1-\imi\sqrt{3}\right)r^2}{3\cdot2^{\sfrac{1}{3}}\theta}+\dfrac{\left(1+\imi\sqrt{3}\right)\theta}{6\cdot2^{\sfrac{1}{3}}}.
    \end{split}
\end{align}
To ensure the realness of the roots in Eq. \eqref{eq:bloch_qutrit_root_full}, we must ensure the criteria below
\begin{equation}\label{eq:bloch_qutrit_real_condition}
    \dfrac{r^2}{2^{\sfrac{1}{3}}\theta} = \dfrac{\theta}{2\cdot2^{\sfrac{1}{3}}},
\end{equation}
which implies
\begin{equation}\label{eq:bloch_qutrit_theta}
    \theta = 2^{\sfrac{1}{3}}r.
\end{equation}
Now we get nice-looking roots for the case of $N=3$ of the Bloch sphere representation as given under
\begin{align}\label{eq:bloch_qutrit_roots}
    \begin{split}
        \lambda_1 &=~ \dfrac{1}{3}\left(1+2r\right),\\
        \lambda_2 &=~ \dfrac{1}{3}\left(1-r\right),\\
        \lambda_3 &=~ \dfrac{1}{3}\left(1-r\right).
    \end{split}
\end{align}
In the following, we consider the $N=4$ case.

\section{The case with $N=4$}\label{sec:bloch_qudit}
We begin by writing down the density matrix using the GGM basis as
\begin{equation}\label{eq:bloch_bloch_4d}
    \rho_4 = \dfrac{1}{4}\left(\mathit{I}_4+\sqrt{6}\bm{r}\cdot\bm{\Gamma}\right).
\end{equation}

In this case, $\bm{r}$ is 15-dimensional, with $r = \abs{\sqrt{\sum_{i=1}^{15}r_i^2}}$.
Also, $\bm{\Gamma}$ has 15 components enumerated and expressed as under. Firstly, the three diagonal matrices,
\begin{align}\label{eq:bloch_qudit_ggm_d}
    \Gamma_1 = \mqty[1 & 0 & 0 & 0 \\0 & -1 & 0 & 0\\ 0 & 0 & 0 & 0\\ 0 & 0 & 0 & 0], ~~&~~~~~ \Gamma_2 = \dfrac{1}{\sqrt{3}}\mqty[1 & 0 & 0 & 0 \\0 & 1 & 0 & 0\\ 0 & 0 & -2 & 0\\ 0 & 0 & 0 & 0], &\Gamma_3 = \dfrac{1}{\sqrt{6}}\mqty[1 & 0 & 0 & 0 \\0 & 1 & 0 & 0\\ 0 & 0 & 1 & 0\\ 0 & 0 & 0 & -3].
\end{align}
And the 12 off-diagonal matrices are,
\begingroup
\allowdisplaybreaks
\begin{align}
    \Gamma_4 = \mqty[0    & 1    & 0 & 0 \\1 & 0 & 0 & 0\\ 0 & 0 & 0 & 0\\ 0 & 0 & 0 & 0], & \Gamma_5 = \mqty[0 & 0 & 1 & 0 \\0 & 0 & 0 & 0\\ 1 & 0 & 0 & 0\\ 0 & 0 & 0 & 0], \Gamma_6 = \mqty[0 & 0 & 0 & 1 \\0 & 0 & 0 & 0\\ 0 & 0 & 0 & 0\\ 1 & 0 & 0 & 0],\notag\\
    \Gamma_7 = \mqty[0    & 0    & 0 & 0 \\0 & 0 & 1 & 0\\ 0 & 1 & 0 & 0\\ 0 & 0 & 0 & 0], & \Gamma_8 = \mqty[0 & 0 & 0 & 0 \\0 & 0 & 0 & 1\\ 0 & 0 & 0 & 0\\ 0 & 1 & 0 & 0], \Gamma_9 = \mqty[0 & 0 & 0 & 0 \\0 & 0 & 0 & 0\\ 0 & 0 & 0 & 1\\ 0 & 0 & 1 & 0],\notag\\
    \Gamma_{10} = \mqty[0 & \imi & 0 & 0 \\-\imi & 0 & 0 & 0\\ 0 & 0 & 0 & 0\\ 0 & 0 & 0 & 0], & \Gamma_{11} = \mqty[0 & 0 & \imi & 0 \\0 & 0 & 0 & 0\\ -\imi & 0 & 0 & 0\\ 0 & 0 & 0 & 0], \Gamma_{12} = \mqty[0 & 0 & 0 & \imi \\0 & 0 & 0 & 0\\ 0 & 0 & 0 & 0\\ -\imi & 0 & 0 & 0],\notag\\
    \Gamma_{13} = \mqty[0 & 0    & 0 & 0 \\0 & 0 & \imi & 0\\ 0 & -\imi & 0 & 0\\ 0 & 0 & 0 & 0], & \Gamma_{14} = \mqty[0 & 0 & 0 & 0 \\0 & 0 & 0 & \imi\\ 0 & 0 & 0 & 0\\ 0 & -\imi & 0 & 0], \Gamma_{15} = \mqty[0 & 0 & 0 & 0 \\0 & 0 & 0 & 0\\ 0 & 0 & 0 & \imi\\ 0 & 0 & -\imi & 0].\label{eq:bloch_qudit_GGM_od}
\end{align}
\endgroup
Using this basis, we can express the following
\begin{equation}\label{eq:bloch_vector_4d_traceless}
    \bm{r}\cdot\bm{\Gamma}={\small
    \begin{pmatrix}
        r_1+\frac{r_2}{\sqrt{3}}+\frac{r_3}{\sqrt{6}} & r_4-\imi r_{10}                                & r_5-\imi r_{11}                             & r_6-\imi r_{12}        \\
        r_4+\imi r_{10}                               & -r_1+\frac{r_2}{\sqrt{3}}+\frac{r_3}{\sqrt{6}} & r_7-\imi r_{13}                             & r_8-\imi r_{14}        \\
        r_5+\imi r_{11}                               & r_7+\imi r_{13}                                & -\frac{2r_2}{\sqrt{3}}+\frac{r_3}{\sqrt{6}} & r_9-\imi r_{15}        \\
        r_6+\imi r_{12}                               & r_8+\imi r_{14}                                & r_9+\imi r_{15}                             & -\frac{\sqrt{6}r_3}{2}
    \end{pmatrix}},
\end{equation}
the reduced density matrices can be found as,
\begin{align}\label{eq:bloch_4d_reduce}
    \begin{split}
        \rho_A &= \left(\mathit{I}_2+\bm{r}_A\cdot\bm{\sigma}\right)/2\,,\\
        \rho_B &= \left(\mathit{I}_2+\bm{r}_B\cdot\bm{\sigma}\right)/2\,,\\
        r_{Aj}&=\sqrt{6}\Tr\left(\sigma_j\PTr{B}{\bm{r}\cdot\bm{\Gamma}}\right)\,,\\
        r_{Bj}&=\sqrt{6}\Tr\left(\sigma_j\PTr{A}{\bm{r}\cdot\bm{\Gamma}}\right)\,,\\
        \bm{r}_A&=	{\small
        \begin{pmatrix}
            4(\sqrt{2}r_2+r_3)        \\
            2 \sqrt{6}(r_5+r_8)       \\
            2 \sqrt{6}(r_{11}+r_{14}) \\
        \end{pmatrix}}\,,\quad
        \bm{r}_B= 	{\small
        \begin{pmatrix}
            2(\sqrt{6}r_1-\sqrt{2}r_2+2r_3) \\
            2 \sqrt{6}(r_4+r_9)             \\
            2 \sqrt{6}(r_{10}+r_{15})       \\
        \end{pmatrix}}	\\
    \end{split}
\end{align}
As it has been obvious by now, it will be quite difficult and frivolous to the current discussion if we explicitly write down the coefficients $a_3$ and $a_4$ like we did in the case of the qutrit. Instead, we will note the following, that $\tr(\rho^2) = \dfrac{1}{N}\left(1+(N-1)r^2\right)$. This gives us upon substituting in Eq. \eqref{eq:bloch_coeffs_trace} the following expression
\begin{equation}\label{eq:bloch_gen_a2}
    a_2 = \dfrac{N-1}{2N}\left(1-r^2\right),
\end{equation}
which for $N=4$ gives, $a_2 = \dfrac{3}{8}\left(1-r^2\right)$. Setting this into the quartic characteristic polynomial of Eq. \eqref{eq:bloch_bloch_4d}, we get the following equation ($a_0=-a_1=1$),
\begin{equation}\label{eq:bloch_the_quartic}
    \lambda^4-\lambda^3+\dfrac{3}{8}\left(1-r^2\right)\lambda^2-a_3\lambda+a_4=0.
\end{equation}
with the help of the Mathematica Software \cite{wolframresearchinc._2022_mathematica} the roots may be put in  the following form,
\begingroup
\allowdisplaybreaks
\begin{align}\label{eq:bloch_quartic_roots}
    4\, \lambda_1 & = 1-\alpha_1- \sqrt{3r^2-\alpha_1^2-\alpha_2}, \notag \\
    4\, \lambda_2 & = 1+\alpha_1+ \sqrt{3r^2-\alpha_1^2+\alpha_2}, \notag \\
    4\, \lambda_3 & = 1+\alpha_1- \sqrt{3r^2-\alpha_1^2+\alpha_2},        \\
    4\, \lambda_4 & = 1-\alpha_1+ \sqrt{3r^2-\alpha_1^2-\alpha_2},\notag
\end{align}
\endgroup
where  $\alpha_1$ and $\alpha_2$ are related to $r$, $a_3$ and $a_4$ as  follows
\begingroup
\allowdisplaybreaks
\begin{align}
    \alpha_1 & = r\sqrt{\alpha_3/2+1}\,
    ,	
    \notag                                                                              \\
    \alpha_2 & = (2^4 a_3+3r^2-1)/\alpha_1\, ,                                          
    \notag                                                                              \\
    \alpha_3 & = \sqrt[3]{\alpha_5 } +\alpha_4/(3\sqrt[3]{\alpha_5 })\,
    ,
    \notag                                                                              \\
    \alpha_4 & = 2^8 a_4-2^6 a_3+3\left(1-r^2\right)^2  ,                               
    \\
    \alpha_5 & = \alpha_6+\sqrt{\alpha_6^2-\alpha_7^3} \, ,                             
    \notag                                                                              \\
    \alpha_6 & =2^8 a_4 r^2+2^8 a_3^2-2^5 a_3 \left(1-r^2\right) + \left(1-r^2\right)^3
    ,
    \notag                                                                              \\
    \alpha_7 & =2^8 a_4/3- 2^6 a_3/3+ \left(1-r^2\right)^2.                             
\end{align}
\endgroup
All the above $\alpha_j$'s vanish in the limit $r\downarrow 0$,
which corresponds to the maximally mixed state $\rho=I_4/4$ for which $r=0$, $a_3=1/2^4$, and $a_4=1/2^8$.
The conditions required for the positivity of the eigenvalues may be written as
\begingroup
\allowdisplaybreaks
\begin{align}\label{eq:bloch_quartic_positivity}
    2^3\,a_2 & =3(1-r^2)\ge 0\,,\notag                                                               \\
    2^4\,a_3 & = \alpha_1\alpha_2+1-3r^3\ge 0\,,                                                     \\
    2^8\,a_4 & = 4\alpha_1^2(\alpha_1^2-3r^2)+4\alpha_1\alpha_2-\alpha_2^2+(3r^2-1)^2 \ge 0\,.\notag
\end{align}
\endgroup
The conditions  required for the $\alpha_j$'s and the eigenvalues to be real may be written as
\begingroup
\allowdisplaybreaks
\begin{align}\label{eq:bloch_quartic_reality}
     & \alpha_3/2+1\ge 0\,,                                                                                \\
     & \alpha_7\le 0\,,                                                                                    \\
     & \alpha_6 ^2-\alpha_7^3\ge 0\,,                                                                      \\
     & \alpha_1^2\le 3r^2\pm \alpha_2\quad\mbox{or}\quad \alpha_1^2-3r^2\le \alpha_2\le 3r^2-\alpha_1^2\,.
\end{align}
\endgroup
Notice also that
\begin{align}\label{eq:bloch_quartic_alpha1alpha2}
    1        & =\lambda_1+\lambda_2+\lambda_3 +\lambda_4                  \\
    r^2      & =(4(\lambda_1^2+\lambda_2^2+\lambda_3^2 +\lambda_4^2)-1)/3 \\
    \alpha_1 & =\lambda_2+\lambda_3-\lambda_1 -\lambda_4                  \\
    \alpha_2 & =2(\lambda_2-\lambda_3)^2-2(\lambda_4 -\lambda_1)^2
\end{align}

\section{The nature of operators in Bloch representation}\label{sec:bloch_operators}
In his work, Beretta provided a unique formalism to understand the nature of operators in Bloch representation for the case of $N=2$ \cite{beretta_1985_entropy}. However, this work has not been picked up by others and we felt the need for a similar formalism for the analytical computation that will be needed in the following chapter. One major obstacle could have been the unavailability of analytical forms of the roots in the Bloch representation for $N\ge3$. To remedy this, here we present Beretta's formalism, followed by our extension of the same in the case of $N=3$ and $N=4$ with the help of the solutions presented in Eq. \eqref{eq:bloch_qutrit_roots} and \eqref{eq:bloch_quartic_roots}, respectively.
\subsection{N=2}\label{secsub:bloch_frho_n2}
For the two-level case, the roots are given as $\lambda_1=\half\left(1+r\right)$, and $\lambda_2=\half\left(1-r\right)$, and $\lambda_1-\lambda_2=r$. We also know that if $\lbrace\lambda_i\rbrace$ are the eigenvalues of a matrix $A$ with the corresponding eigenvectors denoted by $\lbrace\ket{\lambda_i}\rbrace$, then using spectral theorem, we can write the following,
\begin{equation}\label{eq:bloch_spectral}
    F(A) = \sum_{i=1}^{N}F(\lambda_i)\dyad{\lambda_i}{\lambda_i},
\end{equation}
where $F$ acts on real parameters $x$. Also, we will use the completeness relation $\sum_{i=1}^{N}\dyad{\lambda_i}{\lambda_i}=\mathit{I}_N$. So, we can write the following steps beginning from Eq. \eqref{eq:bloch_spectral} for $N=2$ \cite{beretta_1985_entropy}
\begin{align}\label{eq:bloch_qubit_operator}
    \begin{split}
        F(\rho) &=~ F(\lambda_1)\dyad{\lambda_1}{\lambda_1}+F(\lambda_2)\dyad{\lambda_2}{\lambda_2},\\
        &=~\dfrac{1}{r}\left[\lambda_1F(\lambda_2)-\lambda_2F(\lambda_1)\right]\mathit{I}_2+\dfrac{1}{r}\left[F(\lambda_1)-F(\lambda_2)\right]\rho,\\
        &=~\half\left[F(\lambda_1)+F(\lambda_2)\right]\mathit{I}_2+\dfrac{1}{2r}\left[F(\lambda_1)-F(\lambda_2)\right]\bm{r}\cdot\bm{\sigma}.
    \end{split}
\end{align}
In deriving the last line of Eq. \eqref{eq:bloch_qubit_operator}, Eq. \eqref{eq:qubit_bloch} has been used. The benefit of this formalism can be seen immediately, the operator $F(\rho)$ has two components. The component involving $\mathit{I}_2$ in the last line of Eq. \eqref{eq:bloch_qubit_operator} is of non-zero trace. Whereas, the other component involving the Pauli operator remains traceless. Thus, for analytical computations of operators such as $\rho\ln(\rho)$ and their respective traces, we find ourselves at an advantage. However, this is not easily generalizable to higher dimensions. We will shortly see that below.
\subsection{N=3}\label{secsub:bloch_frho_n3}
For the $N=3$ case, we use Eq. \eqref{eq:bloch_qutrit_roots} to see that $2\lambda_1-\lambda_2-\lambda_3 = 2r$, and $\lambda_2=\lambda_3$. As before, we begin with the spectral theorem and use the definition in Eq. \eqref{eq:bloch_bloch_3d}
\begingroup
\allowdisplaybreaks
\begin{align}
    F(\rho_3) & = F(\lambda_1)\dyad{\lambda_1}{\lambda_1}+F(\lambda_2)\dyad{\lambda_2}{\lambda_2}+F(\lambda_3)\dyad{\lambda_3}{\lambda_3},\notag                                                                               \\
              & = \dfrac{2\lambda_1-\lambda_2-\lambda_3}{2r}\left[F(\lambda_1)\dyad{\lambda_1}{\lambda_1}+F(\lambda_2)\dyad{\lambda_2}{\lambda_2}+F(\lambda_3)\dyad{\lambda_3}{\lambda_3}\right],\notag                        \\
              & = \dfrac{1}{2r}\left[\left(2\lambda_1-\lambda_3-r\right)F(\lambda_2)+\left(2\lambda_1-\lambda_2-r\right)F(\lambda_3)-\left(\lambda_2+\lambda_3\right)F(\lambda_1)\right]\mathit{I}_3\notag                     \\
              & +\dfrac{1}{2r}\left[2F(\lambda_1)-F(\lambda_2)-F(\lambda_3)\right]\rho_3,\notag                                                                                                                                \\
              & = \dfrac{1}{3}\left[F(\lambda_1)+F(\lambda_2)+F(\lambda_3)\right]\mathit{I}_3+\dfrac{1}{2\sqrt{3}r}\left[2F(\lambda_1)-F(\lambda_2)-F(\lambda_3)\right]\bm{r}\cdot\bm{\Gamma}.\label{eq:bloch_qutrit_operator}
\end{align}
\endgroup

\subsection{N=4}\label{secsub:bloch_frho_n4}
We observe that there is various class of degeneracy when four eigenvalues are involved. Either one is different, and all three are the same, or there is a pairwise degeneracy, or all four are the same. The first class is
\begin{align}\label{eq:bloch_class_degen_1}
    \begin{split}
        \lambda_4&=\lambda_3=\lambda_1=(1-\alpha_1)/4,\qquad  \lambda_1\le 1/3 \\  \lambda_2&=1-3\lambda_1=(1+3\alpha_1)/4,\\
        \alpha_1&=1-4\lambda_1,\qquad  -1/3 \le\alpha_1\le 1\\
        \alpha_2&=2\alpha_1^2=2(1-4\lambda_1)^2,\qquad
        r^2=\alpha_1^2
    \end{split}
\end{align}
which includes the one-dimensional pure states, $\alpha_1=1$ ($\lambda_1=\lambda_3=\lambda_4=0$, $\lambda_2=1$), the maximally mixed state,  $\alpha_1=0$ ($\lambda_1=\lambda_2=\lambda_3=\lambda_4=1/4$), the  maximally mixed states with three-dimensional support, $\alpha_1=-1/3$ ($\lambda_1=\lambda_3=\lambda_4=1/3$, $\lambda_2=0$), and the separable ($-1/3<\alpha_1\le 1/3$) and the entangled ($1/3<\alpha_1< 1$)  Werner states.

A second class is
\begin{align}\label{eq:bloch_class_degen_2}
    \begin{split}
        \lambda_4&=\lambda_1=(1-\alpha_1)/4,\qquad  \lambda_1\le 1/2 \\  \lambda_3&=\lambda_2=(1-2\lambda_1)/2=(1+\alpha_1)/4\\
        \alpha_1&=1-4\lambda_1,\qquad  -1 \le\alpha_1\le 1\\
        \alpha_2&=0,\qquad
        r^2 =\alpha_1^2/3
    \end{split}
\end{align}
which includes  the maximally mixed state,  $\alpha_1=0$ ($\lambda_1=\lambda_2=\lambda_3=\lambda_4=1/4$), and the  maximally mixed states with two-dimensional support, $\alpha_1=\pm 1$ ($\lambda_1=\lambda_4=1/2$, $\lambda_2=\lambda_3=0$).

Consider a function $F(x)$ such that $x\in\mathbb{R}$. Now to find the analytical expression for the operator $F(\rho)$ in view of the degeneracy of the states in these two classes, when $\lambda_2\ne\lambda_1$ [i.e., excluding the trivial case $\rho=\mathit{I}_4/4$ for which $ F(\rho)=\mathit{I}_4F(1/4)$] we may write ($\mathrm{P}_i$ is the projector onto the degenerate subspace $i$)
\begin{align}\label{eq:bloch_degenerate_projection}
    \begin{split}
        \rho&=\lambda_1 \mathrm{P}_1+\lambda_2 \mathrm{P}_2, \qquad  \mathit{I}_4=\mathrm{P}_1+ \mathrm{P}_2\\
        \mathrm{P}_2&=\mathit{I}_4-\mathrm{P}_1,\qquad \mathrm{P}_1=\frac{1}{\lambda_1 -\lambda_2}\rho-\frac{\lambda_2}{\lambda_1 -\lambda_2} \mathit{I}_4.
    \end{split}
\end{align}
From this, we have
\begin{align}\label{eq:bloch_class_degen_Frho}
    \begin{split}
        F(\rho)&=F(\lambda_1) \mathrm{P}_1+F(\lambda_2) \mathrm{P}_2 \\
        &=\frac{\lambda_1F(\lambda_2)-\lambda_2 F(\lambda_1)}{\lambda_1 -\lambda_2} \mathit{I}_4+\frac{F(\lambda_1)-F(\lambda_2)}{\lambda_1 -\lambda_2} \rho \\
        &=\frac{(4\lambda_1-1)F(\lambda_2)-(4\lambda_2-1) F(\lambda_1)}{4(\lambda_1 -\lambda_2)} \mathit{I}_4+\dfrac{\sqrt{6}\left[F(\lambda_1)-F(\lambda_2)\right]}{4(\lambda_1 -\lambda_2)} \bm{r}\cdot\bm{\Gamma}\\
    \end{split}
\end{align}
where for the first class, the eigenprojectors of $\rho$ are such that $\Tr(\mathrm{P}_1)=3$ and  $\Tr(\mathrm{P}_2)=1$, whereas for the second class $\Tr(\mathrm{P}_1)=\Tr(\mathrm{P}_2)=2$, and in the last equation we used Eq. (\ref{eq:bloch_bloch_Nd}) for $N=4$.

We can conjecture from above that given there exists a relation between the roots of the density matrix $\rho_N$ in $N$-level Bloch representation of the form $f(\{\lambda_i\}) = cr$, where $c$ is an arbitrary constant and $f(\{\lambda_i\})$ is an algebraic relation between the $N$ eigenvalues, there exist some eigenvalues for which the following relation holds
\begin{equation}\label{eq:bloch_gen_operator}
    F(\rho_N) = \dfrac{1}{N}\left[\sum_{i=1}^{N}F(\lambda_i)\right]\mathit{I}_N+\dfrac{1}{N\cdot cr}\sqrt{\dfrac{N(N-1)}{2}}f(\{F(\lambda_i)\})\bm{r}\cdot\bm{\Gamma}.
\end{equation}

\section*{Summary}
To conclude, we have discussed the Bloch parametrization in the context of finite-level density matrices. We have characterized the positivity of the eigenvalues by imposing positivity on the coefficients of the characteristic polynomial. We have further used trace invariants to simplify the general expressions for the coefficients of the polynomial. Eventually, we solved the cubic for $N=3$, and the quartic for $N=4$ and present the analytic expressions of the roots. Finally, we analyzed the general operator acting on these density matrices into zero trace and non-zero trace contributions.\conditionalBlankPage

\chapter{SEA in a Composite System}\label{ch:comSEA}
\blankpage
\newpage
\thispagestyle{empty}
\begin{figure}[H]
  \centering
  \captionsetup{justification=centering}

  \includegraphics[height=0.75\textheight, width=0.75\textwidth]{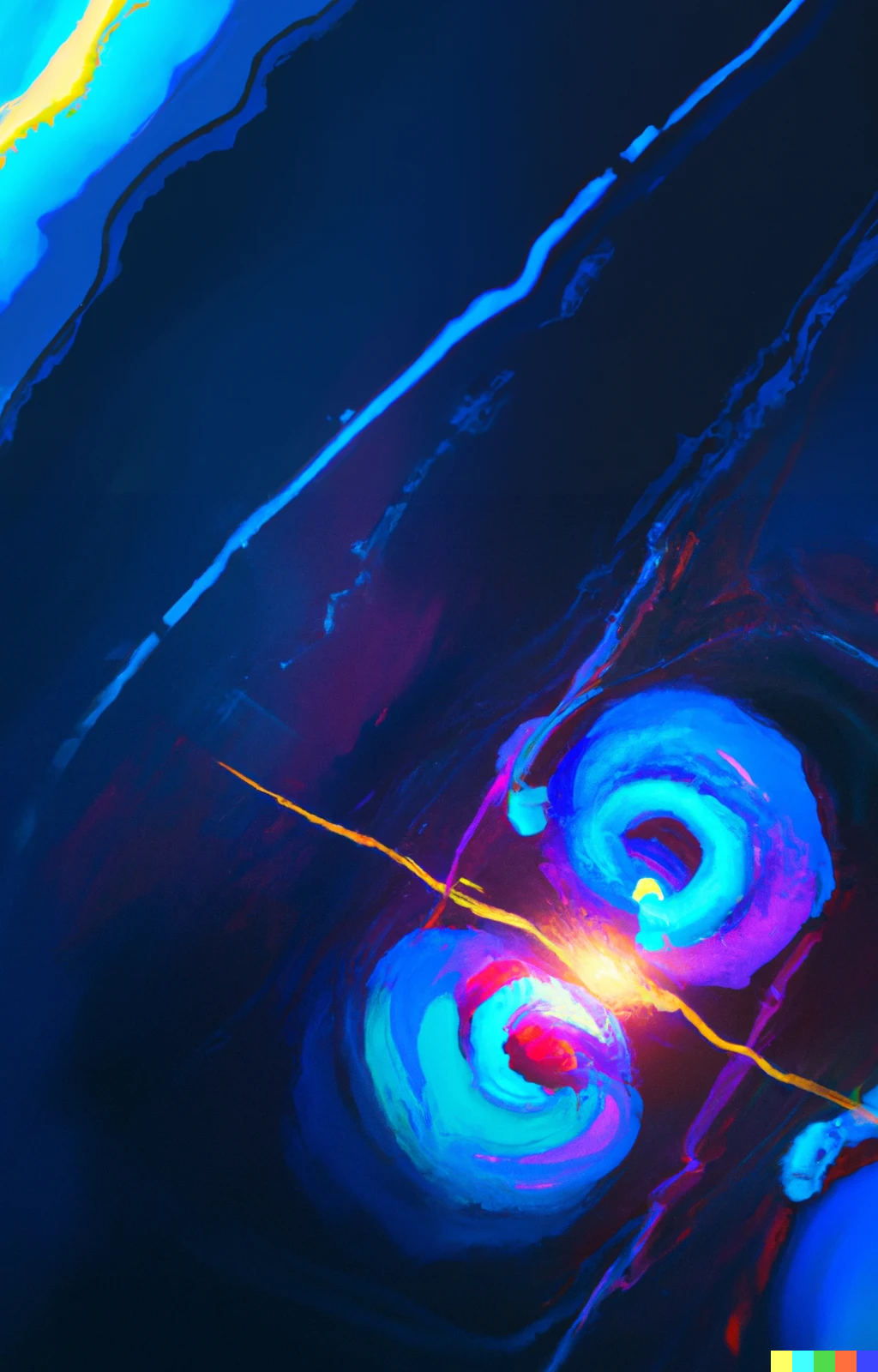}
  \caption*{\emph{I dove into my foundations,}\\
    \emph{had limitless communication}\\
    \emph{to reforge my directions,}\\
    \emph{my enthusiasm rebounding.}}
\end{figure}
\newpage
\blankpage
\lettrine[lines=3]{H}{aving} discussed the implications of the application of the steepest entropy ascent ansatz for the case of a single particle quantum system with finite energy levels, we turn our attention to study the composition of the same. In the preceding chapter \ref{ch:bloch_SEA} we presented the necessary mathematics required to proceed with our discussion. This chapter will focus on a fundamental issue that arises whenever one considers a nonlinear extension of quantum mechanics and its application to composite systems. It has been noted in literature \cite{gisin_1990_weinberg,polchinski_1991_weinberg} that a nonlinear extension of quantum mechanics attracts the possibility of a faster-than-light communication (signaling) between two noninteracting subsystems of a composite. As a consequence, such nonlinear extensions of QM are dreaded and nontrivial. Despite various attempts, a truly nonlinear extension of QM satisfying the criteria of no-signaling as set by Gisin and colleagues has not been discussed in the literature so far \cite{gisin_1990_weinberg,gisin_1995_relevant,ferrero_2004_nonlinear,rembielinski_2020_nonlinear,rembielinski_2021_nonlinear}. Considering the strong nonlinearity present in the evolution under SEA as we saw in the preceding chapters \ref{ch:litsurv}, \ref{ch:qubitSEA}, and \ref{ch:qwalker}, it is prudent that before committing to solving the composite system under SEA, we must address the issue of signaling in the context of SEA, and show that such an ` \ac{EPR} telephone line' is precluded in the SEA theory by construction.  We begin by constructing the SEA equation of motion for a composite system. And we will find the equation of motion for the reduced density matrices belonging to the (noninteracting) subsystems of the composite. The case of the interacting subsystems and further discussion about the nature of the SEA evolutions are beyond the scope of this thesis.
\section{No-signaling in nonlinear QM theroy}\label{sec:comSEA_nosignal}
Weinberg proposed that despite the prevalent linearity in formalism, it has not been concluded conclusively that QM is indeed a linear theory \cite{weinberg_1989_testing}. He also suggested some nonlinear extensions in the Hamiltonian operators that could be experimentally verified in some high-precision measurements. However, in the following year in separate works, Gisin \cite{gisin_1990_weinberg} and Polchinski \cite{polchinski_1991_weinberg} showed such a nonlinear extension through operators will result in the establishment of supraluminal communication (signaling), and one can have an `EPR telephone' line, which strongly violates causality. Here, we present Gisin's \textit{Gedankenexperiment} for completeness from Ref. \cite{gisin_1990_weinberg}.\\

\subsection{Gisin's \emph{Gedankenexperiment}}\label{secsub:comSEA_Gisin}
In this thought experiment, there is a source sitting midway between two observers Alice (A) and Bob (B). A pair of entangled qubits in a singlet state (Bell pair) is emitted from the source (S) along the $y$-axis. A and B receive each part of the pair, and when both of them have received parts of the same entangled pair, we say a channel has been established. S is emitting such pairs continuously maintaining the channel for a sufficiently long time so that particles are available to both observers for multiple measurements. A has two detectors of the Stern-Gerlach type, one is oriented along the $z$-axis, while the other is rotated $45^\circ$ to it in the $z-x$ plane, respectively. Let us consider Alice's system. To encode a message to be sent over the channel, she is free to choose any one of the detectors and the corresponding measurement outcomes would be her single-bit messages. Note that this local operation by Alice does not change the reduced density matrix corresponding to her subsystem.

Now on the side of Bob, we have a stream of spin-half particles one-half in the up state and one-half in the down state in either of the detector basis of A and the order depending on her choice of operation. However, just using local linear operation B is unable to distinguish which basis the measurements were made, and thus no communication takes place using linear Hamiltonians on the maximally entangled channel. Gisin argued, instead, if Bob would incorporate a non-bilinear Hamiltonian locally to act on his pair, he can in principle know which detector setting was used by Alice in encoding her message. This decoding of Alice's message upon the nonlinear operation of the Hamiltonian by Bob creates a scenario where the instantaneous state preparation due to Bell inequality violation becomes a resource for communication. And signaling takes place.

\subsection{No-signaling condition}\label{secsub:comSEA_nosignaling_condition}
In the language of quantum mechanics, signaling implies supraluminal communication. During the evolution of a composite system, Polchinski \cite{polchinski_1991_weinberg} argued that the necessary and sufficient condition for no-signaling is when the observables acting on a given noninteracting subsystem depend only on the reduced density matrix of the same. However, Ferrero \emph{et. al.,} \cite{ferrero_2004_nonlinear} showed that the probability distribution and observables of a particular subsystem also have to be independent of the effects of the remaining subsystems for signaling to be precluded. These definitions stem from imposing causality arguments on the linear structure of QM. Hence, a nonlinear theory by this logic should end up signaling \emph{ex vi termini}.

As noted in \cite{ferrero_2004_nonlinear}, the no-signaling condition is imposed by requiring that in the mutually non-interacting subsystems A and B, the evolution of the local observables of A should only depend on its own reduced state. The SEA formalism, however, allows us to take a less restrictive view \cite{beretta_2005_nonlinear}: the only requirement is that, if A and B are non-interacting, the law of evolution must not allow local unitary operations within B to affect the time evolution of local (reduced, marginal) state of A. Thus, the condition  $\rho_A=\rho_A'$, such as for the two different states $\rho\ne\rho_A\otimes\rho_B$ and $\rho'=\rho_A\otimes\rho_B$, does not require that $\rm{d}\rho_A/\rm{d}t=\rm{d}\rho_A'/\rm{d}t$, because local memory of past interactions, i.e., existing entanglement and/or correlations, may well influence the local evolutions without violating no-signaling. This incorporates the idea that (1) by studying the local evolutions we can disclose the existence of correlations, but only of the type that can be classically communicated between the subsystems, and (2) in the absence of interactions the nonlinear dynamics may produce the fading away of correlations (spontaneous decoherence) but cannot create new correlations.

In linear QM, the system's composition is specified by declaring: (1) the Hilbert space structure as direct product   $\Hil = \bigotimes_{\J=1}^M\Hil_\J$  of the subspaces of the $M$ component subsystems,
and (2) the overall Hamiltonian operator $H  = \sum_{J=1}^M H_\J\otimes I_\Jbar + V$ where $H_\J$ (on $\Hil_\J$) is the local Hamiltonian of the $J$-th subsystem, $I_\Jbar$ the identity on the direct product $\Hil_\Jbar= \bigotimes_{K\ne J}\Hil_K$ of all the other subspaces, and $V$ (on $\Hil$) is the interaction Hamiltonian. The linear law of evolution, $\dot\rho= -i[H,\rho]/\hbar$, has a universal structure and entails the local evolutions through partial tracing, $\dot\rho_\J= -i[H_\J,\rho_\J]/\hbar -i\Tr_\Jbar([V,\rho])/\hbar $. Thus, we recover the universal law $\dot\rho_\J= -i[H_\J,\rho_\J]/\hbar$ for the local density operator $\rho_J=\Tr_\Jbar(\rho)$ if  subsystem $\J$ does not interact with the others (i.e., if $V=I_\J\otimes V_\Jbar$).

As we can see, we cannot have a similar uniform structure for a fully nonlinear QM, because the subdivision into subsystems must explicitly concur to the structure of the dynamical law (see Ref. \cite{beretta_2010_maximum} for more on this). Each new partition of the Hilbert space introduces new equations of motion. Thus one ends up paying a high price for abandoning linearity in QM. But in result makes the theory compatible with the compelling constraint that correlations should not build up and signaling between subsystems should not occur other than per effect of the interaction Hamiltonian $V$ through the standard Schr\"{o}dinger term  $-i[H,\rho]/\hbar$  in the evolution law.

We also impose that the physical observables to be considered in composite quantum dynamics analysis are the `local perception' operators (on $\Hil_\J$). First defined in \cite{beretta_1985_quantuma} together with their `deviation from the local mean value' operators and covariance functionals as \cite{ray_2023_nosignalinga},
\begin{align}
  (X)^J_\rho        & = \Tr_\Jbar[(I_\J\otimes\rho_\Jbar)X] \,,\label{eq:comSEA_local_perception_operators}             \\
  \Delta(X)^\J_\rho & =(X)^\J_\rho-I_\J\Tr[\rho_\J(X)^\J_\rho]\,,\label{eq:comSEA_deviation_X}                          \\
  (X,Y)^\J_\rho     & = \half\Tr[\rho_\J\acomm{\Delta(X)^\J_\rho}{\Delta(Y)^\J_\rho}]\,,\label{eq:comSEA_covariance_XY}
\end{align}
where $\rho_\Jbar=\Tr_\J(\rho)$. For a bipartite system AB, the local perception operators $(X)^A_\rho$ (on $\Hil_A$) and $(X)^B_\rho$  (on $\Hil_B$)  are the unique operators  that for a given $X$ on $\Hil_{AB}$ satisfy for all states $\rho$  the identity
\begin{equation}\label{eq:comSEA_local_perception_operators_id}
  \Tr[\rho_A(X)^A_\rho] =\Tr[(\rho_A\otimes\rho_B)X] =\Tr[\rho_B(X)^B_\rho]\,,\\
\end{equation}
which shows that they represent all that A and B can say about the overall observable X by classically sharing their local states. Operator $(X)^A_\rho$ can be viewed as the projection onto $\Hil_A$ of the operator $X$ weighted by the local state $\rho_B$ of subsystem $B$. It is a local observable for subsystem A which, however, depends on the overall state $\rho$ and overall observable $X$. Its local mean value $\Tr_A[\rho_A(X)^A_\rho]$ differs from the mean value $\Tr(\rho X)$ for the overall system AB, except when A and B are uncorrelated ($\rho=\rho_A\otimes\rho_B$). It was dubbed  `local perception' because even if B performs a local tomography and sends the measured $\rho_B$ to A by classical communication, the most that A can measure locally about the overall observable $X$ is $(X)^A_\rho$.

The overall energy and entropy of the composite system are locally perceived
within subsystem $J$ through the operators $(H)^\J_\rho$ and $(S(\rho))^\J_\rho$ defined on $\Hil_\J$ by Eq. (\ref{eq:comSEA_local_perception_operators}), respectively with $X=H$, the overall Hamiltonian, and $X=S(\rho)=-\Boltz\mathit{B}\ln(\rho)$, that we  call the overall entropy operator, where $\mathit{B}\ln(x)$ denotes the discontinuous function $\mathit{B}\ln(x)=\ln(x)$ for $0<x\le 1$ and $\mathit{B}\ln(0)=0$. It must be carefully observed that the `locally perceived overall entropy' operator $(S(\rho))^\J_\rho$ is different from the `local entropy' operator $S(\rho_\J)=-\Boltz{\rm B}_J{\rm ln}(\rho_\J)$. Not only that, their mean values  $\Tr[\rho_\J(S(\rho))^\J_\rho]=-\Boltz\Tr[(\rho_\J\otimes\rho_\Jbar){\rm Bln}(\rho)]$ and $\Tr[\rho_\J S(\rho_\J)]= -\Boltz\Tr[\rho_\J\ln (\rho_\J)]$ are also different. Only when $\rho=\rho_\J\otimes\rho_\Jbar$ they are related by $\Tr[\rho_\J(S(\rho))^\J_\rho]=\Tr[\rho_\J S(\rho_\J)]+\Tr[\rho_\Jbar S(\rho_\Jbar)]= -\Boltz\Tr[\rho\ln (\rho)]$.  Likewise, for the `locally perceived overall Hamiltonian' operator $(H)^\J_\rho$ we can make similar statements. When the overall observable $X$ is `separable for subsystem J', in the sense that  $X=X_\J\otimes I_\Jbar+I_\J\otimes X_\Jbar$ then, even if $\rho\ne\rho_\J\otimes\rho_\Jbar$,  the deviations and covariances reduce to their local versions (special case),
\begin{align}
  \Delta(X)^\J_\rho & =\Delta X_\J = X_\J-I_\J\Tr[\rho_\J X_\J]\,,\label{eq:comSEA_deviation_separableJ}       \\
  (X,Y)^\J_\rho     & =\Tr[\rho_\J\acomm{\Delta X_\J}{\Delta Y_\J}]/2\,.\label{eq:comSEA_covariance_separable}
\end{align}

Now, to formalize the no-signaling definition following \cite{beretta_2005_nonlinear} as discussed above, we impose that if A and B are non-interacting, a local unitary operation on B should not affect the evolution of A. So, consider the composite AB in the state $\rho$, where a local arbitrary unitary operation $U_B$ on B ($U_B^\dagger U_B=I_B$) changes $\rho$ to
\begin{equation}\label{eq:comSEA_partial_unitary_op_on_rho}
  \rho'=(I_A\otimes U_B)\,\rho\, (I_A\otimes U_B^\dagger)\,.
\end{equation}
Using the partial cyclic properties of the partial trace,
\begin{align}
   & \Tr_B[(I_A\otimes X_B)Z_{AB}]=\Tr_B[Z_{AB}(I_A\otimes X_B)]\,,\notag         \\
   & \Tr_A[(I_A\otimes X_B)Z_{AB}(I_A\otimes Y_B)]= X_B\Tr_A(Z_{AB})Y_B \,,\notag
\end{align}
we obtain the identities
\begin{align}
  \rho_B  & =\Tr_A[(I_A\otimes U_B^\dagger)\,\rho'\, (I_A\otimes U_B)]=U_B^\dagger\rho'_BU_B,
  \label{eq:comSEA_no_signaling_unit_rhoB}                                                                            \\
  \rho'_A & =\Tr_B[(I_A\otimes U_B)\,\rho\, (I_A\otimes U_B^\dagger)]=\Tr_B[(I_A\otimes U_B^\dagger U_B)\,\rho]\notag \\&=\Tr_B[(I_A\otimes I_B)\,\rho]=\rho_A. \label{eq:comSEA_rhoAprime}
\end{align}
This establishes that a local operation on B does not affect the local state $\rho_A$ of A, hence the usual idea \cite{ferrero_2004_nonlinear} that for no-signaling it is sufficient that the dynamical model implies evolutions of local observables that depend only on $\rho_A$. However, this must be noted that it is a sufficient condition and not a necessary one. We prove next that not only the local reduced state $\rho_A$ but also the local perception operators $(F(\rho))^A$ of any well-defined nonlinear function $F(\rho)$ of the overall state (such as the function $S(\rho)$ defined above for entropy) are not affected by local operations on B according to Eq. (\ref{eq:comSEA_partial_unitary_op_on_rho}). And since the SEA formalism is based on such local perception operators, this is an important lemma in the proof that SEA is no-signaling.

So, let us apply Eq. (\ref{eq:comSEA_partial_unitary_op_on_rho}) to a function of $F(\rho)$ as locally perceived by A represented, according to definition Eq. (\ref{eq:comSEA_local_perception_operators}), by its partial trace weighted  with respect to $\rho_B$,
\begin{equation}\label{eq:comSEA_no_signaling_on_f5}
  (F(\rho))^A = \Tr_B[(I_A\otimes\rho_B)F(\rho)].
\end{equation}
A function of $\rho$ is defined from its eigenvalue decomposition by
$ F(\rho) = VF(D)V^\dagger=\sum_jF(\lambda_j)\dyad{\lambda_j}$, where  $\rho=VDV^\dagger$, $D=\sum_j\lambda_j\dyad{j}$, and $V=\sum_j\dyad{\lambda_j}{j}$.
Since unitary transformations do not alter the eigenvalues,
\begin{equation}\label{eq:comSEA_no_signaling_on_f7}
  F(\rho') = V'F(D)V'^\dagger \mbox{ where } V'=(I_A\otimes U_B)V \,,
\end{equation}
and therefore, using Eq. (\ref{eq:comSEA_no_signaling_unit_rhoB}) in the last step, we obtain
\begin{align}\label{eq:comSEA_no_signaling_on_f8}
   & (F(\rho'))^A = \Tr_B[(I_A\otimes\rho'_B)F(\rho')]\notag \\ &=\Tr_B[(I_A\otimes \rho'_B)\,(I_A\otimes U_B)VF(D)V^\dagger(I_A\otimes U_B^\dagger)]\notag\\ &=\Tr_B[(I_A\otimes U_B^\dagger\rho'_B U_B)\,VF(D)V^\dagger]\notag\\
   & =\Tr_B[(I_A\otimes \rho_B)\,F(\rho)]=(F(\rho))^A\,.
\end{align}
This confirms that local operations on B do not affect the local perception operators of A and, therefore, their proper use in nonlinear QM does not cause signaling issues.

Considering all of this, we can formally write the no-signaling condition as
\begin{equation}\label{eq:comSEA_no_signaling_condition}
  \dv{\rho_J}{t} = f(\rho_J,(C_k)^J).
\end{equation}

\section{The composite EoM}\label{sec:comSEA_EoM}
We are now ready to introduce the last but not least essential ingredient of a general composite-system nonlinear QM, namely, the system's structure-dependent expressions of the separate contribution of each subsystem to the dissipative term of the equation of motion for the overall state $\rho$. As discussed above (and clearly recognized in the early SEA literature \cite{beretta_1985_quantuma,beretta_2005_nonlinear,beretta_2010_maximum}), the composite-system nonlinear evolution should reflect explicitly the internal structure of the system, essentially by declaring which subsystems are to be prevented from nonphysical effects such as signaling,  exchange of energy, or build-up of correlations between non-interacting subsystems. In terms of the notation introduced above, the structure proposed in \cite{beretta_1985_quantuma,beretta_2010_maximum} for the dissipative term of the dynamics to be added to the usual Hamiltonian term is as follows
\begin{equation}\label{eq:comSEA_SEA_Composite_simp}
  \dv{\rho}{t} = -\frac{\imi}{\hbar}\comm{\mathit{H}}{\rho}-\sum_{J=1}^M\acomm{\mathcal{D}^\J_\rho}{\rho_\J}\otimes\rho_\Jbar\,,
\end{equation}
where the `local dissipation operators' $\mathcal{D}^\J_\rho$ (on $\Hil_\J$) may be nonlinear functions of the local observables of $J$, the reduced state $\rho_\J$, and the local perception operators of overall observables.  For the dissipative term to preserve $\Tr(\rho)$, operators $\acomm{\mathcal{D}^\J_\rho}{\rho_\J}$ must be traceless. To preserve $\Tr(\rho H)$ [and possibly other conserved properties $\Tr(\rho C_k)$], operators  $\acomm{\mathcal{D}^\J_\rho}{\rho_\J}(H)^\J_\rho$ [and $\acomm{\mathcal{D}^\J_\rho}{\rho_\J}(C_k)^\J_\rho$] must also be traceless. The rate of change of the overall system entropy $s(\rho)=-\Boltz\Tr[\rho\ln(\rho)]$ is
\begin{equation}\label{eq:comSEA_SEA_entropy_prod}
  \dv{s(\rho)}{t}= -\sum_{J=1}^M\Tr[\acomm{\mathcal{D}^\J_\rho}{\rho_\J}(S(\rho))^\J_\rho]\,,
\end{equation}
and the local nonlinear evolution of subsystem $J$ is obtained by partial tracing over $\Hil_\Jbar$, in general,
\begin{equation}\label{eq:comSEA_SEA_local}
  \dv{\rho_\J}{t} = -\frac{\imi}{\hbar}\comm{H_\J}{\rho_\J} -\frac{\imi}{\hbar}\Tr_\Jbar(\comm{V}{\rho}) -\acomm{\mathcal{D}^\J_\rho}{\rho_\J}\,.
\end{equation}

One can notice, for $N=1$, Eq. \eqref{eq:comSEA_SEA_Composite_simp} reduces to Eq. \eqref{eq:litsurv_SEA_withD} \cite{beretta_1985_quantuma,beretta_2009_nonlinear,ray_2023_nosignalinga}. Finally, to introduce the SEA assumption in the spirit of the fourth law of thermodynamics \cite{beretta_2020_fourth,beretta_2014_steepest}, one way is employing a variational principle. We first observe from Eq. (\ref{eq:comSEA_SEA_entropy_prod})  that the rate of entropy change contributed by subsystem $J$ is directly proportional to the norm of operator  $\mathcal{D}^\J_\rho$, so there is no maximum entropy production rate because we can trivially increase it indefinitely by simple multiplication of $\mathcal{D}^\J_\rho$ by a positive scalar. But we can fix that norm, and maximize against the direction in operator space, to identify, for each given state $\rho$, the operators $\mathcal{D}^\J_\rho$ that point in the direction of steepest entropy ascent. To this end, to recover the original SEA formulation \cite{beretta_1985_quantuma} let us maximize Eq. (\ref{eq:comSEA_SEA_entropy_prod}) subject to the conservation constraints  $\Tr[\acomm*{\mathcal{D}^\J_\rho}{\rho_\J}(C_k)^\J_\rho]=0$ where $C_1=I$, $C_2=H$, and $C_k$ are other conserved properties (if any), together with the fixed weighted norm constraints $\Tr [\rho_\J(\mathcal{D}^\J_\rho)^2]=\rm{const}$ (for more general SEA formulations in terms of a different metric as necessary to incorporate Onsager reciprocity see \cite{beretta_2014_steepest,beretta_2020_fourth}). Introducing Lagrange multipliers  $\beta^\J_k$ and $\tau_\J$ for the conservation and norm constraints, respectively, and imposing vanishing variational derivatives with respect to operators $\mathcal{D}^\J_\rho$ at fixed $\rho$ and $\rho_\J$'s (derivation details in \cite{beretta_2010_maximum,beretta_2014_steepest}, and in chapter \ref{ch:litsurv}) yields
\begin{equation}\label{eq:comSEA_DJ_Defi}
  2\tau_\J\mathcal{D}^\J_\rho =(\mathit{B}\ln(\rho))^\J_\rho+ {\textstyle \sum_\ell}\beta^\J_\ell(C_\ell)^\J_\rho.
\end{equation}
where the multipliers $\beta^\J_\ell$ must solve the system of equations obtained by substituting these  maximizing expressions of the $\mathcal{D}^\J_\rho$'s into the conservation constraints,
\begin{equation}\label{eq:comSEA_betakJ_system}
  \sum_\ell\beta^\J_\ell\Tr[\rho_\J\acomm{(C_\ell)^\J_\rho}{(C_k)^\J_\rho}]=-\Tr[\rho_\J\acomm{(\mathit{B}\ln(\rho))^\J_\rho}{(C_k)^\J_\rho}]\, .
\end{equation}
When $C_1=I$ and $C_2=H$ determine the conserved properties and Eqs. (\ref{eq:comSEA_betakJ_system}) are linearly independent, using Cramers' rule, properties of determinants, and definitions (\ref{eq:comSEA_deviation_X}) and
(\ref{eq:comSEA_covariance_XY}) we can compute $\beta_i^J$ associated with each local subsystem. We first figure out the $\Omega^J$ by extending the definition used in Eq. \eqref{eq:litsurv_betaBarOm1} and rescale it appropriately,
\begin{equation}\label{eq:comSEA_omegaJ}
  \Omega^J =  \vmqty{\tr(\tfrac{\rho_J}{2}\acomm{(\mathit{C}_1)^J}{(\mathit{C}_1)^J}) &  \tr(\tfrac{\rho_J}{2}\acomm{(\mathit{C}_1)^J}{(\mathit{C}_2)^J})\\
    \tr(\tfrac{\rho_J}{2}\acomm{(\mathit{C}_2)^J}{(\mathit{C}_1)^J}) & \tr(\tfrac{\rho_J}{2}\acomm{(\mathit{C}_2)^J}{(\mathit{C}_2)^J})}.
\end{equation}
Upon expressing this, we can consider the $\beta_k^\J$ as under,

\begin{equation}\label{eq:comSEA_betaJ1}
  \beta_1^\J =\dfrac{1}{\Omega^J}\vmqty{\tr(\tfrac{\rho_J}{2}\acomm{(\mathit{C}_1)^J}{(B\ln(\rho))^J}) &  \tr(\tfrac{\rho_J}{2}\acomm{(\mathit{C}_1)^J}{(\mathit{C}_2)^J})\\
    \tr(\tfrac{\rho_J}{2}\acomm{(\mathit{C}_2)^J}{(B\ln(\rho))^J}) & \tr(\tfrac{\rho_J}{2}\acomm{(\mathit{C}_2)^J}{(\mathit{C}_2)^J})},
\end{equation}
\begin{equation}\label{eq:comSEA_betaJ2}
  \beta_2^J= \dfrac{1}{\Omega^J}\vmqty{\tr(\tfrac{\rho_J}{2}\acomm{(\mathit{C}_1)^J}{(B\ln(\rho))^J}) &\tr(\tfrac{\rho_J}{2}\acomm{(\mathit{C}_1)^J}{(\mathit{C}_1)^J})\\
    \tr(\tfrac{\rho_J}{2}\acomm{(\mathit{C}_2)^J}{(B\ln(\rho))^J}) &\tr(\tfrac{\rho_J}{2}\acomm{(\mathit{C}_2)^J}{(\mathit{C}_1)^J})}.
\end{equation}
After finding these $\beta_k^\J$s we are essentially done with the construction of EoM for composite SEA. The local dissipation operators can be explicitly written as
\begin{align}\label{eq:comSEA_DJ_explicit}
  \begin{split}
    &\acomm{\mathcal{D}_\rho^\J}{\rho_J} =\\
    & \dfrac{1}{2\tau_J}\tfrac{\vmqty{\rho_J(B\ln(\rho))^J & \half\acomm{(\mathit{C}_1)^J}{\rho_J} &  \half\acomm{(\mathit{C}_2)^J}{\rho_J}\\
        \tr(\tfrac{\rho_J}{2}\acomm{(\mathit{C}_1)^J}{(B\ln(\rho))^J}) & \tr(\rho_J((\mathit{C}_1)^J)^2) &  \tr(\tfrac{\rho_J}{2}\acomm{(\mathit{C}_1)^J}{(\mathit{C}_2)^J})\\ \tr(\tfrac{\rho_J}{2}\acomm{(\mathit{C}_2)^J}{(B\ln(\rho))^J}) &\tr(\tfrac{\rho_J}{2}\acomm{(\mathit{C}_2)^J}{(\mathit{C}_1)^J}) & \tr(\rho_J((\mathit{C}_2)^J)^2)}}{\vmqty{\tr(\tfrac{\rho_J}{2}\acomm{(\mathit{C}_1)^J}{(\mathit{C}_1)^J}) &  \tr(\tfrac{\rho_J}{2}\acomm{(\mathit{C}_1)^J}{(\mathit{C}_2)^J})\\
        \tr(\tfrac{\rho_J}{2}\acomm{(\mathit{C}_2)^J}{(\mathit{C}_1)^J}) & \tr(\tfrac{\rho_J}{2}\acomm{(\mathit{C}_2)^J}{(\mathit{C}_2)^J})}}.
  \end{split}
\end{align}

\section{SEA in a two-qubit composite}\label{sec:comSEA_composite}
Gisin \emph{et. al.,} showed that although stochastic formalism using Lindbladian formalism can accommodate nonlinearity while respecting no-signaling \cite{gisin_1995_relevant}, the formalism itself is riddled with many conceptual shortcomings (see Section (\ref{sec:litsurv_decoherence}) for a short survey of the same). Fererro \emph{et. al.,} showed that a nonlinear formalism can be accommodated only if the said characteristic exists in the time evolution of the quantum system and not in the state space or in the structure of the operators \cite{ferrero_2004_nonlinear}. A few years later it was shown in Refs. \cite{rembielinski_2020_nonlinear,rembielinski_2021_nonlinear} that the least nonlinearity that can be accepted in the QM formalism is if it respects the convex quasilinear mapping. In this section, we write the SEA EoM for two-qubit composites in the form of noninteracting separable, maximally-entangled, and Werner state cases. In the process, we will see Eq. \eqref{eq:comSEA_no_signaling_condition} is being respected \cite{ray_2023_nosignalinga}.
\subsection{Noninteracting separable two-qubit composite}\label{secsub:comSEA_separable}
\subsubsection{A. The case with $\rho=\rho_A\otimes\rho_B$:}
Let us consider a bipartite system, whose Hilbert space is given as $\mathcal{H}=\mathcal{H}_A\otimes\mathcal{H}_B$. We call the system non-interacting if,
\begin{equation}\label{eq:comSEA_non_interacting}
  \mathit{C}_m   = (\mathit{C}_m)^A\otimes\mathit{I}_B+\mathit{I}_A\otimes(\mathit{C}_m)^B,
\end{equation}
and uncorrelated if $\rho=\rho_A\otimes\rho_B$ so that,
\begin{equation}\label{eq:comSEA_un_correlated}
  \mathit{S}(\rho)  = \mathbf{s}(\rho)_A\otimes\mathit{I}_B+\mathit{I}_A\otimes\mathbf{s}(\rho)_B.
\end{equation}
We have used $\mathit{S}(\rho)= \mathit{B}\ln(\rho)$, and $\mathbf{s}(\rho)_J = \mathit{B}^J\ln(\rho_J)$ \cite{beretta_1985_quantuma}. Using these, we can write the Eq. \eqref{eq:comSEA_SEA_Composite_simp} as,
\begin{equation}\label{eq:comSEA_sepBipartite}
  \dv{\rho}{t} = -\imi\comm{\mathit{H}}{\rho}-\acomm{\mathcal{D}^A}{\rho_A}\otimes\rho_B-\rho_A\otimes\acomm{\mathcal{D}^B}{\rho_B}.
\end{equation}
Which implies \cite{beretta_1985_quantuma},
\begin{align}
  \PTr{B}{\dv{\rho}{t}} = \dv{\rho_A}{t} & = -\imi\comm{\mathit{H}_A}{\rho_A}-\acomm{\mathcal{D}^A}{\rho_A},\label{eq:comSEA_PartialA} \\
  \PTr{A}{\dv{\rho}{t}} = \dv{\rho_B}{t} & = -\imi\comm{\mathit{H}_B}{\rho_B}-\acomm{\mathcal{D}^B}{\rho_B}\label{eq:comSEA_PartialB}.
\end{align}
We see for the strong separability \cite{beretta_2005_nonlinear} considered here, each subsystem evolves as if it were strictly isolated, and its evolution equation reduces to the form of Eq. \eqref{eq:litsurv_SEA_nice} for an indivisible system. The definition of no-signaling as in Eq. \eqref{eq:comSEA_no_signaling_condition} is also satisfied trivially.

\subsubsection{B. The case with $\rho=\sum_ip_i\rho_A^i\otimes\rho_B^i$ for $p_i\ge 0$:}
Consider the following separable density matrix,
\begin{equation}\label{eq:comSEA_mixed_density_matrix}
  \rho_m = \mu(\rho_A\otimes\rho_B)+\dfrac{1-\mu}{4}\mathit{I}_4,
\end{equation}
achieved by introducing white noise to the separable state $\rho_A\otimes\rho_B$ and a mixing parameter $0\le\mu\le 1$. We can use the general Bloch sphere representation for $N$ level systems using generators ($\left\lbrace \Gamma_i\right\rbrace$) of the SU$(N)$ group as given in previous chapter, Eq. \eqref{eq:bloch_bloch_Nd} \cite{park_1971_general,band_1971_general,bruning_2012_parametrizations}. For the case of a composite two-qubit system, the density matrix can be represented by points in a 15-dimensional sphere of unit radius. However, the mapping is not bijective, as not every point in such a sphere represents a valid density matrix. Hence, treating the problem generally is quite difficult. We restrict our attention to the mixed states resulting in the eigenvalues of the four-level system being of the form $\lambda_1 = \lambda_2= \dfrac{1-\sqrt{3}r}{4}$, and $\lambda_3= \lambda_4 = \dfrac{1+\sqrt{3}r}{4}$ as in Eq. \eqref{eq:bloch_class_degen_2} with $\alpha_1=\sqrt{3}$. We can expand the expression of $\rho_m$ in the following manner
\begin{equation}\label{eq:comSEA_mixed_state}
  \rho_m = \dfrac{1}{4}\Big[\mathit{I}_4+\mu\Big((\bm{r}_A\cdot\bm{\sigma}_A)\otimes\mathit{I}_2+\mathit{I}_2\otimes(\bm{r}_B\cdot\bm{\sigma}_B)+(\bm{r}_A\cdot\bm{\sigma}_A)\otimes(\bm{r}_B\cdot\bm{\sigma}_B)\Big)\Big]
\end{equation}
The states as expressed above are contained in $\rho_4$ of Eq. \eqref{eq:bloch_bloch_4d} with eigenvalues given by Eq.\eqref{eq:bloch_quartic_roots} if we consider $r_{J,3}=0$. We also consider $\rho_B = \half\mathit{I}_2$. So we make the following assumptions to ease our computation. The non-zero components of the Bloch vector giving real eigenvalues of the form considered here are $r_5 = r_8 = \dfrac{\tbar{r}_{A,1}}{\sqrt{6}}$, and $r_{11} = r_{14} = \dfrac{\tbar{r}_{A,2}}{\sqrt{6}}$, where, $\tbar{r}_{A,i}$ is the $i^\text{th}$ component of the qubit-Bloch vector of the system $A$ with $\tbar{\bm{r}}_A = \mu\bm{r}_A$. Without any loss in generality, for this computation, we have considered $r_{A,3}=0$. And we have $r_B =0$ by construction. This implies, we get $r = \sqrt{2\dfrac{\tbar{r}_{A,1}^2}{6}+2\dfrac{\tbar{r}_{A,2}^2}{6}} = \dfrac{\tbar{r}_A}{\sqrt{3}}$. $\rho$ and in extension $\rho_m$ will have support on all eigenvalues if $r<\dfrac{1}{\sqrt{3}}$ implying $\tbar{r}_A <1$.

As seen in the expression of Eq. \eqref{eq:comSEA_DJ_explicit}, we need to compute various trace functions as well as commutation and anti-commutation relations. To perform this, we make use of the operator formalism introduced in Eq. \eqref{eq:bloch_class_degen_Frho} and find the following relations. We e compute the operator $\ln(\rho_m)$ as under,
\begin{align}
  \begin{split}
    \ln(\rho_m) & = \dfrac{1}{4}\left[\ln(\lambda_1)+\ln(\lambda_2)+\ln(\lambda_3)+\ln(\lambda_4)\right]\mathit{I}_4\\
    &+\dfrac{\sqrt{6}}{4\sqrt{3}r}\left[\ln(\lambda_4)+\ln(\lambda_3)-\ln(\lambda_2)-\ln(\lambda_1)\right]\bm{r}\cdot\bm{\Gamma},
  \end{split}\label{eq:comSEA_log_rho_def} \\
  \begin{split}
    \ln(\rho_m) &= \half\left[\ln(\dfrac{1-3r^2}{16})\right]\mathit{I}_4+\dfrac{\sqrt{6}}{2\sqrt{3}r}\left[\ln(\dfrac{1+\sqrt{3}r}{1-\sqrt{3}r})\right]\bm{r}\cdot\bm{\Gamma}.
  \end{split}\label{eq:comSEA_log_rho}
\end{align}
It is to be noted since $\rho_m$ has support on all the eigenvalues, $B=\mathit{I}_4$. From Eq. \eqref{eq:comSEA_mixed_state}, we have $(\rho_m)_A = \half(\mathit{I}_2+\tbar{\bm{r}}_A\cdot\bm{\sigma}_A)$, with $\tbar{\bm{r}}_A=\mu\bm{r}_A$, and $(\rho_m)_B=\half\mathit{I}_2$. The operators $(\ln(\rho_m))^J$ are as given,
\begin{align}
  \begin{split}
    (\ln(\rho_m))^A &=~ \PTr{2}{(\mathit{I}\otimes(\rho_m)_2)\ln(\rho_m)},\\
    &=~\half\ln(\dfrac{1-\tbar{r}_A^2}{16})\mathit{I}_2+\dfrac{1}{2\tbar{r}_A}\ln(\dfrac{1+\tbar{r}_A}{1-\tbar{r}_A})\tbar{\bm{r}}_A\cdot\bm{\sigma}_A,
  \end{split}\label{eq:comSEA_partial1_log_rho} \\
  \begin{split}
    (\ln(\rho_m))^B &=~ \PTr{1}{((\rho_m)_1\otimes\mathit{I}_2)\ln(\rho_m)},\\
    &=~\half\left[\ln(\dfrac{1-\tbar{r}_A^2}{16})+\tbar{r}_A\ln(\dfrac{1+\tbar{r}_A}{1-\tbar{r}_A})\right]\mathit{I}_2.
  \end{split}\label{eq:comSEA_partial2_log_rho}
\end{align}
For the rest of the computation, let us use the following shorthand
\begin{align}
  \mathrm{A} & = \ln(\dfrac{1-\tbar{r}_A^2}{16})+\tbar{r}_A\ln(\dfrac{1+\tbar{r}_A}{1-\tbar{r}_A})            \\
  \mathrm{B} & = \ln(\dfrac{1-\tbar{r}_A^2}{16})+\dfrac{1}{\tbar{r}_A}\ln(\dfrac{1+\tbar{r}_A}{1-\tbar{r}_A})
\end{align}
Using, Eqs. \eqref{eq:comSEA_partial1_log_rho} - \eqref{eq:comSEA_partial2_log_rho} and expressions for reduced $(\rho_m)_J$ we can get the following expressions for locally perceived entropy operators $\left(B\ln(\rho_m)\right)^J$ for each of the subsystems
\begin{align}
  (\rho_m)_A(B\ln(\rho_m))^A & =~ \dfrac{1}{4}\mathrm{A}\mathit{I}_2+\dfrac{1}{4}\mathrm{B}\tbar{\bm{r}}_A\cdot\bm{\sigma}_A,\label{eq:comSEA_lo_entA} \\
  (\rho_m)_B(B\ln(\rho_m))^B & =~ \dfrac{1}{4}\mathrm{A}\mathit{I}_2.\label{eq:comSEA_lo_entB}
\end{align}
Now let us consider the Hamiltonian representative of noninteracting systems. Denoting, $\mathit{H}_J = \omega_J\bm{h}_J\cdot\bm{\sigma}_J$, we can write
\begin{equation}\label{eq:comSEA_hamilton_total}
  \mathit{H} = \mathit{H}_A\otimes\mathit{I}_2+\mathit{I}_2\otimes\mathit{H}_B.
\end{equation}
We express the locally perceived Hamiltonians ($(\mathit{C}_2)^J$ in the Section (\ref{sec:comSEA_EoM})) using $(r_e)_A = \bm{h}_A\cdot\tbar{\bm{r}}_A$ as,
\begin{align}
  (\mathit{H})^A & =~ \omega_A\bm{h}_A\cdot\bm{\sigma}_A,\label{eq:comSEA_hamil_partialA}                              \\
  (\mathit{H})^B & =~ \omega_A(r_e)_A\mathit{I}_2+\omega_B\bm{h}_B\cdot\bm{\sigma}_B. \label{eq:comSEA_hamil_partialB}
\end{align}
Using these, we express the following trace relations which are required to compute $\beta_i^J$ from Eqs. \eqref{eq:comSEA_omegaJ} - \eqref{eq:comSEA_betaJ2}
\begingroup
\allowdisplaybreaks
\begin{align}
   & \tr((\rho_m)_A(\mathit{H})^A)                                       =~ \omega_A(r_e)_A,\label{eq:comSEA_trace_rh_a}                        \\
   & \tr((\rho_m)_B(\mathit{H})^B)                                       =~ \omega_A(r_e)_A,\label{eq:comSEA_trace_rh_b}                        \\
   & \tr((\rho_m)_A((\mathit{H})^A)^2)                                     =~ \omega_A^2,\label{eq:comSEA_trace_rh2_a}                          \\
   & \tr((\rho_m)_B((\mathit{H})^B)^2)                                     =~ \omega_A^2(r_e)_A^2+\omega_B^2,\label{eq:comSEA_trace_rh2_b}      \\
   & \tr(\dfrac{(\rho_m)_A}{2}\acomm{(\mathit{H})^A}{(B\ln(\rho_m))^A})  =~ \dfrac{\omega_A}{2}\mathrm{B}(r_e)_A,\label{eq:comSEA_trace_rhln_a} \\
   & \tr(\dfrac{(\rho_m)_B}{2}\acomm{(\mathit{H})^B}{(B\ln(\rho_m))^B})  =~ \dfrac{\omega_A}{2}\mathrm{A}(r_e)_A,\label{eq:comSEA_trace_rhln_b} \\
   & \tr((\rho_m)_A(B\ln(\rho_m))^A)                                     =~ \dfrac{1}{2}\mathrm{A},\label{eq:comSEA_trace_rln_a}                \\
   & \tr((\rho_m)_B(B\ln(\rho_m))^B)                                     =~ \dfrac{1}{2}\mathrm{A}.\label{eq:comSEA_trace_rln_b}
\end{align}
\endgroup
Thereafter, putting these expressions into the Eqs. \eqref{eq:comSEA_omegaJ} -\eqref{eq:comSEA_betaJ2}, we get the following expressions
\begingroup
\allowdisplaybreaks
\begin{align}
  \Omega^A  & =~ \omega_A^2(1-(r_e)_A^2),\label{eq:comSEA_OmegaA}                                                  \\
  \Omega^B  & =~ \omega_B^2,\label{eq:comSEA_OmegaB}                                                               \\
  \beta_1^A & =~ \dfrac{\omega_A^2}{2\Omega^A}\left(\mathrm{A}-\mathrm{B}(r_e)_A^2\right),\label{eq:comSEA_beta1A} \\
  \beta_1^B & =~ \dfrac{\omega_B^2}{2\Omega^B}\mathrm{A},\label{eq:comSEA_beta1B}                                  \\
  \beta_2^A & =~ \dfrac{\omega_A(r_e)_A}{2\Omega^A}\left(\mathrm{A}-\mathrm{B}\right),\label{eq:comSEA_beta2A}     \\
  \beta_2^B & =~ 0.\label{eq:comSEA_beta2B}
\end{align}
\endgroup
Let us write down the expressions for the anti-commutation relations as under
\begin{align}
  \acomm{(\mathit{H})^A}{(\rho_m)_A} & =~\omega_A\left(\bm{h}_A\cdot\bm{\sigma}_A+(r_e)_A\mathit{I}_2\right),\label{eq:comSEA_acomm_rh_a}         \\
  \acomm{(\mathit{H})^B}{(\rho_m)_B} & =~\left(\omega_B\bm{h}_B\cdot\bm{\sigma}_B+\omega_A(r_e)_A\mathit{I}_2\right).\label{eq:comSEA_acomm_rh_b}
\end{align}
Therefore, using the multipliers in Eqs. \eqref{eq:comSEA_OmegaA} - \eqref{eq:comSEA_beta2B}, the anti-commutations of Eqs. \eqref{eq:comSEA_acomm_rh_a} -\eqref{eq:comSEA_acomm_rh_b} into the expression for $\mathcal{D}^J$ in Eq. \eqref{eq:comSEA_DJ_Defi} to compute $\acomm{\mathcal{D}^J}{(\rho_m)_J}$. However, since $\rho_m$ is separable, we use $\mathcal{D}^J$ and we get after some tedious algebra,
\begin{align}
  \acomm{\mathcal{D}^A}{(\rho_m)_A} & =~ \dfrac{\mathrm{A}-\mathrm{B}}{4\tau_A(1-(r_e)_A^2)}\left[(r_e)_A\bm{h}_A-\tbar{\bm{r}}_A\right]\cdot\bm{\sigma}_A,\label{eq:comSEA_acomm_dr_a} \\
  \acomm{\mathcal{D}^B}{(\rho_m)_B} & =~ 0.\label{eq:comSEA_acomm_dr_b}
\end{align}

Clearly, as is evident, here also, the reduced density dynamics in Eq. \eqref{eq:comSEA_acomm_dr_b} follow the form of Eq. \eqref{eq:comSEA_no_signaling_condition}. The EoM Eq. \eqref{eq:comSEA_acomm_dr_a} gives rise to the solution of the form in Eq. \eqref{eq:qubit_analytic_sol}, which eventually mixes the subsystem $A$, and drives it towards the local maximally mixed state, which in effect turns the composite towards the global maximal mixed state. Moreover, as it is also evident, despite the `locally perceived' operators having contributions from the other subsystems, the overall system dynamics does not contain that contribution. And signaling is prevented \cite{ray_2023_nosignalinga}. To see whether the separate energy conservation conditions are satisfied, we can show from straightforward computation, that this indeed, also holds. Similar computations involving $\rho_A = \half\mathit{I}_2$ and $\rho_B = \half\left(\mathit{I}_2+\bm{r}_B\cdot\bm{\sigma}_B\right)$ with $r_{B,3} =0$ can be carried out to reach the analogous conclusion, however, in this case, the nonzero components of the Bloch vector will be $r_4=r_9 = \dfrac{\tbar{r}_{B,1}}{\sqrt{6}}$, and $r_{10}=r_{15} = \dfrac{\tbar{r}_{B,2}}{\sqrt{6}}$. The case with $\rho_J = \half\left(\mathit{I}_2+\bm{r}_J\cdot\bm{\sigma}_J\right)$ with $r_{J,3}=0$ leads to eigenvalues of the composite of the form $\lambda_1 = \dfrac{1-\sqrt{6}r}{4}$, $\lambda_2=\lambda_3=\dfrac{1}{4}$, and $\lambda_4= \dfrac{1+\sqrt{6}r}{4}$, where the nonzero components of the Bloch vector are $r_4 = r_9 = \dfrac{\tbar{r}_{B,1}}{\sqrt{6}}$, $r_5 = r_8= \dfrac{\tbar{r}_{A,1}}{\sqrt{6}}$, $r_{10}=r_{15}=\dfrac{\tbar{r}_{B,2}}{\sqrt{6}}$, and $r_{11} = r_{14} = \dfrac{\tbar{r}_{A,2}}{\sqrt{6}}$. Carrying out similar analytical computation with eight nonzero Bloch vector components is quite non-trivial and is beyond the scope of this thesis. In what follows we will consider the maximally mixed-state to complete our discussion on finding SEA EoM for two-qubit composites.
\subsection{Noninteracting two-qubit mixed composite}\label{secsub:comSEA_mixed}
Bell diagonal states \cite{horodecki_1996_informationtheoretic,lang_2010_quantum} with Bell vector $\bm{b}=\left(b_x,b_y,b_z\right)$
\begin{align}\label{eq:comSEA_Bell_diagonal_states}
  \rho^{\mbox{\tiny Bell}} & =\dfrac{1}{4}\left[\mathrm{I}_4+\sum_{i=\{x,y,z\}}b_i\cdot\sigma_i\otimes\sigma_i\right],                             \\
                           & =\frac{1}{4}\begin{pmatrix} 1+b_z&0&0&b_x-b_y\\0&1-b_z&b_x+b_y&0\\0&b_x+b_y&1-b_z&0\\b_x-b_y&0&0&1+b_z \end{pmatrix},
\end{align}
are identified in the notation of the previous chapter by the three parameters
\begin{align}\label{eq:comSEA_Bell_diagonal_parameters}
  \begin{split}
    \alpha_1&=b_z\\
    \alpha_2&=-2b_xb_y\\
    r^2&=b^2/3 \text{ where } b^2=b_x^2+b_y^2+b_z^2\\
    4\, \lambda_1 & = 1-b_x-b_y-b_z, \\
    4\, \lambda_2 & =  1-b_x+b_y+b_z,  \\
    4\, \lambda_3 & = 1+b_x-b_y+b_z, \\
    4\, \lambda_4 & =  1+b_x+b_y-b_z. \\
  \end{split}
\end{align}
The states thus introduced can be used to classify a large class of states. These include the standard Bell states for $b_x=b_y=b_z=-1$, or $-b_x=b_y=b_z=1$. They can be used to denote Werner states \cite{werner_1989_quantum,czerwinski_2021_quantifying} for $b_x=b_y=b_z=-w$. These Bell diagonal states represent maximally entangled pure composite (the Bell states), can represent mixed entangled states (Werner states) and can represent states with color noise with support only on two eigenbases of $\rho_4$, thus providing a plethora of flavor to choose from. Proving no-signaling for general Bell diagonal states thus becomes quite useful. Since $\rho_A^{\mbox{\tiny Bell}}=\rho_B^{\mbox{\tiny Bell}}=\mathrm{I}_2/2$, we readily obtain
\begin{align}
  (\mathit{S}(\rho^{\mbox{\tiny Bell}}))^A & =~ -\dfrac{1}{2}B\mathrm{L}\mathrm{I}_2,\label{eq:comSEA_lo_entA_} \\
  (\mathit{S}(\rho^{\mbox{\tiny Bell}}))^B & =~ -\dfrac{1}{2}B\mathrm{L}\mathrm{I}_2.\label{eq:comSEA_lo_entB_}
\end{align}
where  $\mathrm{L}$ equals the logarithm of the product of the nonzero eigenvalues of $ \rho^{\mbox{\tiny Bell}}$, e.g., for a non-singular  $ \rho^{\mbox{\tiny Bell}}$,
\begin{align}
  \mathrm{L} & = \ln(\lambda_1\lambda_2\lambda_3\lambda_4).
\end{align}
Thus the expressions $\rho^{\mbox{\tiny Bell}}_J(\mathit{S}(\rho^{\mbox{\tiny Bell}}))^J$ can be evaluated as below,
\begin{align}
  (\rho^{\mbox{\tiny Bell}})_A(B\ln(\rho^{\mbox{\tiny Bell}}))^A & =~ -\dfrac{1}{4}B\mathrm{L}\mathrm{I}_2,\label{eq:comSEA_Bell_lo_entA} \\
  (\rho^{\mbox{\tiny Bell}})_B(B\ln(\rho^{\mbox{\tiny Bell}}))^B & =~ -\dfrac{1}{4}B\mathrm{L}\mathrm{I}_2.\label{eq:comSEA_Bell_lo_entB}
\end{align}
Now let us consider the Hamiltonian representative of noninteracting systems.
We find the locally perceived Hamiltonians $(\mathit{C}_2)^J$ as using Eq. \eqref{eq:comSEA_hamilton_total},
\begin{align}
  (\mathit{H})^A & =~ \mathit{H}_A,\label{eq:comSEA_Bell_hamil_partialA}  \\
  (\mathit{H})^B & =~ \mathit{H}_B. \label{eq:comSEA_Bell_hamil_partialB}
\end{align}
Using these, we express the following trace relations which are required to compute $\beta_i^J$ as
\begingroup
\allowdisplaybreaks
\begin{align}
   & \tr\left((\rho^{\mbox{\tiny Bell}})_A(\mathit{H})^A\right)                                       =~ 0,\label{eq:comSEA_Bell_trace_rh_a}                                           \\
   & \tr\left((\rho^{\mbox{\tiny Bell}})_B(\mathit{H})^B\right)                                       =~ 0,\label{eq:comSEA_Bell_trace_rh_b}                                           \\
   & \tr\left((\rho^{\mbox{\tiny Bell}})_A((\mathit{H})^A)^2\right)                                     =~ \omega_A^2,\label{eq:comSEA_Bell_trace_rh2_a}                               \\
   & \tr\left((\rho^{\mbox{\tiny Bell}})_B((\mathit{H})^B)^2\right)                                     =~ \omega_B^2,\label{eq:comSEA_Bell_trace_rh2_b}                               \\
   & \tr\left(\dfrac{(\rho^{\mbox{\tiny Bell}})_A}{2}\acomm{(\mathit{H})^A}{(B\ln(\rho^{\mbox{\tiny Bell}}))^A}\right)  =~ 0,\label{eq:comSEA_Bell_trace_rhln_a}                       \\
   & \tr\left(\dfrac{(\rho^{\mbox{\tiny Bell}})_B}{2}\acomm{(\mathit{H})^B}{(B\ln(\rho^{\mbox{\tiny Bell}}))^B}\right)  =~ 0,\label{eq:comSEA_Bell_trace_rhln_b}                       \\
   & \tr\left((\rho^{\mbox{\tiny Bell}})_A(B\ln(\rho^{\mbox{\tiny Bell}}))^A\right)                                     =~ -\dfrac{1}{2}B\mathrm{L},\label{eq:comSEA_Bell_trace_rln_a} \\
   & \tr\left((\rho^{\mbox{\tiny Bell}})_B(B\ln(\rho^{\mbox{\tiny Bell}}))^B\right)                                     =~ -\dfrac{1}{2}B\mathrm{L}.\label{eq:comSEA_Bell_trace_rln_b}
\end{align}
\endgroup
Thereafter, putting these expressions into the defining Eqs. (\ref{eq:comSEA_omegaJ}-\ref{eq:comSEA_betaJ2}), we get the following expressions
\begingroup
\allowdisplaybreaks
\begin{align}
  \Omega^A  & =~ \omega_A^2,\label{eq:comSEA_Bell_OmegaA}               \\
  \Omega^B  & =~ \omega_B^2,\label{eq:comSEA_Bell_OmegaB}               \\
  \beta_1^A & =~ -\dfrac{1}{2}B\mathrm{L},\label{eq:comSEA_Bell_beta1A} \\
  \beta_1^B & =~ -\dfrac{1}{2}B\mathrm{L},\label{eq:comSEA_Bell_beta1B} \\
  \beta_2^A & =~ 0,\label{eq:comSEA_Bell_beta2A}                        \\
  \beta_2^B & =~ 0.\label{eq:comSEA_Bell_beta2B}
\end{align}
\endgroup
Let us write down the expressions for the anti-commutation relations as under
\begin{align}
  \acomm{(\mathit{H})^A}{(\rho^{\mbox{\tiny Bell}})_A} & =~\omega_A\bm{h}_A\cdot\bm{\sigma}_A,\label{eq:comSEA_Bell_acomm_rh_a} \\
  \acomm{(\mathit{H})^B}{(\rho^{\mbox{\tiny Bell}})_B} & =~\omega_B\bm{h}_B\cdot\bm{\sigma}_B.\label{eq:comSEA_Bell_acomm_rh_b}
\end{align}
Therefore, using the multipliers in Eqs. (\ref{eq:comSEA_Bell_OmegaA}-\ref{eq:comSEA_Bell_beta2B}), the anti-commutations of Eqs. (\ref{eq:comSEA_Bell_acomm_rh_a}-\ref{eq:comSEA_Bell_acomm_rh_b}) into the expression for $\mathcal{D}^J$ in Eq. (\ref{eq:comSEA_DJ_Defi}) to compute $\acomm{\mathcal{D}^J}{(\rho^{\mbox{\tiny Bell}})_J}$. However, since $\rho^{\mbox{\tiny Bell}}$ is nonseparable, we use $\mathcal{F}^J$ instead of $\mathcal{D}^J$ and we get after some algebra,
\begin{align}
  \acomm{\mathcal{F}^A}{(\rho^{\mbox{\tiny Bell}})_A} & =~ 0,\label{eq:comSEA_Bell_acomm_dr_a} \\
  \acomm{\mathcal{F}^B}{(\rho^{\mbox{\tiny Bell}})_B} & =~ 0.\label{eq:comSEA_Bell_acomm_dr_b}
\end{align}
These lead us to compute the expression for separate energy conservation as required by no-signaling. We have $\mathcal{D}= \acomm{\mathcal{F}^A}{(\rho^{\mbox{\tiny Bell}})_A}\otimes(\rho^{\mbox{\tiny Bell}})_B+(\rho^{\mbox{\tiny Bell}})_A\otimes\acomm{\mathcal{F}^B}{(\rho^{\mbox{\tiny Bell}})_B} $. For sub-system $A$ using Eqs. (\ref{eq:comSEA_Bell_hamil_partialA}-\ref{eq:comSEA_Bell_hamil_partialB}), and (\ref{eq:comSEA_Bell_acomm_dr_a}-\ref{eq:comSEA_Bell_acomm_dr_b}),
\begin{equation}
  \left((\mathit{H})^A\otimes\mathrm{I}_2\right)\mathcal{D} = 0.
\end{equation}
The same is trivially true for subsystem B. We see that the solution for the Bell diagonal states is non-dissipative and energy-conserving. Thus, the SEA is no-signaling for all these cases \cite{ray_2023_nosignalinga}.

\section*{Summary}
In brief, we have discussed the concept of signaling in the framework of non-linearity in quantum mechanics. We presented Gisin's argument. We further discussed the concept of no-signaling in the context of SEA and showed, how by relaxing and broadening the definition of no-signaling, we can accommodate a larger class of non-linear formalism than previously anticipated. We then used this definition to construct the SEA EoM for a composite system. Finally, we considered the cases of separable and non-separable two-qubit composites to write the EoM and show that SEA respects no-signaling.\conditionalBlankPage

\chapter{Conclusions and Future Scope}\label{ch:conclusion}

\blankpage
\newpage
\thispagestyle{empty}
\begin{figure}[H]
    \centering
    \captionsetup{justification=centering}

    \includegraphics[height=0.75\textheight, width=0.75\textwidth]{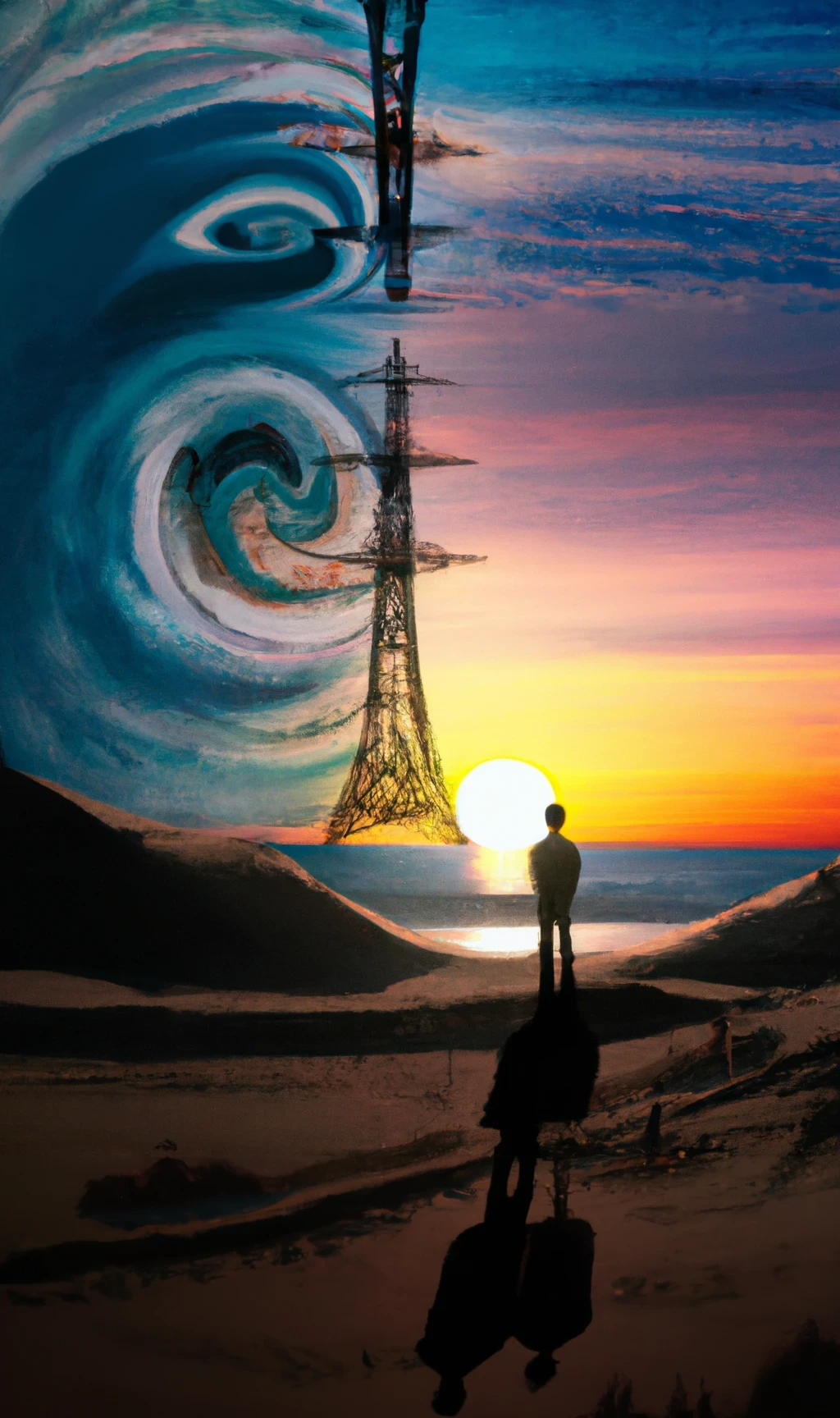}
    \caption*{\emph{I reappear by that sea,}\\
        \emph{with the soothing waves beneath me,}\\
        \emph{my shadows embed my new reality,}\\
        \emph{preapred for new learnings.}}
\end{figure}
\newpage
\blankpage
\lettrine[lines=3]{T}{his} is where we conclude our story. We have explored some of the intricacies of the SEA theory as the fourth law of thermodynamics and its application to finite-level quantum systems. And now we must present a summary of the key aspects that we have garnered from this discussion. In this concluding chapter, we present our findings. Then we will comment on the limitations of the thesis and present some ideas which represent the future scope of this work.

\section{In conlcusion}\label{sec:conclusion_conclusion}
We have begun this thesis by discussing the motivation, foundation, and derivation of the BSEA. There in chapter \ref{ch:litsurv}, we saw the underlying geometric structure of the theory, and we understood the crucial role played by the concept of stability in the formalism presented in this thesis. This chapter serves both as a review of the literature and as an introduction to the tenets of SEA and thus plays an important role in the development of the thesis.

In the vein of the above, one must also address the cases of mesoscopic system evolution, and local entropy decrease observations. As it can be argued that while the steepest entropy ascent formalism imposes a non-negative rate of entropy generation, there are cases in nature that behave to the contrary, at least for a short amount of time. So, it may be said that from this point of view, SEA is not a general law of dynamics. However, it can be also argued that considering a quantum observable - upon making a single measurement and we get an outcome, which may be above or below the mean value. On repeated measurements on an ensemble of identically prepared systems in the same state, we get statistics of outcomes. The mean value may be positive, but some parts of the statistical distribution of the outcomes may well be in the negative. In that state, that observable has some uncertainty. There is no surprise, of course, in any state there are always observables that have some uncertainty. Now assume that this observable happens to be such that its mean value is the rate of change of the entropy functional. Since it is the meanvalue that is always positive, we can say this is tantamount to claiming the rate of entropy production is always positive. And thus the generality of SEA formalism can be retained.

Following this, chapter \ref{ch:qubitSEA} presents the first finite-level quantum system, the trivial case of a qubit or a two-level system to be studied under the SEA formalism. We presented the exact analytical result due to Beretta in this chapter. However, the important role played by this chapter lies in the development and application of our key approximation scheme, the FLM method. We deployed FLM for the qubit case and showed that depending on the choice of $\rho$, the scheme has the capacity to provide a good approximation to the exact result. A justification for such a method is that $\beta_i$'s are not always rapidly changing, and it is sometimes more intriguing to know the nature of a dynamics rather than an accurate depiction of the same. FLM also simplifies SEA nonlinearity which makes the theory easily tractable and the EoM looks more appealing. In this chapter, using Bloch representation we were also able to demonstrate the concept of `steepest entropy ascent' by showing the rise of the trajectory in the state space over the entropy hill, which is an interesting by-product of our analysis.

But aesthetics and ease of access aside, FLM must provide some real benefit over the full numerical solution, otherwise one may justly question the point of using such a scheme. To answer this, we present chapter \ref{ch:qwalker} to the readers. where we see FLM in its full glory. We find out that FLM is not only a fairly good approximation for higher dimensional systems but also it is very reliable in representing the key aspects of the dynamics. Our plots of comparison of probability amplitude for a CTQW on a cycle graph with the full numerical solution are testaments to our claim. Not only that, a similar comparison of results for studying entropy and entropy generation confirms that FLM works fairly well in initial and later times during the dynamics too.\\
There is another set of conclusions to be made from this chapter. The use of CTQW to model SEA evolution for a $N-$level system, or the modeling of decoherence in CTQW using SEA has been a novel approach that opens up a plethora of research avenues. We have for the first time used a first-principle approach to decoherence in the study of CTQW. This leads to the introduction of SEA in the arena of quantum computation and quantum information science. For a given \ac{tau}, one can now analytically study mixing in \ac{QW}, without resorting to studying the limiting value of the time-averaged decay of unitarily evolving amplitudes, or usage of master equations of the Lindbladian type. We have seen through the dependence of the diagonal elements of the Graph Hamiltonian in the FLM solution that the decay to uniform distribution is highly reliant on the degree of the graph, which implies dependence on the dimensionality of the lattice on which the walk is being performed. Also, we see that $\tau$ can be used to speed up the decay or to slow it down, taking forever to decay in the case of large values.  Our results show that by properly tuning $\tau$, one can swing between localization and complete delocalization. \\
Through the results of the chapters \ref{ch:qubitSEA} and \ref{ch:qwalker}, we fulfill the first two objectives of this thesis as discussed in the section \ref{sec:intro_motiv}.

Motivated by the success of FLM in studying single finite-level systems under SEA evolution, it is only natural that we would want to extend our formalism to study composite system evolution. However, as was the case presented by Beretta in the case of a qubit, we wanted to figure out a way to analytically study the simplest of composite, namely the two-qubit composite. The results in chapter \ref{ch:bloch_SEA} serve a very important purpose in this regard. We have attempted to find analytical root expressions for Bloch representation for levels $N=3$ and $N=4$. These results are important on their own and can be used outside the context of this thesis. It is very well understood how the analytical roots of $N=2$ help us characterize the Bloch vector representation of a qubit. Similarly, the geometry of Bloch vector representation for a qutrit or a qudit with $d=4$ had been studied in the literature, but not having a parametric form for finding analytical roots was a handicap to the analytical studies of the density matrices evolution in those cases. Our results are well suited for that purpose. Consider the qudit case, for example. We have found the roots in the form of three parameters ($r$, $\alpha_1$, and $\alpha_2$). By appropriately tuning these, one can get all the roots of the case of a four-level system. This result in particular opens up the possibility of numerous studies similar to what the qubit case has done. \\
Our second contribution from this chapter is in finding the analytical expressions for the behavior of $F(\rho)$. In our approach, we have exploited the degeneracy involved in the roots. The formalism allows us to find the analytical expression for the real functions be it $\rho^2$, $\ln(\rho)$, or $\exp(\rho)$ without being bothered about the eigenvector expressions. Moreover, this formalism allows us to take analytical traces over these functions, as the expressions are decomposed into trace-less and non-zero trace parts, which can be easily computed. Besides, our formalism allows us to take the products of the form $\rho\ln(\rho)$ easily by exploiting the relations between the generalized Gell-Mann matrices, which is an added advantage.

As discussed as objective 3 in the section \ref{sec:intro_motiv}, we do stumble upon a philosophical problem of composite system studies via SEA. As it is oft-mentioned in the literature, a general nonlinear theory of \ac{QM} signals. Thus, it became imperative that we address this issue with SEA and check that it does not signal. We thus scrutinize the construction of the SEA EoM and especially \ac{BCSEA}. We begin by understanding the conventional definition of no-signaling, and then show that while it is necessary, this definition is not sufficient. We relax the no-signaling criteria to accommodate a larger class of nonlinear evolution under which the local operation performed in one of the noninteracting subsystems does not affect the outcome of the other. We show that by construction, SEA is no-signaling. In the process, we cast some more light on the details of the composite EoM formalism hitherto missing from the relevant literature. We then proceed to consider some examples. Which are also novel results in regard to SEA. We study the trivial separable case and a nontrivial one. We explicitly show for the second case how via the application of perceived functionals and their maximization, the separable states are locally driven to maximally mixed states. We further considered the case of non-separable correlated composites. We present some novel results concerning the use of Bell diagonal states under SEA and show how they form non-dissipative limit cycles in the local evolution under SEA. Our results conclusively show that SEA is no-signaling, and establish that SEA and theories of the same class are no-signaling.\\
Thus we complete the three main objectives of this thesis and in the process as by-products get some helpful results of the Bloch representation.

\section{Limitations and future scope}\label{sec:conclusion_lim_scope}
Nothing is perfect, neither is this thesis. There are some serious limitations of this thesis that have existed because of temporal bound or resource bound on our part. It is also true, even if we addressed those issues some other thing would have been mentioned in this section. So without further ado, we state the following.

Firstly, chapter \ref{ch:litsurv} does not do service to the recent works in the SEA literature. Especially the works by von Spakovsky and group is left undiscussed. The main reason for such exclusion is that those works are beyond the scope of this thesis, which is why we have mentioned them in section \ref{sec:intro_second_age} but not discussed in detail. However, those results are quite intriguing, especially von Spaovsky's work on the typicality connection \cite{vonspakovsky_2014_trends}. It is one of the future scope of this thesis, as the results stated here can be applied to compare the outcomes of fluctuation theorem-related results.

In chapter \ref{ch:qubitSEA}, we could have considered more cases and provided a non-equatorial view of the evolution also. However, since Beretta's original exact solution pertained to that region, for ease of comparison, we also restrained ourselves in the same domain.

In chapter \ref{ch:qwalker}, we believe we could have considered various graphs and not only the cycle graph. In fact, we performed numerical analysis on the hypercube graph and found similar results. We have retained those results from the discussion because we think it would be prudent to present the same in the context of more general graphs. This is also why we could not discuss mixing in detail here.

Our work in chapter \ref{ch:bloch_SEA} could have been augmented by some discussion on the geometric aspects of the representations given that we already presented some in the context of SEA interpretation. However, being restricted by time and resources, we thought it would be better to leave the discussion of the roots and their relation to the geometry of these higher-dimensional Bloch spheres for future work.

Finally, in chapter \ref{ch:comSEA} we could not actually show the analytical results for the composites and had to restrain ourselves within the discussion of the EoMs only. We were seriously restricted by the limitation of time, and the computations of chapters \ref{ch:bloch_SEA} along with the current one took more time than we had allotted ourselves to. However, these works have presented us with the following ideas that we wish to move forward with in detail study.

\begin{enumerate}
    \item The application of SEA in CTQW has provided a gateway to the domain of quantum information and computation sciences. The effect of decoherence in spatial search applications, mixing of walks, and application of SEA in general random graphs such as Erd\"{o}s-Renyi random graphs.
    \item Our results of Bloch representation can be applied to study the two-qubit composite evolution analytically, along with studying the rich geometry of those higher-dimensional objects under similar evolutions. We wish to perform similar studies as we did in qubit evolution through the Bloch sphere.
    \item Finally, the EoM for the composite opens up a plethora of research opportunities. As discussed above, we wish to analytically solve the two-qubit cases using the EoMs derived in this thesis.
    \item We can use the BCSEA to solve many body QWs. And may also enquire whether FLM or similar approximations can be used there as well.
    \item The BCSEA, along with mean-field approximations can be used to attempt to find the equation for spontaneous decoherence in larger quantum systems, for example, the \ac{bec}.
\end{enumerate}\conditionalBlankPage
\onehalfspacing
\printbibliography
\end{document}